\documentclass[review]{elsarticle}
\usepackage[margin=2.0cm]{geometry}
\usepackage{lineno,hyperref}
\usepackage{subfig}
\usepackage{longtable}
\usepackage{caption}
\usepackage{hyperref}
\usepackage{enumitem}
\usepackage{xcolor}
\usepackage{multirow}
\usepackage{subfloat}
\usepackage[linesnumbered,ruled,vlined]{algorithm2e}

\usepackage{float}
\usepackage{amsmath}
\usepackage{setspace}
\usepackage{pdflscape} 
\usepackage{geometry} 

\journal{Journal of \LaTeX\ Templates}
\makeatletter \def\ps@pprintTitle{  \let\@oddhead\@empty  \let\@evenhead\@empty  \def\@oddfoot{\hfill\thepage}  \def\@evenfoot{\thepage\hfill}} \makeatother
\begin{document}

\begin{frontmatter}

\title{A Dual-Tier Adaptive One-Class Classification IDS for Emerging Cyberthreats}


\author[1]{Md. Ashraf Uddin}\ead{ashraf.uddin@deakin.edu.au}
\author[1]{Sunil Aryal}\ead{sunil.aryal@deakin.edu.au}  
\author[1]{Mohamed Reda Bouadjenek}
\author[1]{Muna Al-Hawawreh}
\author[2]{Md. Alamin Talukder} \ead{alamin.cse@iubat.edu}

\address[1]{School of Information Technology, Deakin University, Geelong, VIC 3125, Australia}
\address[2]{Department of Computer Science and Engineering, International University of Business Agriculture and Technology, Dhaka, Bangladesh}

\cortext[mycorrespondingauthor]{Corresponding authors: Sunil Aryal and Md Ashraf Uddin}

\begin{abstract}
In today's digital age, our dependence on IoT (Internet of Things) and IIoT (Industrial IoT) systems has grown immensely, which facilitates sensitive activities such as banking transactions and personal, enterprise data, and legal document exchanges. Cyberattackers consistently exploit weak security measures and tools. The Network Intrusion Detection System (IDS) acts as a primary tool against such cyber threats. However, machine learning-based IDSs, when trained on specific attack patterns, often misclassify new emerging cyberattacks. Further, the limited availability of attack instances for training a supervised learner and the ever-evolving nature of cyber threats further complicate the matter. This emphasizes the need for an adaptable IDS framework capable of recognizing and learning from unfamiliar/unseen attacks over time. In this research, we propose a one-class classification-driven IDS system structured on two tiers. The first tier distinguishes between normal activities and attacks/threats, while the second tier determines if the detected attack is known or unknown. Within this second tier, we also embed a multi-classification mechanism coupled with a clustering algorithm. This model not only identifies unseen attacks but also uses them for retraining them by clustering unseen attacks. This enables our model to be future-proofed, capable of evolving with emerging threat patterns. Leveraging one-class classifiers (OCC) at the first level, our approach bypasses the need for attack samples, addressing data imbalance and zero-day attack concerns and OCC at the second level can effectively separate unknown attacks from the known attacks. Our methodology and evaluations indicate that the presented framework exhibits promising potential for real-world deployments. 

\end{abstract}


\end{frontmatter}

\section{Introduction}
\label{Introduction}

The growing reliance on IoT/IIoT (Industrial IoT) networks has inadvertently increased the likelihood of new cyberattack occurrences. Conventional  IDS struggle to effectively detect modern threats due to their limited ability to adapt and counter the sophisticated techniques used in these attacks.  
Development of a practical IDS for critical applications is still challenging despite the considerable amount of research conducted in this field. Most existing IDS systems do not have the capability of detecting the continuously emerging new security threats and retraining the model using the novel attack samples \cite{roshan2018adaptive}. The dynamic nature of cyberattacks necessitates regular updates to IDS to effectively detect and respond to emerging attack patterns. The detection of new malicious attacks has garnered significant interest among researchers. Contemporary IDSs are not effective in detecting unseen attacks as those IDSs generate a significant number of false negatives  \cite{folino2016ensemble, hubballi2014false}.

Attackers frequently evade a company's security measures by exploiting vulnerabilities that have not been disclosed or seen by security personnel. These kinds of attacks are termed as zero-day vulnerabilities, a vendor's undetected or unaddressed threat for software or applications. Bilge et al. \cite{bilge2012before} presented the substantial impacts of zero-day attacks on both consumers and systems. Their research revealed that organizations encounter considerable challenges in detecting zero-day attacks, with an average detection time ranging from 312 days to as long as 30 months. This indicates the critical need for the development of IDS capable of promptly identifying and mitigating such attacks as soon as they attack the system. Most traditional security methods utilize either signatures or machine learning models trained for predefined normal and attack sample patterns to detect vulnerabilities. However, these methods are ineffective in identifying zero-day vulnerabilities due to the uncertainty of signature data. Detecting zero-day exploits using conventional IDS is highly challenging \cite{joshi2016vibration} because such vulnerability could potentially remain undisclosed to the general public for an extended period, ranging from several months to several years, owing to the advanced capabilities of the attackers \cite{yang2016improving}.

 To improve the performance of IDS by minimizing false-negative rates, it is important for an IDS to effectively and accurately detect new/zero-day attacks. This can be achieved by regularly detecting unknown attacks and updating IDS in real-time \cite{hossain2003adaptive}. Regarding this, Masdari and Khezri \cite{masdari2020survey} highlighted the importance of adaptability in an IDS for the successful detection of novel threats. An Adaptive IDS refers to a classification model that is dynamically updated to identify emerging attack instances.
 
The precondition of building an adaptive model is to correctly detect unseen/new attacks and label them for retraining the model. Traditional IDS typically identify new attacks offline through manual or semi-automated processes \cite{joseph2010carrads}. This task necessitates significant exertion from experts. Automated models are faster and more efficient compared to manually coded models. An adaptive model facilitates collecting and promptly integrating into current detection models to swiftly mitigate potential harm caused by such previously unseen threats. In recent years, there has been an increased focus on the utilization of data mining and machine learning methods for building intrusion detection models \cite{talukder2024securing, talukder2024mlstl}. The models are developed based on normal behavior and known threats to detect previously unseen threats. In this process, we first require to distinguish attack samples from the normal samples. Next, unseen/unknown attacks are collected and labeled for training a multi-classifier model so that it can identify such attacks next time. Many modern IDS systems require the availability of extensive, balanced datasets containing both normal and attack instances as they utilize mostly supervised machine learning models. In real-world scenarios, procuring a substantial number of attack instances is difficult, given that malicious network traffic doesn't occur at the same frequency as regular traffic. In addition, attack patterns evolve swiftly, meaning a supervised model, trained on a specific attack distribution, might fail in recognizing novel or zero-day attacks \cite{talukder2024machine, talukder2023dependable}. To address this issue, Some existing studies  \cite{bezerra2019iotds, fahad2017applying, anand2023efficient, dini2022design} have proposed one-class classification (OCC) methods that train exclusively on normal instances, identifying attack instances as anomalies. Our findings show that this strategy efficiently differentiates between normal and attack instances with higher accuracy. However, when it comes to pinpointing specific attack families—like DoS, Ransomware, Trojan Horse, and Spywar, existing solutions seek multi-class supervised learning. Such models can only identify attack types they've been trained on, leading them to misclassify new attacks as known variants. One potential solution involves examining the prediction probability generated by a multi-class classifier when it assesses an instance's likelihood of belonging to a known attack type. By setting a threshold probability, we might be able to distinguish between known and unknown attacks. Yet, determining an optimal threshold remains a challenge since multi-classifiers often assign a high probability to one of the known attack categories even for unseen attack instances.

In the domain of IDS, a semi-supervised learning technique called a one-class classifier (OCC) is intended to find patterns in data that deviate from the normal behaviour of traffic instances \cite{al2017real}. An OCC is trained just on data that represents regular behaviour, as opposed to typical classifiers that use both normal and malicious instances to categorize new data points. Because of this, it is especially well-suited for identifying hidden attacks, or zero-day or unique threats that do not correspond with any known attack signatures.

OCC like OCSVM (One Class SVM) has been applied in some recent works  \cite{al2017real, singh2019framework, al2020unknown, hindy2020utilising} to detect attack and normal instances. Nevertheless, a limited number of studies  \cite{al2017real, hindy2020utilising} investigated the SVM or OCSVM model to distinguish unseen attacks from known attack types. While it is more usual in the literature to utilise the OCC model to distinguish between unseen attacks and regular network traffic, there has been very little research done in the literature to distinguish unseen attacks from seen attack categories using either the supervised learner or semi-supervised learning. For security administration purposes, it is necessary to examine the effectiveness of the most recently developed OCC models, such as usfAD, in distinguishing unseen attacks from seen attack categories. It is also necessary to accurately identify the family types of cyberattacks. This can enable administrators for the appropriate respond appropriately to emerging attack types. 

Most of the machine learning-based literature has primarily focused on distinguishing normal network traffic from attacks but has not prioritized the categorization of specific attack families or the detection of unseen attacks. Consequently, when employing these models, it becomes challenging to generate comprehensive reports on the nature of the attacks. Such reports could provide valuable insights for security administrators to formulate and enforce effective security policies.  Another limitation of this approach is its inability to differentiate the behaviour of novel attacks from benign or previously known attack categories.

Research in IDS has partially addressed the challenge of unknown network threats. Some studies  \cite{bezerra2019iotds, fahad2017applying, anand2023efficient, dini2022design, hindy2020utilising, al2020unknown} have recognized these threats but haven't provided methods for systems to learn and adapt to them. Others \cite{soltani2023adaptable, roshan2018adaptive} have outlined retraining processes with new, unseen threats but fell short in the initial detection of such attacks. Literature suggests network administrators manually detect and categorize these novel attacks to enhance supervised learning models. Therefore, there needs a robust, adaptive IDS that not only detects new kinds of attacks at multiple levels but also can incorporate these into its retraining process so that it can anticipate and recognize future threats more effectively. 

Soltani et al. \cite{soltani2023adaptable} suggested deep learning models to identify novel attacks and a semi-supervised clustering method for updating the model. However, the model did not produce promising outcomes and IDS datasets are high-dimensional which demands a significant computational cost for the deployment and retraining of deep learning models. Therefore, the utilization of machine learning (ML) models is the preferred choice for constructing a resilient IDS capable of detecting and retraining unknown attacks.

To counteract this limitation, we introduce a two-staged integrated OCC-based technique for detecting unknown attacks and retraining the model. In this work, we have suggested a two-level hierarchical structure for detecting known and unknown attacks. To the best of our knowledge, our research is the first utilization of a hierarchical structure to construct an IDS that effectively distinguishes between known, unknown, and benign network traffic instances. This novel approach leverages recently developed OCC techniques( usfAD) in the context of network security. Generally, machine learning (ML) models exhibit proficiency in creating distinct decision boundaries for binary classifications, like differentiating between normal and attack traffic. Based on the presumption that initially separating attacks results in higher accuracy compared to concurrently classifying normal, known, and unknown attacks, we propose a hierarchical approach. In this approach, the first level handles the classification of normal and attacks, while the second level specializes in distinguishing between unknown and known attacks. This design enables the model to concentrate specifically on establishing decision boundaries between these two distinct categories of network traffic. 

In this structure, we place a usfAD algorithm at the first level to distinguish benign and attack samples. At the second level, another usfAD approach is placed to recognize known and unknown attacks. The usfAD at the first level is trained using only normal data whereas the usfAD at the second level is trained using known attack samples. By training solely on known attack types, our method can effectively categorize unfamiliar attacks as anomalies. We also advocate for a parallel supervised learning model, trained on known attack types. When the OCC method tags an instance as a known attack, the supervised model then identifies its specific family. If the OCC labels an attack instance as unknown, the instance is set aside in a designated bucket. Once we accumulate a significant number of such unknown instances, a clustering technique (DBSCAN/DPC) helps group similar unknown attack types. We then consider the largest cluster which might consist of several unknown attack samples depending on the purity of the cluster. We extract the samples with dominating unknown attack types to retrain, both the supervised model and the second-level OCC to ensure future attacks of this nature are correctly detected. Finally, We rigorously trained and assessed our proposed IDS system using ten distinct IDS datasets. This evaluation is executed using stratified 5-fold cross-validation to ensure its robustness. We measured the model's effectiveness by measuring average accuracy, precision, recall, and the f1-score. 

The structure of this paper is as follows: Section \ref{Literature Review} presents a review of related literature. Section \ref{Proposed} details the hierarchical architecture and methodology of the adaptive model. In Section \ref{RESULTS AND DISCUSSION}, we present the results of our experiments and analyze the outcome to show the effectiveness of the proposed model. Finally, Section \ref{Conclusion} summarizes the paper and outlines potential future research directions.

\section{Related works} 
\label{Literature Review}

Roshan et al. \cite{roshan2018adaptive} targeted to improve the adaptability of IDS using Extreme Learning Machines (ELM). Their proposed system detected known and unknown threats and included an efficient update mechanism to integrate new data patterns suggested by security experts. Their results showed that the proposed solution could detect emerging threats while maintaining a satisfactory level of false alarm rates. Their IDS enabled human experts to intervene and update with minimal computational overhead. These updates might involve modifying existing data, adding new labeled or unlabeled data, or introducing new data categories. Literature also suggested the ensemble method for creating an IDS that can effectively identify and respond to new threats  \cite{cuzzocrea2015distributed}.

Conventional adaptive IDS is computationally expensive and time-consuming due to the need for retraining with both known and unknown data. There needs a cost-effective adaptive IDS for promptly identifying unseen attacks. Al-Yaseena et. Al. \cite{al2017real} examined audit data and network traffic and presented a real-time detection system that utilized multi-agent strategies. The system integrated multi-level hybrid SVM and ELM classification models to effectively detect normal and known intrusions. By utilizing parallel multi-agent system (MAS) processes, the system acquired knowledge of new attacks in real-time, thereby decreasing the costs associated with training. The performance analysis conducted on the KDDCup'99 dataset demonstrated that the model surpassed conventional models in detecting targeted attacks. 

Singh et al. \cite{singh2019framework} presented an integrated framework for detecting and mitigating cyber threats. Their proposed framework employed a probabilistic methodology to detect zero-day attack paths and assess the severity of identified vulnerabilities. This technique was a hybrid detection-based approach that identified previously unidentified network vulnerabilities.
IDS is a critical security solution in cloud networks to protect them against cyber-attacks. Existing IDSs have limitations such as the inability to adapt to changing attack patterns, identify new attacks, require significant computational resources, and lack a balance between accuracy and false-positive rates (FPR). The deficiencies of current IDSs limit their effectiveness in cloud-based application systems. Furthermore, most IDS researchers assess their systems using traditional network benchmark datasets like NSL-KDD. However, these datasets fail to accurately represent the performance of these systems in real-world cloud environments. To combat these problems, Sethi et. Al.  \cite{sethi2020robust} offered an adaptive intrusion detection system (IDS) model based on a Double Deep Q-Network (DDQN) and prioritized experience replay, which is ideally suited for identifying sophisticated attacks in the cloud. They evaluated the proposed model using two datasets: ISOT-CID, which is a practical cloud-specific intrusion dataset, and NSL-KDD, which is a conventional network-based benchmark dataset.

The main challenge encountered in IDS is the detection and differentiation of new (zero-day) attacks from both regular network traffic and existing types of attacks. This challenge persists even in the latest generation of IDSs, which use deep learning to automatically extract high-level features and are thus free from the time-consuming and costly signature extraction procedure. Soltani et al. \cite{soltani2023adaptable} presented a deep learning-based framework for building adaptive IDSs and detect mitigate emerging attacks. They combined deep novelty-based classifiers and traditional clustering using a specialized layer of deep structures. In the preprocessing phase, they utilized the Deep Intrusion Detection (DID) framework to improve the effectiveness of deep learning algorithms in identifying content-based attacks. They used four algorithms: DOC (Deep Open Classification), DOC++, OpenMax, and AutoSVM) to identify zero-day attacks. Cluster groups the similar kinds of unknown attacks. They suggested a security expert to label the groups of known attacks and then retrain a supervised deep learning model. They used both the CIC-IDS2017 and CSE-CIC-IDS2018 datasets for evaluation. Their research indicated that DOC++ is the most effective implementation for the open-set recognition module. Furthermore, the clustering and post-training phases of this model exhibited high levels of completeness and uniformity, indicating its suitability for the supervised labeling and updating phases. Our work differs from the current body of work in several ways: 1) To differentiate between known, unknown, and normal traffic, we used a hierarchical structure. The first level separates attack samples from normal samples, and the second level makes the distinction between known and unknown attacks. We trained the model using a large number of IDS benchmark datasets. 3) We proposed to consider the largest cluster and automatically retrain the model until it completes the training using all of the unknown attack categories. 4) Our methodology is more realistic to fit in real-time applications in terms of binary classification and multi-classification. 5) We applied an efficient advanced semi-OCC method called usfAD, which does not require attack samples at the first level and unknown attack samples at the second level to train the model.    

When a zero-day or an unseen attack occurs, it does not match the known patterns of attacks. As this kind of attack deviates from the normal behavior profile, the OCC can flag it as suspicious. Further, OCC models can be updated continuously or periodically with new normal behaviors to adapt to the changing environment, although this might not always be the case.

As the variety and frequency of cyberattacks continue to escalate, traditional IDS solutions, reliant on databases of known attack patterns, struggle to keep up. This has spurred demand for IDSs that can effectively identify zero-day attacks, which are not yet recorded in signature databases. However, the effectiveness of ML-based OCC-based methods for detecting these new threats, including techniques like OCSVM and Local Outlier Factor (LOF), is often compromised by a high rate of false negatives, undermining their practical deployment and reliability. Focusing on this, Hindy et al. \cite{hindy2020utilising} suggested utilizing an autoencoder to identify zero-day attacks. They created an IDS model that achieved a high recall rate while maintaining an acceptable miss rate, or false-negative rate. The model used two widely recognized IDS datasets, namely CICIDS2017 and NSL-KDD. To assess the effectiveness of their model, they conducted a comparative analysis with a Support Vector Machine (SVM) utilizing the One-Class approach. They highlighted the effectiveness of a One-Class SVM in detecting zero-day attacks with atypical behaviour. Their proposed model greatly benefited from the encoding and decoding capabilities of autoencoders. The results suggested that autoencoders are effective in identifying complex zero-day attacks. 


Al-Zewairi et al. \cite{al2020unknown} introduced a novel classification system for unknown attacks, consisting of two classes: Type-A, encompassing completely new types of unknown attacks, and Type-B, encompassing unknown attacks within existing categories of known attacks. They conducted experiments using two widely recognized benchmark datasets for network intrusion detection. They evaluated the performance of modern intrusion detection systems that utilize shallow and deep artificial neural network models in detecting Type-A and Type-B attacks. The research problem was examined using both binary classification and multi-class classification approaches. The findings showed that the models exhibited a high generalization error, as evidenced by a classification error rate of 50.09\% across 92 experiments in detecting various unknown attacks. This underscores the necessity for innovative strategies and methods to tackle this problem.

Xianwei et al. \cite{gao2019adaptive} proposed a MultiTree algorithm which is created by modifying the training data proportion and implementing multiple decision trees. To enhance the detection performance, they employed multiple base classifiers, such as decision tree, random forest, kNN, and DNN. They developed an ensemble adaptive voting algorithm. The NSL-KDD Test+ dataset was utilized to evaluate the approach. 


Mike et al.  \cite{nkongolo2022cloud} selected important features from three IDS datasets: UNSW-NB15, CAIDA and UGRansome1819 using a genetic algorithm. They formed an ensemble model of three machine learning algorithms: Naive Bayes (NB), Random Forest (RF), and Support Vector Machine (SVM) to detect unknown attacks. They also tuned their models using genetic algorithms to increase performance. However, supervised learners usually cannot detect unseen attacks as their detection capability depends on the training data. Ali et al.  \cite{ali2022comparative} presented a comparative study of the stat-of-the-art methods for detecting zero-day attacks. They discussed the well-known algorithms and the challenges associated with their implementation. They examined nearly ten distinct methodologies for generic attacks identified by intrusion detection systems (IDS). 


Ahmet et al.  \cite{topcu2023social} analyzed data from the Twitter platform and utilized machine learning techniques, specifically word categorization, to promptly identify vulnerabilities and effectively counteract zero-day attacks. They employed TensorFlow Natural Language Toolkit (NLTK) tool to process Twitter data to identify terms related to zero-day attacks. Their model achieved an 80\% success rate in identifying terms related to zero-day attacks. Elfeshawy et al. \cite{elfeshawy2013divided} introduced a two-part adaptive intrusion detection system to reduce the rate of false alerts. They suggested placing IDS system in the middle of the Firewall and the ISP in the network system. Their IDS included an adaptive self-learning neural network based on a Radial Basis Functions (RBF) neural network. The model addressed the issue of false positives by utilizing an adaptation strategy with the RBF neural network.  They claimed an accuracy of 88.89\% in detecting changes in the behavior of newly added hosts to the network during the first learning iteration.  

From the above review, we can observe that some existing literature focused on either binary or multi-classification for detecting both known and novel attack types, they often overlooked the potential benefits of building a hierarchical structure. In our work, we first employ binary classification to efficiently filter out benign samples for the second level. This not only streamlines the subsequent multi-classification process but also bolsters the overall detection accuracy at the second level. We present the overview of the existing related works in Table \ref{tab:overview1} and \ref{tab:overview2}.

Most existing literature has suggested relying on human expertise to differentiate between known and novel attack categories. Given the current infrastructure where IDS systems frequently encounter a vast number of new attack types, especially during critical moments, manual separation becomes both time-intensive and costly. Our system addresses this issue promptly using an advanced semi-supervised model termed as usfAD, ensuring optimal user protection. Only a handful of studies have ventured into using One Class SVM to distinguish between familiar and unfamiliar attacks, but this method hasn't consistently achieved high accuracy. In contrast, we leverage a cutting-edge OCC detection technique called usfAD that can classify both known and unknown threats with higher accuracy.

We've developed a comprehensive adaptive IDS framework that incorporates clustering, semi-supervised learning, and supervised learning. Our approach, validated through stratified cross-validation using 10 different IDS datasets, ensures a more cohesive and robust system. Our work offers a more integrated, efficient, and advanced approach to threat detection, setting a new benchmark in the field.

\begin{table}[!htbp]
\centering
\caption{Overview of the state-of-the-art works in detecting unseen attacks}
\label{tab:overview1}
\resizebox{\columnwidth}{!}{
\begin{tabular}{|l|l|l|l|l|}
\hline
Ref. & Datasets & Model & Contributions & Remarks \\ \hline
1 & NSL-KDD & \begin{tabular}[c]{@{}l@{}}CLUS-\\ ELM\end{tabular} & \begin{tabular}[c]{@{}l@{}}Proposed   modified ELM that clusters\\ new attacks and retrain the model with\\ new attack   samples. The target was to\\ update the model with lower\\  computational cost.\end{tabular} & \begin{tabular}[c]{@{}l@{}}The   model can accommodate human\\  expert to verify the label of the novel\\  attack types.\end{tabular} \\ \hline
2 & \begin{tabular}[c]{@{}l@{}}UKM-IDS20,\\ KDD99, \\ UNSW-NB15\end{tabular} & \begin{tabular}[c]{@{}l@{}}DANN, \\ HOE-\\ DANN\end{tabular} & \begin{tabular}[c]{@{}l@{}}Suggested  an ensemble model to detect\\ known attack and retrain a novel attack\\ using a   new model every time a new \\ attack is detected by human experts.\end{tabular} & \begin{tabular}[c]{@{}l@{}}The suggestion of training a new\\ model for every type of attack \\ category is not  practical  approach\\ as new attack emerges at higher rate. \\ The model is fully dependent on the\\ human experts to detect unknown\\ attacks\end{tabular} \\ \hline
3 & KDDCup’99 & \begin{tabular}[c]{@{}l@{}}ELM, \\ SVM\end{tabular} & \begin{tabular}[c]{@{}l@{}}The model recommended to utilize ELM\\ -SVM at multiple levels to detect normal,\\   known and unknown instances.\end{tabular} & \begin{tabular}[c]{@{}l@{}}Different   agents collect unknown \\ attack and finally a coordinating \\ agents retrain the model.\end{tabular} \\ \hline
4 & NSL-KDD & SVM & \begin{tabular}[c]{@{}l@{}}They used Moth-Flame Optimization (MFO)\\ for determining optimal hyper-parameter\\ for SVM\end{tabular} & \begin{tabular}[c]{@{}l@{}}The   model did not address the issue\\ of unseen attack and updating the \\ model.\end{tabular} \\ \hline
5 & \begin{tabular}[c]{@{}l@{}}Real time\\ datasets\end{tabular} & \begin{tabular}[c]{@{}l@{}}Database \\ Signature\end{tabular} & \begin{tabular}[c]{@{}l@{}}They   captured real time network traffic \\ and extracted the dependency to match  \\ signature stored in the database.\end{tabular} & \begin{tabular}[c]{@{}l@{}}They   set a likelihood of a network\\ traffic being new attacks by \\ considering  pre-stored signature\\ which is ineffective in IDS to \\ detect new attacks\end{tabular} \\ \hline
6 & \begin{tabular}[c]{@{}l@{}}ISOT-CID,  \\ NSL-KDD\end{tabular} & DDQN & \begin{tabular}[c]{@{}l@{}}The   model used DDQN in cloud \\ environment to detect normal and \\ attack network   instances.\end{tabular} & \begin{tabular}[c]{@{}l@{}}They   model did not include any \\ module to detect unseen attacks \\ and retrain the model despite\\  labelling it adaptive model.\end{tabular} \\ \hline
7 & \begin{tabular}[c]{@{}l@{}}CIC-IDS2017,  \\ CIC-IDS2018\end{tabular} & \begin{tabular}[c]{@{}l@{}}DOC,   \\ DOC++, \\ OpenMax,   \\ AutoSVM\end{tabular} & \begin{tabular}[c]{@{}l@{}}The   author claimed their model first to\\ use deep model to detect unknown  attacks.\\ The model combined clustering  algorithms\\ with deep models, offering a new DOC++\\ open set classifier. The model has four parts:\\ Open set recognition, Clustering/post-training,\\ Supervised labelling, and updating.\end{tabular} & \begin{tabular}[c]{@{}l@{}}The   overall structure is ideal for an \\ adaptive model. However, to detect \\ unknown, they need to train their \\ model using some unknown samples\\ and they do not consider the normal\\ vs attack categorization.\end{tabular} \\ \hline
8 & DARPA’99 & BN & \begin{tabular}[c]{@{}l@{}}A Bayesian   network was employed to \\ develop learning datasets containing signatures\\ of   normal activities and various attack types.\\ When a new instance's signature  is absent in \\ these datasets and its probability is lower  than \\ that of normal  instances, it is classified as an\\ attack  instance. Subsequently, the learning \\ datasets are updated with this information.\end{tabular} & \begin{tabular}[c]{@{}l@{}}While   the adaptive model effectively \\ differentiates  between normal and \\ attack instances, it is  not equipped \\ to identify the specific family types\\ of new   attack instances.\end{tabular} \\ \hline
\end{tabular}
}
\end{table}

\begin{table}[!htbp]
\centering
\caption{Overview of the state-of-the-art works in detecting unseen attacks}
\label{tab:overview2}
\resizebox{\columnwidth}{!}{
\begin{tabular}{|l|l|l|l|l|}
\hline
Ref. & Datasets & Model & Contributions & Remarks \\ \hline
9 & NSL-KDD & \begin{tabular}[c]{@{}l@{}}SMO, BN, \\ RF\end{tabular} & \begin{tabular}[c]{@{}l@{}}Network   instances were categorized as\\ either normal or attack using SMO, BN, \\ and RF techniques.  Instances  not present \\ in the current datasets were added to these\\ datasets for the purpose of retraining and \\ enhancing the model.\end{tabular} & \begin{tabular}[c]{@{}l@{}}The   model dynamically \\ incorporates new instances into \\ the existing datasets, showcasing\\ adaptability. However,  it lacks the\\ capability to distinguish between\\ known and  unknown types of \\ attack instances.\end{tabular} \\ \hline
10 & UNSW-NB15 & \begin{tabular}[c]{@{}l@{}}SVM, DT,\\ LR,  DBN\end{tabular} & \begin{tabular}[c]{@{}l@{}}Every IoT node created a local profile of \\ normal data utilizing OCSVM, and  these\\ profiles from all nodes were collectively \\ analysed(forming global profiles) using\\ K-means algorithms.This analysis trained \\ a  DBN for each  cluster.  An ensemble of\\ DBNs then assesses whether an instance \\ is an attack. A specific local to global ratio\\ was developed to identify  and initiate the\\  model's retraining process.\end{tabular} & \begin{tabular}[c]{@{}l@{}}While   this model may be effective \\ for binary classification,  it is  not \\ suitable for multi-class classification\\  tasks.\end{tabular} \\ \hline
11 & NSL-KDD & \begin{tabular}[c]{@{}l@{}}DT, SVM, \\ LR, KNN\end{tabular} & \begin{tabular}[c]{@{}l@{}}The   suggested model employs an adaptive\\ voting-based ensemble method, where the\\ weight assigned to each DBN is based on \\ the accuracy of individual models like DT, \\ SVM and KNN.\end{tabular} & \begin{tabular}[c]{@{}l@{}}The model fails to tackle the \\ challenge of recognizing\\ unseen attacks and lacks a\\ mechanism for updating itself\\ accordingly.\end{tabular} \\ \hline
12 & \begin{tabular}[c]{@{}l@{}}CICIDS2017,   \\ NSL-KDD\end{tabular} & OCSVM & \begin{tabular}[c]{@{}l@{}}Autoencoder   and One-Class SVM are \\ compared to detect zero day attacks.\end{tabular} & \begin{tabular}[c]{@{}l@{}}They   investigated OCSVM for \\ binary classification only.\end{tabular} \\ \hline
13 & \begin{tabular}[c]{@{}l@{}}CAIDA,  \\ UNSWNB-15, \\ UGRansome1819\end{tabular} & \begin{tabular}[c]{@{}l@{}}SVM, NB,\\  RF\end{tabular} & \begin{tabular}[c]{@{}l@{}}Genetic   algorithm was used to select \\ features  and tuning  hyper parameter \\ of used supervised learning algorithms.\\ Ensemble of SVM, NB and RF were\\ employed to predict  zero day attacks\end{tabular} & \begin{tabular}[c]{@{}l@{}}Zero-day   attacks are not effectively\\  detected by supervised  learning\\  algorithms.\end{tabular} \\ \hline
14 & \begin{tabular}[c]{@{}l@{}}UNSW-NB1, \\ Bot-IoT\end{tabular} & \begin{tabular}[c]{@{}l@{}}DANN, \\ SANN\end{tabular} & \begin{tabular}[c]{@{}l@{}}The authors identified two varieties of \\ unknown attacks: Type-A for completely\\ new categories of unknown  attacks, and \\ Type-B for unknown attacks within \\ existing categories. They utilized both\\ shallow and deep neural networks to\\ discern between these Type-A and \\ Type-B attack categories.\end{tabular} & \begin{tabular}[c]{@{}l@{}}Nonetheless,   the module lacks any \\ component for  incorporating \\  zero-day  attacks into the model.\end{tabular} \\ \hline
 & \begin{tabular}[c]{@{}l@{}}NSL-KDD,\\ UNSW-NB15, \\ DDoS2018, \\ DDoS2019, \\ Malmem2022,\\  ISCXURL, \\ Darknet2020, \\ ToN-IoT-Network, \\ ToN-IoT-Linux,\\  XIIOTID\end{tabular} & \begin{tabular}[c]{@{}l@{}}usfAD,\\ LOF,  \\ OCSVM, \\ IOF, \\ DBSCAN, \\ RF\end{tabular} & \begin{tabular}[c]{@{}l@{}}We developed and executed a   two-tier \\ adaptive IDS framework. The first level \\ differentiates normal and   attack instances \\ using usfAD,  while the second level \\ identifies  unknown  attacks with another\\ usfAD,  determines the attack family class \\ using RF,  and innovatively retrains \\ both usfAD and RF  through  DBSCAN\\  clustering.\end{tabular} & \begin{tabular}[c]{@{}l@{}}We first employ  usfAD in a \\ hierarchical manner to detect zero-\\ day  attacks, followed by introducing\\ a  new methodology for retraining \\ the IDS model.\end{tabular} \\ \hline
\end{tabular}
}
\end{table}


\section{Methodology of Adaptive Hierarchical Model}
\label{Proposed}

\begin{figure}[!htbp]
    \centering
    \includegraphics[scale = .65]{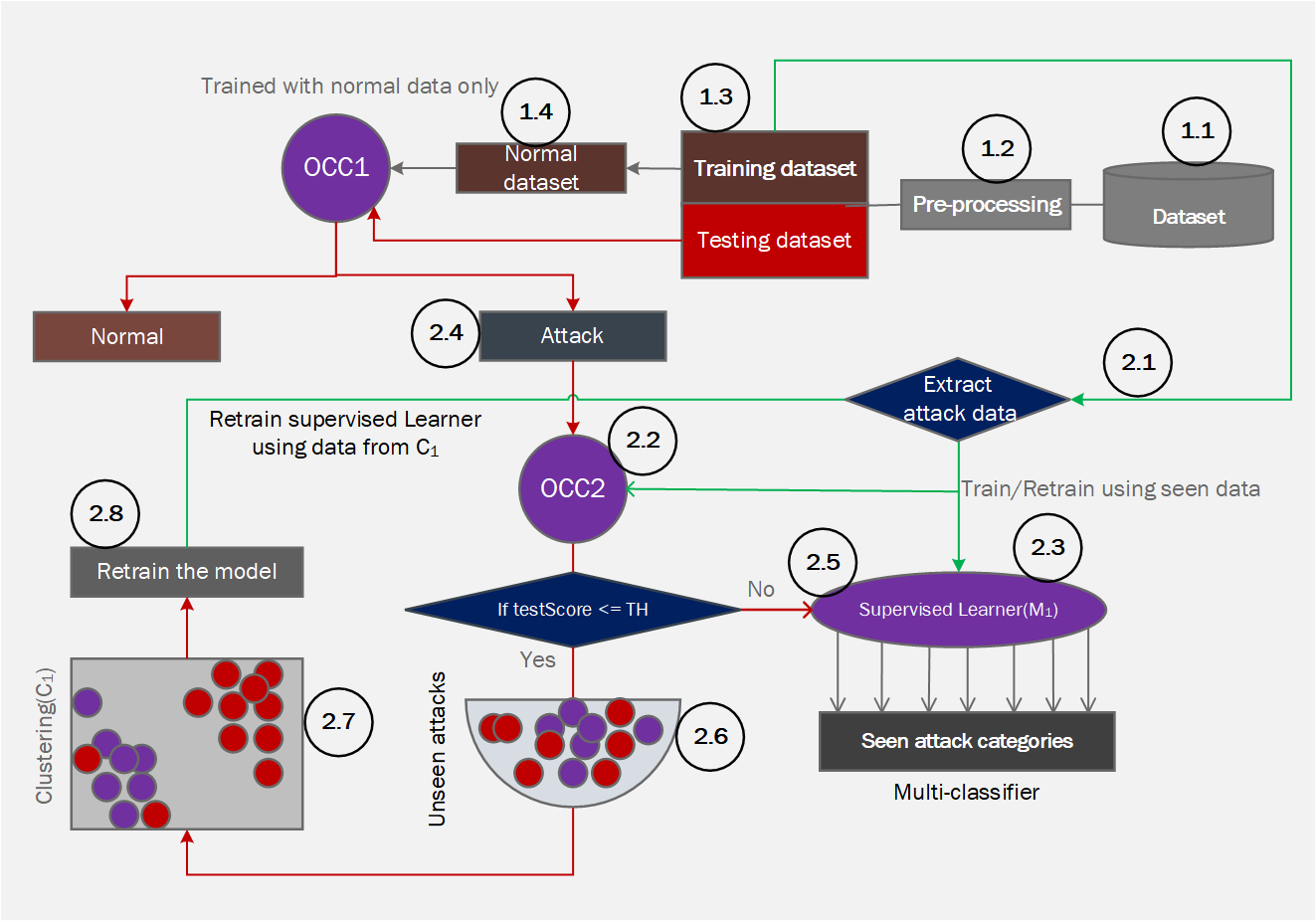}
    \caption{Overview of adaptive IDS framework}
    \label{fig:1}
\end{figure}

In this work, we aim to ensure that future instances of novel attacks are correctly recognized and classified by the IDS, fostering a more resilient and adaptive cybersecurity infrastructure. We introduce a dual-layered framework for the IDS. At the first level, we employ an OCC detection method to discern between benign and malicious network instances. In the next tier, another OCC mechanism categorizes detected attack samples as either known or unknown. Within this level, a supervised algorithm classifies the family of known attacks, while unknown attacks are stored separately. These unknown attacks are then grouped using clustering, and their respective groups are labeled with the assistance of expert judgment. Figures \ref{fig:1} visually represent the comprehensive structure of our IDS framework, which we detail in the sections below.

\begin{itemize}

    \item [1.1] We collect ten different publicly accessible benchmarks IDS datasets: NSL-KDD, UNSW-NB15, CIC-DoS2017, CIC-DDoS2019, Darknet 2020, MalMem 2022, XIIOTID, ToN-IoT-Network, ToN-IoT-Linux and ISCXURL2016.
    
    \item [1.2] The pre-processing unit performs minimal pre-processing including converting categorical values to numerical values using label encoding and standardizing the features' value using min-max techniques.
    
    \item [1.3] We apply $5$-fold stratified cross-validation for training and testing the model for every dataset. In every fold of the stratified cross-validation, the dataset is split into training and testing sets.
    \item [1.4] To train the OCC1 (usfAD, Local Outlier Factor (LOF), Isolation Forest (IOF), and Auto Encoder(AE)) at the first level, we extract the normal instances from the training set.  
    \item [2.1][2.2][2.3] To train the OCC2 (usfAD, LOF, IOF, and AE) at the second level, we extract only attack instances from the training datasets. In this case, we deliberately create different combinations of unknown and known attacks. If the attack categories in the training set are $a_1, a_2, a_3, and a_4$. We create the following combination of unknown and known categories: $[a_1], [a_2,a_3,a_4]$ ([unknown], [known], $[a_1]$ is unknown set and $[a_2,a_3,a_4]$ is known attack categories), $([a_2], [a_1, a_3, a_4])$, $([a_3], [a_1, a_2, a_4])$, $([a_4], [a_1,a_2,a_3])$, $([a_1, a_2], [a_3, a_4])$, $([a_1,a_3], [a_2, a_4])$, $([a_2, a_3], [a_1, a_4])$, $([a_1, a_2, a_3], [a_4])$ and so on. We train the second OCC2 and the supervised learner ($M_1$) using the known attacks from each set of unknown and known attack combinations. 
    
    \item  [2.4] In the testing phase, for the usfAD model, we set a threshold to differentiate between normal and attack instances. We obtain scores after training the usfAD for training and testing instances. The formula for the threshold is TH = mean(scores of training instances)- 3* standard deviation (scores of training instances). If the score of a testing instance is less than the threshold, we consider the instance as an attack, otherwise, it is labeled as a normal instance. For all the OCC models including usfAD, we collect all the instances that are predicted as attacks by the OCC1 as a testing set for the second OCC2 model.  
    \item [2.5]  If a data point is predicted as an attack class from the OCC1 model, then this is predicted by the OCC2 to identify if the attack point is known or unknown. If an instance is identified as a known attack, this is passed to the supervised model ($M_1$) at the second level to reveal the specific attack family type.  

    \item [2.6] After collecting a certain number of unknown attacks, we utilize DBSCAN and DPC clustering to group similar kinds of unknown attack samples. The number of similar groups in the unknown samples is not known. For this reason, we choose DBSCAN/DPC as we do not need to determine the number of clusters in DBSCAN/DPC. 
    
    \item [2.7] In our methodology, for both known and novel attack scenarios, we focus on the largest cluster identified in the data, which likely includes samples from various attack classes. Specifically, we target the predominant unknown attack type within this cluster for retraining our second-level OCC and Random Forest models (RF). This enables the models to better identify such attacks in future instances. The training set does not contain particular types of attacks. In our simulation, we ensure that the test set includes all kinds of attack instances. In an ideal case, our second-level OCC detects such attack samples that are not used in the training set as unknown attacks. 
    
    Upon accumulating a sufficient quantity of such unknown samples (e.g., 1000), we apply clustering algorithms like DBSCAN to these samples. For instance, if the majority cluster corresponds to attack 1 initially, we retrain OCC2 and RF with this new data. Consequently, in subsequent iterations, the model starts recognizing attack 1. We repeat this process, collecting another set of 1000 unknown samples for retraining with the next prevalent attack type, and so on. This iterative retraining process aims to progressively enhance the model's capability to autonomously learn and recognize emerging attack categories over time.

    \item [2.8] The known attack instances are used to train a supervised model (RF) for identifying individual categories of known attacks. Finally, performance is measured using various performance metrics such as accuracy, precision, recall, and f1-score. 

\end{itemize}

In the next section, we describe the retraining of the proposed model in detail. 

\subsection{Simulation of Retraining Semi-supervised and Supervised Learner}

\begin{figure}[!htbp]
    \centering
    \includegraphics[scale = .70]{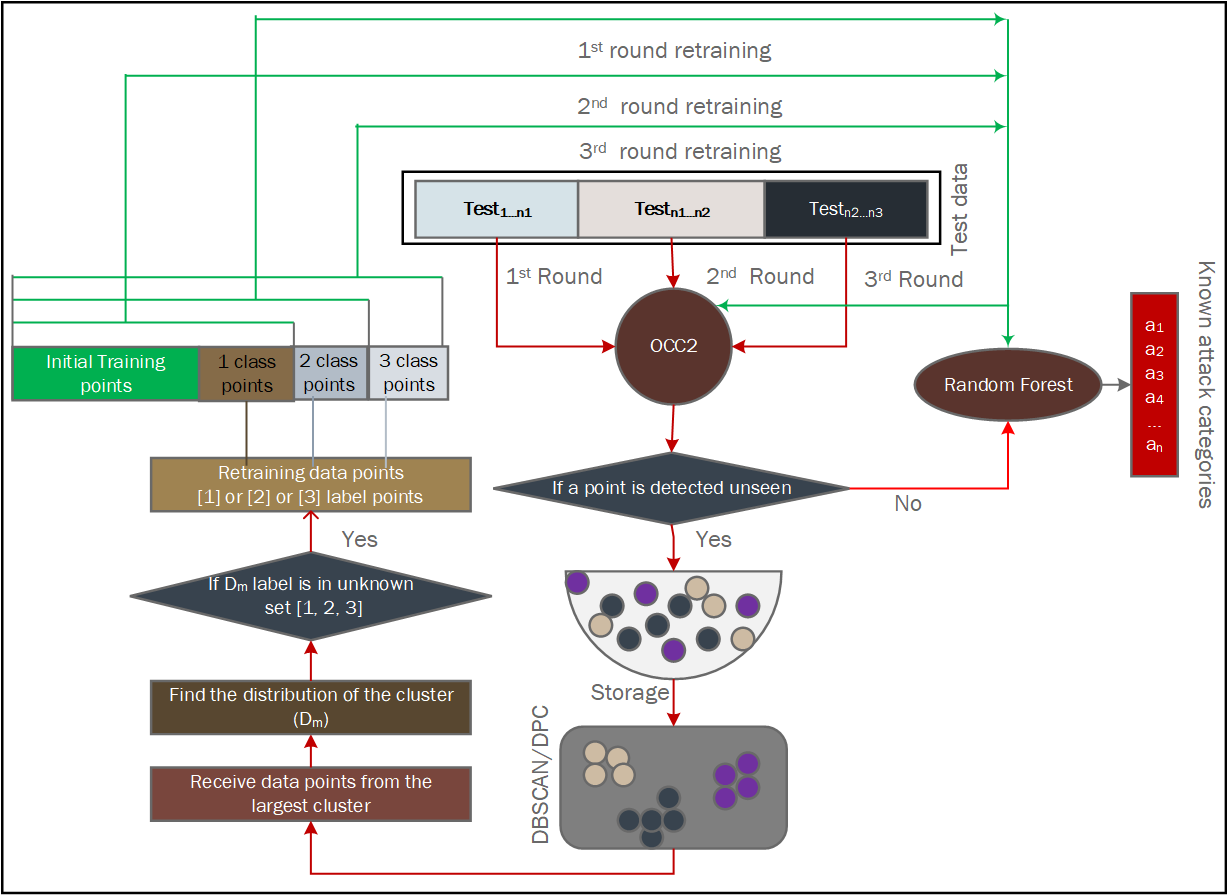}
    \caption{Simulation of retraining the adaptive IDS}
    \label{fig:2}
\end{figure}

Figure \ref{fig:3} presents a methodology for updating training processes in the face of unidentified attack categories. Initially, an OCC algorithm at the second level detects an instance as either a known or unknown attack category. Known attacks are forwarded to a RF classifier ($M_1$) to detect their specific family types such as DoS, Ransomware, or Trojan Horse. On the contrary, unknown attacks are temporarily stored in a bucket in every test. Upon accumulating a substantial number of unknown threats/attacks, we apply DBSCAN on unknown instances to make clusters based on similarities. We find the distribution pattern of the largest cluster. We extract the instances of the largest unknown attacks in the cluster. These new instances are added to the original training dataset to retrain both the OCC and RF at the second level. Figure \ref{fig:2} shows the retraining phases aimed at incorporating a different new category of unknown attack. The sequential approach of retraining the model with different new attack types to enable it to detect them in the future.

\begin{itemize}
    \item \textbf{Initial Setup:} In Figure \ref{fig:3}, the test data is divided into different segments $(Test_1,...,n_1, Test_1,...,n_2, Test_2,...,n_3)$. The length of these segments depends on the number of unknown attacks detected by the first-level OCC. Figure \ref{fig:3} shows three rounds of retraining the model. The process starts with initial training points, indicating that the system has been trained on a dataset with known attack categories.

    \item \textbf{Detection and Classification:} The OCC2 checks incoming unknown attacks from the OCC1. If an attack is recognized as known, it is passed to a multi-classifier ($M_1$) which determines the specific category of the attack $(a_1, a_2, ... a_n)$. If the attack is unknown, it's stored for later processing.
    
    \item \textbf{Handling Unknown Attacks:} As unknown attack data accumulates, clustering algorithms like DBSCAN are used to group similar attack patterns. This implies that the system has some unsupervised learning capability to handle novel threats. The process focuses on the largest cluster, assessing its distribution to decide if it represents a new type of attack. If confirmed as a new attack type, the data points from this cluster are added to the initial training set.
    
    \item \textbf{ Retraining:} With new data points added, both the OCC2 and $M_1$ model are retrained to improve their detection and classification abilities. Figure \ref{fig:3} suggests that multiple rounds of retraining may be necessary to capture all unknown attack categories. Each round incorporates new findings from the largest cluster of unknown attacks at that time.
    
    \item \textbf{ Iterative Process:} The process is iterative, with the potential for several retraining rounds (1st round retraining, 2nd round retraining, 3rd round retraining). This iterative process likely continues until the system no longer encounters new types of unknown attacks, or until all unknown attacks have been classified into new categories and added to the known list.
\end{itemize}

Our model employs both supervised learning (RF for known attack classification) and unsupervised learning (DBSCAN/DPC for clustering unknown attacks). The system is designed to adapt over time, learning from new types of attacks and expanding its classification abilities through retraining. Our retraining process is presented in Algorithm \ref{alg1:retraining}.

\begin{algorithm}[H]
\setstretch{1} 

\caption{Adaptive Attack Detection and Classification}
\label{alg1:retraining}

\SetAlgoLined
\SetKwInOut{Input}{Input}
\SetKwInOut{Output}{Output}

\Input{Initial dataset with known attack categories, Test data}
\Output{Updated detection and classification models}

\BlankLine
\BlankLine

\SetKwComment{Comment}{\# }{}
\Comment{\textbf{Initial Setup}}
Train system on the initial dataset with known attack categories\;
Divide test data into segments: $Test_1,...,n_1, Test_1,...,n_2, Test_2,...,n_3$\;
Form the different combination of unknown and known attack sets: $([a_1], [a_2, a_3]), ..., ([a_1,a_2..a_{n-1}],[a_n])$\;

\For{ each unknown-known attack set}{
    \For{each set of test data}{
        \Comment{\textbf{Detection and Classification}}
        \For{each data point in a testing set}{
            Use OCC2 to predict attack class\;
            \If{attack is known}{
                Classify attack using multi-classifier $M_1$\;
            }
            \Else{
                Store unknown attack data\;
            }
        }
        
        \Comment{\textbf{Handling Unknown Attacks}}
        Accumulate data on unknown attacks\;
        Apply clustering algorithms to group unknown attacks\;
        Focus on the largest cluster\;
        \If{largest cluster represents a new attack type}{
            Add data points from cluster to training set\;
        }
        
        \Comment{\textbf{Retraining}}
        Update training set with new data points\;
        Retrain OCC2 and $M_1$ model\;
        \If{no new unknown attacks are identified}{
            \textbf{break} \Comment{Exit the loop if no new unknown attacks}\;
        }
            
    }
}
\end{algorithm}

\subsection{Dataset and Pre-Processing}

In this section, we provide a short description of the 10 different IDS benchmark datasets that we have used in this work. 

\begin{itemize}
    \item NSL-KDD dataset \cite{su2020bat} was designed to overcome the issues with KDD’99 dataset. This updated version of the KDD data set is still regarded as an effective benchmark dataset for researchers to compare different intrusion detection approaches. The NSL-KDD training and testing sets have a balanced quantity of records for benign and attack samples. The shape of the datasets is (148517, 44). 

    \item UNSW-NB15 dataset: The Network Security Research Lab at the University of New South Wales, Australia, built the UNSW-NB15 dataset by capturing network traffic in a realistic setting using a high-speed network sniffer and various tools and techniques such as packet flooding, port scanning, and SQL injection. The original dataset contains 257,673 records and 45 fields. 

    \item  Canadian Institute for Cybersecurity released CIC-IDS2017 dataset \cite{jazi2017detecting} which is a benchmark dataset for Intrusion Detection Systems. The dataset includes user behaviour models that are protocol-agnostic through HTTP, HTTPS, FTP, SSH, and email. The dataset consists of 222914,  and 78 features having four classes:  benign samples,  DoS SlowLoris samples,  DoS Slow Httptest samples,  DoS Hulk samples, DoS GoldenEye samples, and Heartbleed samples in the output class label.

    \item CIC-DDoS2019: The Canadian Centre for Cybersecurity at the University of New Brunswick created a dataset of DDoS attacks called CIC-DDoS2019. This data set contains both normal traffic patterns and a wide variety of distributed denial of service (DDoS) assaults, such as UDP flood, HTTP flood, and TCP SYN. The shape of the dataset is(431371, 79) where attack instances are 333540 and benign instances are 97831. 

    \item Malmem2022: Obfuscated malware hides them to avoid detection and elimination using conventional anti-malware software. Malmem 2022 \cite{carrier2022detecting} is a simulated obfuscated dataset designed to be as realistic as possible to train and test machine learning algorithms to detect obfuscated malware. The dataset is balanced one having level 2 categories:   Spyware, Ransomware, and Trojan Horse.
    
    \item ToN-IoT-Network and ToN-IoT-Linux: ToN-IoT was extracted from a realistic large-scale IoT simulated environment at the Cyber Range Lab led by ACCS in 2019. The dataset contains heterogeneous telemetry IoT services, traffic flows, and logs of the operating system. Later, Bro-IDS known as Zeek having 44 features was formed from the original dataset considering the network traffic flows. Label encoding is used to convert its categorical features into numerical features following  \cite{moustafa2021new} and  \cite{guo2023iot}. These datasets contain IP addresses. We can treat each unique IP address as a category and perform one-hot encoding. Although this is theoretically possible, it's usually not practical for real-world IDS systems due to the vast number of unique IP addresses, which leads to extremely high-dimensional data.
    
   \item ISCXURL2016: In WWW web, URLs serve as the primary mode of transport and attackers insert malware into users’ computer systems through URLs. The researchers focus on developing methods for blacklisting malicious URLs. Mamun et al.  \cite{mamun2016detecting} formed a modern URL dataset that contains the following categories of URLs: benign URLs, spam URLs, phishing URLs, malware URLs and defacement URLs. The shape of the original datasets is (36707, 80). 
   
  \item CIC-Darknet2020: The CIC-Darknet dataset has 141530 records with 85 columns features and was labeled in two ways. We apply label encoding to convert its categorical features into numerical values. 

  \item XIIoTID: The XIIoTID dataset \cite{9504604} has an initial shape of (596017, 64).  The dataset has features from network traffic, system logs, application logs, device resources (CPU, input/Output, Memory, and others), and commercial Intrusion detection systems' logs (OSSEC and Zeek/Bro).

\end{itemize}

 In this work, we apply label encoding to convert categorical features into numerical features for all of the datasets. We perform the normalization to scale features' values in the range of 0 and 1. Normalization refers to a data pre-processing technique applied in machine learning to standardize the scale of all dataset's attributes. This entails converting the original data to a new range, typically between 0 and 1 or -1 and 1. The min-max normalization formula is as follows:
    
    $X_{norm} = \frac{X - X_{min}}{X_{max} - X_{min}}$
    
    Where $X_norm$ is the normalized value of $X$, $X_min$ is the minimum value of $X$, and $X_max$ is the maximum value of $X$.

\subsection{Experimental Setup and Implementation}

Our experimental study was performed on an Intel Xeon E5-2670 CPU (8 cores, 16 threads), 128GB DDR3 RAM and 2x Nvidia GTX 1080 Ti. Python 3.9 was used to execute our code. The study utilized ten different machine learning models and primarily relied on Pandas and NumPy libraries for data pre-processing. Since the framework was developed using Python, the widely recognized Scikit-learn toolkit was utilized to leverage its wide range of algorithms and resources for data scientists, including effective accuracy and precision estimation metrics.

\subsection{Performance Metrics}

In this study, we use accuracy, precision, and F1 scores that are essential for assessing the performance of an IDS model. However, their significance can vary depending on the system's specific objectives and requirements. Accuracy quantifies the proportion of accurate classifications made by the IDS. However, relying solely on accuracy is not the most suitable performance metric for IDS, as this might not accurately reflect the system's capability to identify attacks, which are a minority class within the dataset. Precision refers to the proportion of genuine positive detection out of all positive detection. High precision is essential in IDS to minimize false positives, which can result in false alarms. Recall measures the system's ability to reliably identify all instances of a particular class of attack. Low recall suggests that the system is missing some attacks, which can pose a significant security risk.

The F1 score is a combination of precision and recall that quantifies the proportion of true positive identification relative to the total number of positive instances in the dataset. F1-score is a valuable metric for IDS because this considers both false positives and false negatives and provides a balanced score between precision and recall. The accuracy, precision, recall and f1-score are calculated as follows.

\[
\text{Accuracy} = \frac{\text{TP + TN}}{\text{TP + TN + FP + FN}} \times 100
\]

\[
\text{Precision} = \frac{\text{TP}}{\text{TP+FP}} \times 100
\]

\[
\text{Recall} = \frac{\text{TP}}{\text{TP+FN}} \times 100
\]

\[
\text{F1-score} = 2\times \frac{\text{Precision $\times$ Recall}}{\text{Precision + Recall}} \times 100
\]

where TP = true positive, TN = true negative, FP = false positive, and FN = false negative.

We apply 5-fold stratified cross-validation to split datasets for training and testing while preserving the balanced proportion of each class in each fold. The stratified cross-validation provides a more accurate estimate of model performance, particularly when working with imbalanced datasets in which one class has more samples than the other. In this study, we trained and evaluated our adaptive hierarchical models using stratified 5-fold cross-validation. We divided the datasets into five folds of equal size, with each fold containing a proportional representation of the different classes. This process is repeated for all five folds. 

\section{Results and Discussion}
\label{RESULTS AND DISCUSSION}

In this section, we evaluate the effectiveness of our hierarchical adaptive IDS system. To assess the performance of the proposed model, we use 10 different benchmark IDS datasets that are publicly available. We analyze the performance of the model using usfAD, LOF, IOF, OCSVM and AE semi-supervised learners in terms of average accuracy and f1-score across 10 IDS datasets. 

\subsection{Performance of Semi-Supervised Learner at the First Level}

An OCC/semi-supervised model is positioned at the root level in our approach, primarily to separate attack samples from normal network traffic. This model is trained on regular network traffic. In this case, attack instances are not required which can address the limited availability of attack samples for training purposes in real-life situations. In hierarchical architecture, the efficacy of the second-level model is dependent on the performance of the root-level OCC model. In this section, we evaluate various OCC models employed at the first level to differentiate between normal and attack categories across multiple datasets. We focus on key metrics such as accuracy and F1-score, which are standard benchmarks for assessing classification models.

Table \ref{Table3:binaryclassification} indicates that usfAD surpasses other OCC detection methods on the majority of the datasets, including NSL-KDD, ToN-IoT-Network, CIC-DDoS2019, UNSW-NB15, ISCXURL2016, ToN-IoT-Linux, XIIOTID, and CIC-DDoS2017. In the case of the Malmem2022 dataset, the Autoencoder (AE) achieves superior accuracy and F1-score compared to other OCC detection techniques. For the Darknet2020 dataset, the Local OCC Factor (LOF) demonstrates enhanced effectiveness in distinguishing between benign or normal samples and attack instances.

\begin{table}[!htbp]
\caption{Normal vs Attack: Root Level Model's Performance}
\label{Table3:binaryclassification}
\begin{tabular}{|l|ll|ll|ll|ll|ll|}
\hline
 & \multicolumn{2}{c|}{usfAD} & \multicolumn{2}{c|}{LOF} & \multicolumn{2}{c|}{IOF} & \multicolumn{2}{c|}{OCSVM} & \multicolumn{2}{c|}{AE} \\ \hline
Datasets & \multicolumn{1}{l|}{ACC} & F1-score & \multicolumn{1}{l|}{ACC} & F1-score & \multicolumn{1}{l|}{ACC} & F1-score & \multicolumn{1}{l|}{ACC} & F1-score & \multicolumn{1}{l|}{ACC} & F1-score \\ \hline
NSL-KDD & \multicolumn{1}{l|}{\textbf{94.90}} & \textbf{94.90} & \multicolumn{1}{l|}{69.52} & 68.15 & \multicolumn{1}{l|}{88.06} & 88.00 & \multicolumn{1}{l|}{73.07} & 71.56 & \multicolumn{1}{l|}{89.02} & 89.00 \\ \hline
ToN-IoT-Network & \multicolumn{1}{l|}{\textbf{98.94}} & \textbf{98.94} & \multicolumn{1}{l|}{89.31} & 88.98 & \multicolumn{1}{l|}{63.52} & 55.32 & \multicolumn{1}{l|}{42.79} & 43.84 & \multicolumn{1}{l|}{60.76} & 52.26 \\ \hline
CIC-DDoS2019 & \multicolumn{1}{l|}{\textbf{98.49}} & \textbf{97.83} & \multicolumn{1}{l|}{89.06} & 85.48 & \multicolumn{1}{l|}{49.30} & 48.86 & \multicolumn{1}{l|}{82.66} & 81.79 & \multicolumn{1}{l|}{79.27} & 80.78 \\ \hline
UNSW-NB15 & \multicolumn{1}{l|}{\textbf{82.08}} & \textbf{82.43} & \multicolumn{1}{l|}{74.65} & 74.97 & \multicolumn{1}{l|}{55.54} & 53.25 & \multicolumn{1}{l|}{75.80} & 74.44 & \multicolumn{1}{l|}{53.96} & 53.40 \\ \hline
Malmem2022 & \multicolumn{1}{l|}{91.48} & 91.42 & \multicolumn{1}{l|}{88.41} & 88.39 & \multicolumn{1}{l|}{89.79} & 89.68 & \multicolumn{1}{l|}{74.97} & 73.30 & \multicolumn{1}{l|}{\textbf{94.93}} & \textbf{94.92} \\ \hline
ISCXURL2016 & \multicolumn{1}{l|}{\textbf{89.57}} & \textbf{88.35} & \multicolumn{1}{l|}{80.37} & 81.89 & \multicolumn{1}{l|}{63.94} & 67.05 & \multicolumn{1}{l|}{80.85} & 80.48 & \multicolumn{1}{l|}{73.81} & 76.16 \\ \hline
Darknet2020 & \multicolumn{1}{l|}{90.84} & 90.39 & \multicolumn{1}{l|}{\textbf{93.13}} & \textbf{93.27} & \multicolumn{1}{l|}{78.53} & 75.48 & \multicolumn{1}{l|}{48.61} & 54.87 & \multicolumn{1}{l|}{76.97} & 74.64 \\ \hline
ToN-IoT-Linux & \multicolumn{1}{l|}{\textbf{97.71}} & \textbf{97.39} & \multicolumn{1}{l|}{95.69} & 95.71 & \multicolumn{1}{l|}{66.87} & 57.37 & \multicolumn{1}{l|}{42.92} & 44.29 & \multicolumn{1}{l|}{67.77} & 67.77 \\ \hline
XIIOTID & \multicolumn{1}{l|}{\textbf{93.39}} & \textbf{93.34} & \multicolumn{1}{l|}{78.51} & 76.94 & \multicolumn{1}{l|}{70.60} & 67.88 & \multicolumn{1}{l|}{68.37} & 67.42 & \multicolumn{1}{l|}{82.66} & 82.45 \\ \hline
CIC-DDoS2017 & \multicolumn{1}{l|}{\textbf{96.98}} & \textbf{97.10} & \multicolumn{1}{l|}{83.89} & 87.37 & \multicolumn{1}{l|}{89.68} & 90.21 & \multicolumn{1}{l|}{49.72} & 61.60 & \multicolumn{1}{l|}{86.47} & 88.26 \\ \hline
\end{tabular}
\end{table}

\subsection{Performance of Semi-Supervised Learner at the Second Level}

The role of the second-level OCC algorithm is to distinguish between known and unknown attacks (also called zero-day attacks) after the first-level OCC model filtered out the normal instances. The second level model is trained with known attack instances, enabling it to identify unknown attacks based on deviations from these known patterns. In this section, we first evaluate the performance of this second-level OCC model, focusing on its ability to distinguish both known and unknown attacks (in a binary classification). Next, we present its performance which focuses on how well the OCC detects the family type of individual attacks that are labeled as known in the testing samples before and after retraining.

To robustly test the model's capabilities, we have established a pool of unknown and unknown attack categories. This includes combinations of 1-attack, 2-attack, and 3- unknown attack types. We present the results for each of these combinations, demonstrating the model's effectiveness across different scenarios of unknown attacks.

Table \ref{Table4:knownunknownaccclassification} and \ref{Table5:unknownknownf1scoreclassification} demonstrate the performance of OCC/semi-supervised models at the second level for binary classification—differentiating between seen and unseen attacks—across various datasets in terms of accuracy and F1-score. The 'C1' and 'C2' columns represent the system's performance for two combinations of varying numbers of unknown attack classes. UA1 represents a scenario with one class of unknown attacks, UA2 denotes the presence of two classes of unknown attacks, and UA3 corresponds to situations involving three classes of unknown attacks.

The usfAD model yields superior results in both accuracy and F1-score for the NSL-KDD, ToN-IoT-Network, CIC-DDoS2019, ToN-IoT-Linux, and XIITOID datasets, outperforming other models like OCSVM, IOF, and AE, which exhibit poorer performance on these datasets. Current research has not yet investigated the ToN-IoT-Network, CIC-DDoS2019, ToN-IoT-Linux, and XIITOID datasets for the detection of known and unknown attacks. This demonstrates the usfAD model's effectiveness in handling the most up-to-date IDS datasets. Nevertheless, with the NSL-KDD dataset, the Autoencoder (AE) achieves relatively improved performance when multiple attack classes are unseen. A comparable pattern is noted for the UNSW-NB15 datasets when employing the IOF model.

Malmem2022 shows a significant challenge with low accuracy and F1-Scores in both C1 and C2, suggesting difficulty in differentiating between seen and unseen attacks. For ISCXURL2016, CIC-DDoS2017, and CIC-Darknet2020 datasets, usfAD and LOF achieves moderately better accuracy and F1-score compared to other datasets.

\begin{table}[!htbp]
\caption{Unknown vs Known: Second level OCC model's performance in terms of accuracy}
\label{Table4:knownunknownaccclassification}
\begin{tabular}{|l|l|ll|ll|ll|ll|ll|}
\hline
\multicolumn{1}{|c|}{\multirow{2}{*}{Datasets}} & \multicolumn{1}{c|}{\multirow{2}{*}{NoUAC}} & \multicolumn{2}{c|}{usfAD} & \multicolumn{2}{c|}{LOF} & \multicolumn{2}{c|}{IOF} & \multicolumn{2}{c|}{OCSVM} & \multicolumn{2}{c|}{AE} \\ \cline{3-12} 
\multicolumn{1}{|c|}{} & \multicolumn{1}{c|}{} & \multicolumn{1}{l|}{C1} & C2 & \multicolumn{1}{l|}{C1} & C2 & \multicolumn{1}{l|}{C1} & C2 & \multicolumn{1}{l|}{C1} & C2 & \multicolumn{1}{l|}{C1} & C2 \\ \hline
\multirow{3}{*}{NSL-KDD} & UA1 & \multicolumn{1}{l|}{\textbf{93.49}} & \textbf{92.99} & \multicolumn{1}{l|}{74.77} & 81.15 & \multicolumn{1}{l|}{74.47} & 86.29 & \multicolumn{1}{l|}{66.10} & 41.32 & \multicolumn{1}{l|}{79.26} & 83.61 \\ \cline{2-12} 
 & UA2 & \multicolumn{1}{l|}{\textbf{94.20}} & \textbf{93.27} & \multicolumn{1}{l|}{79.82} & 75.30 & \multicolumn{1}{l|}{90.06} & 74.01 & \multicolumn{1}{l|}{67.29} & 60.13 & \multicolumn{1}{l|}{89.92} & 78.37 \\ \cline{2-12} 
 & UA3 & \multicolumn{1}{l|}{\textbf{94.66}} & \textbf{94.14} & \multicolumn{1}{l|}{71.44} & 80.02 & \multicolumn{1}{l|}{90.21} & 90.68 & \multicolumn{1}{l|}{67.62} & 67.20 & \multicolumn{1}{l|}{89.31} & 89.82 \\ \hline
\multirow{3}{*}{ToN-IoT-Network} & UA1 & \multicolumn{1}{l|}{\textbf{96.70}} & \textbf{96.25} & \multicolumn{1}{l|}{84.69} & 87.97 & \multicolumn{1}{l|}{30.12} & 14.68 & \multicolumn{1}{l|}{6.43} & 0.66 & \multicolumn{1}{l|}{7.07} & 1.22 \\ \cline{2-12} 
 & UA2 & \multicolumn{1}{l|}{\textbf{96.91}} & \textbf{96.92} & \multicolumn{1}{l|}{86.89} & 86.28 & \multicolumn{1}{l|}{30.63} & 33.30 & \multicolumn{1}{l|}{8.73} & 13.59 & \multicolumn{1}{l|}{8.35} & 10.63 \\ \cline{2-12} 
 & UA3 & \multicolumn{1}{l|}{\textbf{97.20}} & \textbf{97.47} & \multicolumn{1}{l|}{88.42} & 85.71 & \multicolumn{1}{l|}{33.58} & 35.10 & \multicolumn{1}{l|}{14.26} & 13.04 & \multicolumn{1}{l|}{12.14} & 21.30 \\ \hline
\multirow{3}{*}{UNSW-NB15} & UA1 & \multicolumn{1}{l|}{\textbf{89.72}} & 68.42 & \multicolumn{1}{l|}{85.04} & 64.33 & \multicolumn{1}{l|}{79.97} & \textbf{85.47} & \multicolumn{1}{l|}{51.52} & 45.81 & \multicolumn{1}{l|}{75.41} & 80.68 \\ \cline{2-12} 
 & UA2 & \multicolumn{1}{l|}{70.62} & 85.81 & \multicolumn{1}{l|}{72.33} & \textbf{87.46} & \multicolumn{1}{l|}{\textbf{80.04}} & 84.98 & \multicolumn{1}{l|}{55.17} & 49.54 & \multicolumn{1}{l|}{79.67} & 78.64 \\ \cline{2-12} 
 & UA3 & \multicolumn{1}{l|}{64.26} & 70.22 & \multicolumn{1}{l|}{69.18} & 76.24 & \multicolumn{1}{l|}{\textbf{86.37}} & \textbf{84.58} & \multicolumn{1}{l|}{57.28} & 58.62 & \multicolumn{1}{l|}{81.72} & 80.56 \\ \hline
\multirow{2}{*}{Malmem2022} & UA1 & \multicolumn{1}{l|}{56.60} & 55.92 & \multicolumn{1}{l|}{54.38} & 54.13 & \multicolumn{1}{l|}{54.86} & 52.87 & \multicolumn{1}{l|}{36.87} & 33.56 & \multicolumn{1}{l|}{\textbf{59.69}} & \textbf{57.63} \\ \cline{2-12} 
 & UA2 & \multicolumn{1}{l|}{\textbf{47.41}} & \textbf{42.22} & \multicolumn{1}{l|}{41.99} & 37.12 & \multicolumn{1}{l|}{34.74} & 30.96 & \multicolumn{1}{l|}{40.85} & 33.40 & \multicolumn{1}{l|}{41.36} & 34.47 \\ \hline
\multirow{2}{*}{ISCXURL2016} & UA1 & \multicolumn{1}{l|}{\textbf{78.60}} & \textbf{80.34} & \multicolumn{1}{l|}{78.97} & 74.49 & \multicolumn{1}{l|}{60.64} & 62.10 & \multicolumn{1}{l|}{43.93} & 55.26 & \multicolumn{1}{l|}{60.89} & 59.33 \\ \cline{2-12} 
 & UA2 & \multicolumn{1}{l|}{\textbf{87.15}} & \textbf{77.05} & \multicolumn{1}{l|}{73.35} & 78.75 & \multicolumn{1}{l|}{52.90} & 47.75 & \multicolumn{1}{l|}{66.41} & 49.10 & \multicolumn{1}{l|}{47.79} & 50.91 \\ \hline
\multirow{3}{*}{CIC-DDoS2019} & UA1 & \multicolumn{1}{l|}{\textbf{89.06}} & \textbf{95.76} & \multicolumn{1}{l|}{84.21} & 83.08 & \multicolumn{1}{l|}{81.33} & 83.94 & \multicolumn{1}{l|}{17.57} & 49.08 & \multicolumn{1}{l|}{41.95} & 88.84 \\ \cline{2-12} 
 & UA2 & \multicolumn{1}{l|}{\textbf{90.63}} & \textbf{93.00} & \multicolumn{1}{l|}{79.92} & 86.88 & \multicolumn{1}{l|}{72.64} & 87.70 & \multicolumn{1}{l|}{25.50} & 8.60 & \multicolumn{1}{l|}{40.00} & 37.32 \\ \cline{2-12} 
 & UA3 & \multicolumn{1}{l|}{\textbf{93.17}} & \textbf{86.72} & \multicolumn{1}{l|}{88.00} & 76.69 & \multicolumn{1}{l|}{91.47} & 62.48 & \multicolumn{1}{l|}{82.06} & 25.44 & \multicolumn{1}{l|}{87.10} & 33.43 \\ \hline
\multirow{3}{*}{CIC-DDoS2017} & UA1 & \multicolumn{1}{l|}{\textbf{62.80}} & \textbf{64.06} & \multicolumn{1}{l|}{40.45} & 27.10 & \multicolumn{1}{l|}{27.59} & 13.64 & \multicolumn{1}{l|}{2.38} & 2.70 & \multicolumn{1}{l|}{34.34} & 17.68 \\ \cline{2-12} 
 & UA2 & \multicolumn{1}{l|}{\textbf{64.05}} & \textbf{62.22} & \multicolumn{1}{l|}{27.07} & 40.11 & \multicolumn{1}{l|}{17.96} & 25.58 & \multicolumn{1}{l|}{1.90} & 2.85 & \multicolumn{1}{l|}{24.87} & 32.19 \\ \cline{2-12} 
 & UA3 & \multicolumn{1}{l|}{\textbf{64.19}} & \textbf{65.61} & \multicolumn{1}{l|}{27.13} & 29.28 & \multicolumn{1}{l|}{16.94} & 15.08 & \multicolumn{1}{l|}{4.60} & 5.56 & \multicolumn{1}{l|}{24.64} & 25.71 \\ \hline
\multirow{3}{*}{ToN-IoT-Linux} & UA1 & \multicolumn{1}{l|}{\textbf{84.04}} & \textbf{82.36} & \multicolumn{1}{l|}{78.58} & 76.12 & \multicolumn{1}{l|}{39.65} & 30.56 & \multicolumn{1}{l|}{10.47} & 9.02 & \multicolumn{1}{l|}{31.48} & 32.34 \\ \cline{2-12} 
 & UA2 & \multicolumn{1}{l|}{\textbf{78.51}} & \textbf{78.76} & \multicolumn{1}{l|}{77.38} & 77.46 & \multicolumn{1}{l|}{29.55} & 30.10 & \multicolumn{1}{l|}{14.80} & 13.91 & \multicolumn{1}{l|}{30.60} & 31.71 \\ \cline{2-12} 
 & UA3 & \multicolumn{1}{l|}{\textbf{87.57}} & \textbf{81.11} & \multicolumn{1}{l|}{84.75} & 78.45 & \multicolumn{1}{l|}{29.80} & 38.56 & \multicolumn{1}{l|}{18.37} & 18.77 & \multicolumn{1}{l|}{30.81} & 36.04 \\ \hline
\multirow{3}{*}{XIIOTID} & UA1 & \multicolumn{1}{l|}{\textbf{94.93}} & \textbf{94.84} & \multicolumn{1}{l|}{85.01} & 92.87 & \multicolumn{1}{l|}{75.12} & 78.07 & \multicolumn{1}{l|}{62.19} & 39.52 & \multicolumn{1}{l|}{71.66} & 79.32 \\ \cline{2-12} 
 & UA2 & \multicolumn{1}{l|}{\textbf{94.91}} & \textbf{94.90} & \multicolumn{1}{l|}{85.09} & 87.15 & \multicolumn{1}{l|}{78.35} & 76.33 & \multicolumn{1}{l|}{47.16} & 61.45 & \multicolumn{1}{l|}{74.94} & 70.43 \\ \cline{2-12} 
 & UA3 & \multicolumn{1}{l|}{\textbf{94.81}} & \textbf{95.08} & \multicolumn{1}{l|}{87.23} & 85.06 & \multicolumn{1}{l|}{78.94} & 79.17 & \multicolumn{1}{l|}{53.13} & 50.26 & \multicolumn{1}{l|}{72.48} & 81.20 \\ \hline
\multirow{3}{*}{CICDarknet2020} & UA1 & \multicolumn{1}{l|}{\textbf{64.12}} & 58.62 & \multicolumn{1}{l|}{60.41} & \textbf{65.06} & \multicolumn{1}{l|}{16.48} & 7.21 & \multicolumn{1}{l|}{9.41} & 3.36 & \multicolumn{1}{l|}{31.63} & 23.79 \\ \cline{2-12} 
 & UA2 & \multicolumn{1}{l|}{61.37} & 67.56 & \multicolumn{1}{l|}{47.28} & \textbf{62.62} & \multicolumn{1}{l|}{39.75} & 16.21 & \multicolumn{1}{l|}{34.36} & 11.56 & \multicolumn{1}{l|}{44.80} & 33.43 \\ \cline{2-12} 
 & UA3 & \multicolumn{1}{l|}{\textbf{65.80}} & 62.31 & \multicolumn{1}{l|}{49.54} & 39.62 & \multicolumn{1}{l|}{39.77} & 47.97 & \multicolumn{1}{l|}{36.88} & 31.51 & \multicolumn{1}{l|}{52.97} & 44.92 \\ \hline
\end{tabular}
\end{table}

\begin{table}[!htbp]
\caption{Unknown vs Known: Second level OCC model's performance in terms of F1-score}
\label{Table5:unknownknownf1scoreclassification}
\begin{tabular}{|l|l|ll|ll|ll|ll|ll|}
\hline
\multicolumn{1}{|c|}{\multirow{2}{*}{Datasets}} & \multicolumn{1}{c|}{\multirow{2}{*}{NoUAC}} & \multicolumn{2}{c|}{usfAD} & \multicolumn{2}{c|}{LOF} & \multicolumn{2}{c|}{IOF} & \multicolumn{2}{c|}{OCSVM} & \multicolumn{2}{c|}{AE} \\ \cline{3-12} 
\multicolumn{1}{|c|}{} & \multicolumn{1}{c|}{} & \multicolumn{1}{l|}{C1} & C2 & \multicolumn{1}{l|}{C1} & C2 & \multicolumn{1}{l|}{C1} & C2 & \multicolumn{1}{l|}{C1} & C2 & \multicolumn{1}{l|}{C1} & C2 \\ \hline
\multirow{3}{*}{NSL-KDD} & UA1 & \multicolumn{1}{l|}{\textbf{91.33}} & \textbf{90.26} & \multicolumn{1}{l|}{73.78} & 82.66 & \multicolumn{1}{l|}{75.27} & 87.71 & \multicolumn{1}{l|}{64.70} & 45.81 & \multicolumn{1}{l|}{79.91} & 85.12 \\ \cline{2-12} 
 & UA2 & \multicolumn{1}{l|}{71.30} & \textbf{90.29} & \multicolumn{1}{l|}{74.20} & 74.17 & \multicolumn{1}{l|}{87.13} & 74.73 & \multicolumn{1}{l|}{58.88} & 58.10 & \multicolumn{1}{l|}{\textbf{87.19}} & 79.13 \\ \cline{2-12} 
 & UA3 & \multicolumn{1}{l|}{48.63} & 70.65 & \multicolumn{1}{l|}{74.03} & 74.43 & \multicolumn{1}{l|}{\textbf{88.80}} & 86.96 & \multicolumn{1}{l|}{58.14} & 58.67 & \multicolumn{1}{l|}{86.79} & \textbf{87.01} \\ \hline
\multirow{3}{*}{ToN-IoT-Network} & UA1 & \multicolumn{1}{l|}{\textbf{96.72}} & \textbf{96.47} & \multicolumn{1}{l|}{85.01} & 89.23 & \multicolumn{1}{l|}{15.49} & 24.62 & \multicolumn{1}{l|}{3.14} & 0.09 & \multicolumn{1}{l|}{1.66} & 0.03 \\ \cline{2-12} 
 & UA2 & \multicolumn{1}{l|}{\textbf{96.91}} & \textbf{96.92} & \multicolumn{1}{l|}{86.81} & 86.21 & \multicolumn{1}{l|}{15.87} & 17.46 & \multicolumn{1}{l|}{6.54} & 3.36 & \multicolumn{1}{l|}{2.03} & 2.75 \\ \cline{2-12} 
 & UA3 & \multicolumn{1}{l|}{\textbf{97.18}} & \textbf{97.46} & \multicolumn{1}{l|}{88.26} & 85.57 & \multicolumn{1}{l|}{17.38} & 19.33 & \multicolumn{1}{l|}{3.73} & 10.08 & \multicolumn{1}{l|}{3.77} & 7.50 \\ \hline
\multirow{3}{*}{UNSW-NB15} & UA1 & \multicolumn{1}{l|}{\textbf{89.61}} & 49.78 & \multicolumn{1}{l|}{85.04} & 62.52 & \multicolumn{1}{l|}{76.97} & \textbf{88.70} & \multicolumn{1}{l|}{51.50} & 49.84 & \multicolumn{1}{l|}{68.98} & 84.70 \\ \cline{2-12} 
 & UA2 & \multicolumn{1}{l|}{67.48} & 85.77 & \multicolumn{1}{l|}{74.85} & \textbf{87.41} & \multicolumn{1}{l|}{\textbf{78.18}} & 82.40 & \multicolumn{1}{l|}{51.97} & 47.40 & \multicolumn{1}{l|}{74.18} & 73.34 \\ \cline{2-12} 
 & UA3 & \multicolumn{1}{l|}{55.99} & 59.43 & \multicolumn{1}{l|}{74.64} & \textbf{79.99} & \multicolumn{1}{l|}{\textbf{84.51}} & 81.29 & \multicolumn{1}{l|}{54.28} & 54.76 & \multicolumn{1}{l|}{77.26} & 74.96 \\ \hline
\multirow{2}{*}{Malmem2022} & UA1 & \multicolumn{1}{l|}{47.70} & 45.65 & \multicolumn{1}{l|}{53.76} & \textbf{54.63} & \multicolumn{1}{l|}{54.26} & 51.96 & \multicolumn{1}{l|}{39.03} & 36.91 & \multicolumn{1}{l|}{\textbf{57.21}} & 54.01 \\ \cline{2-12} 
 & UA2 & \multicolumn{1}{l|}{47.28} & 42.08 & \multicolumn{1}{l|}{41.49} & 34.88 & \multicolumn{1}{l|}{32.22} & 26.53 & \multicolumn{1}{l|}{37.49} & 31.51 & \multicolumn{1}{l|}{38.98} & 28.54 \\ \hline
\multirow{2}{*}{ISCXURL2016} & UA1 & \multicolumn{1}{l|}{\textbf{79.75}} & \textbf{80.41} & \multicolumn{1}{l|}{79.36} & 74.42 & \multicolumn{1}{l|}{54.45} & 59.95 & \multicolumn{1}{l|}{46.17} & 54.19 & \multicolumn{1}{l|}{54.78} & 56.45 \\ \cline{2-12} 
 & UA2 & \multicolumn{1}{l|}{\textbf{86.10}} & \textbf{76.88} & \multicolumn{1}{l|}{75.10} & 78.73 & \multicolumn{1}{l|}{56.97} & 39.36 & \multicolumn{1}{l|}{65.11} & 48.63 & \multicolumn{1}{l|}{51.04} & 43.83 \\ \hline
\multirow{3}{*}{CIC-DDoS2019} & UA1 & \multicolumn{1}{l|}{89.32} & 95.79 & \multicolumn{1}{l|}{84.16} & 84.51 & \multicolumn{1}{l|}{81.34} & 89.56 & \multicolumn{1}{l|}{17.60} & 58.64 & \multicolumn{1}{l|}{44.26} & 92.35 \\ \cline{2-12} 
 & UA2 & \multicolumn{1}{l|}{91.24} & 93.17 & \multicolumn{1}{l|}{80.53} & 86.70 & \multicolumn{1}{l|}{74.49} & 86.90 & \multicolumn{1}{l|}{28.36} & 7.62 & \multicolumn{1}{l|}{41.87} & 43.18 \\ \cline{2-12} 
 & UA3 & \multicolumn{1}{l|}{93.94} & 88.30 & \multicolumn{1}{l|}{87.64} & 78.37 & \multicolumn{1}{l|}{90.81} & 67.57 & \multicolumn{1}{l|}{77.20} & 29.83 & \multicolumn{1}{l|}{87.67} & 37.94 \\ \hline
\multirow{3}{*}{CIC-DDoS2017} & UA1 & \multicolumn{1}{l|}{\textbf{64.23}} & \textbf{68.53} & \multicolumn{1}{l|}{45.88} & 36.38 & \multicolumn{1}{l|}{42.99} & 17.98 & \multicolumn{1}{l|}{4.05} & 1.50 & \multicolumn{1}{l|}{50.72} & 24.96 \\ \cline{2-12} 
 & UA2 & \multicolumn{1}{l|}{\textbf{61.07}} & \textbf{61.38} & \multicolumn{1}{l|}{22.99} & 44.52 & \multicolumn{1}{l|}{24.07} & 37.88 & \multicolumn{1}{l|}{2.22} & 3.69 & \multicolumn{1}{l|}{33.86} & 46.09 \\ \cline{2-12} 
 & UA3 & \multicolumn{1}{l|}{\textbf{58.12}} & \textbf{60.83} & \multicolumn{1}{l|}{21.60} & 28.20 & \multicolumn{1}{l|}{20.83} & 19.46 & \multicolumn{1}{l|}{5.65} & 8.87 & \multicolumn{1}{l|}{31.77} & 34.56 \\ \hline
\multirow{3}{*}{ToN-IoT-Linux} & UA1 & \multicolumn{1}{l|}{\textbf{82.27}} & \textbf{77.77} & \multicolumn{1}{l|}{78.96} & 76.64 & \multicolumn{1}{l|}{45.67} & 41.94 & \multicolumn{1}{l|}{6.40} & 9.95 & \multicolumn{1}{l|}{37.04} & 41.14 \\ \cline{2-12} 
 & UA2 & \multicolumn{1}{l|}{\textbf{77.72}} & \textbf{78.02} & \multicolumn{1}{l|}{77.36} & 77.44 & \multicolumn{1}{l|}{32.33} & 30.30 & \multicolumn{1}{l|}{8.55} & 7.54 & \multicolumn{1}{l|}{31.63} & 33.28 \\ \cline{2-12} 
 & UA3 & \multicolumn{1}{l|}{\textbf{88.15}} & \textbf{81.67} & \multicolumn{1}{l|}{84.58} & 78.57 & \multicolumn{1}{l|}{28.36} & 35.56 & \multicolumn{1}{l|}{10.28} & 9.18 & \multicolumn{1}{l|}{29.09} & 32.85 \\ \hline
\multirow{3}{*}{XIIOTID} & UA1 & \multicolumn{1}{l|}{\textbf{95.02}} & \textbf{94.93} & \multicolumn{1}{l|}{84.61} & 94.09 & \multicolumn{1}{l|}{80.96} & 76.03 & \multicolumn{1}{l|}{65.56} & 40.25 & \multicolumn{1}{l|}{72.44} & 79.75 \\ \cline{2-12} 
 & UA2 & \multicolumn{1}{l|}{\textbf{94.81}} & \textbf{94.90} & \multicolumn{1}{l|}{84.79} & 87.25 & \multicolumn{1}{l|}{76.66} & 80.33 & \multicolumn{1}{l|}{38.28} & 63.29 & \multicolumn{1}{l|}{73.92} & 70.59 \\ \cline{2-12} 
 & UA3 & \multicolumn{1}{l|}{\textbf{94.50}} & \textbf{94.93} & \multicolumn{1}{l|}{87.46} & 85.07 & \multicolumn{1}{l|}{76.38} & 77.54 & \multicolumn{1}{l|}{43.27} & 38.72 & \multicolumn{1}{l|}{72.21} & 78.81 \\ \hline
\multirow{3}{*}{CIC-Darknet2020} & UA1 & \multicolumn{1}{l|}{\textbf{63.03}} & \textbf{67.70} & \multicolumn{1}{l|}{60.35} & 67.70 & \multicolumn{1}{l|}{23.34} & 12.04 & \multicolumn{1}{l|}{3.55} & 4.57 & \multicolumn{1}{l|}{44.60} & 37.51 \\ \cline{2-12} 
 & UA2 & \multicolumn{1}{l|}{\textbf{58.51}} & \textbf{64.27} & \multicolumn{1}{l|}{47.39} & 62.47 & \multicolumn{1}{l|}{52.41} & 16.98 & \multicolumn{1}{l|}{42.68} & 5.51 & \multicolumn{1}{l|}{57.58} & 44.57 \\ \cline{2-12} 
 & UA3 & \multicolumn{1}{l|}{60.22} & \textbf{58.02} & \multicolumn{1}{l|}{50.31} & 39.83 & \multicolumn{1}{l|}{49.55} & 59.12 & \multicolumn{1}{l|}{45.58} & 38.51 & \multicolumn{1}{l|}{\textbf{62.68}} & 56.95 \\ \hline
\end{tabular}
\end{table}

\subsection{Performance of Semi-Supervised Learner at the Second Level in Retraining Process}

\begin{figure}[!htbp]
    \centering
    \includegraphics[scale = .90]{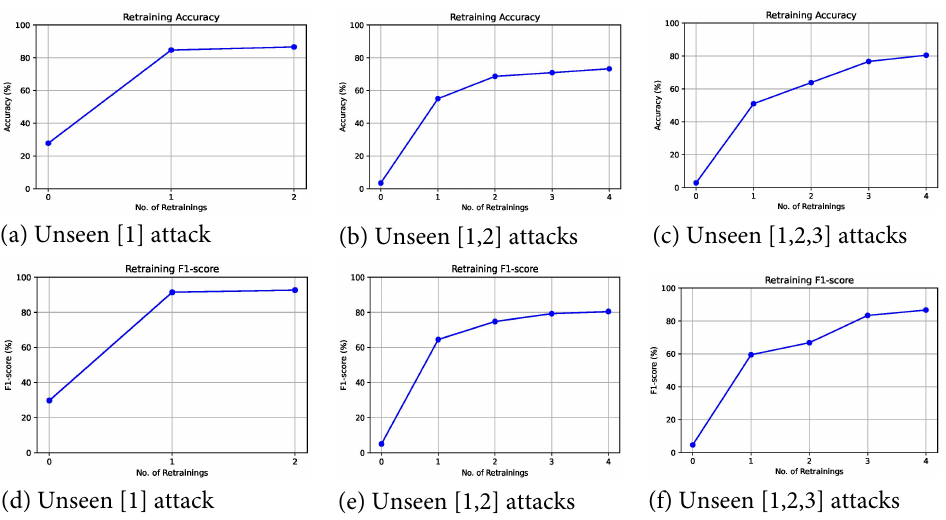}
    \caption{Impact of retraining on accuracy and F1-score on NSL-KDD datasets}
    \label{fig:3}
\end{figure}

\begin{figure}[!htbp]
    \centering
    \includegraphics[scale = .90]{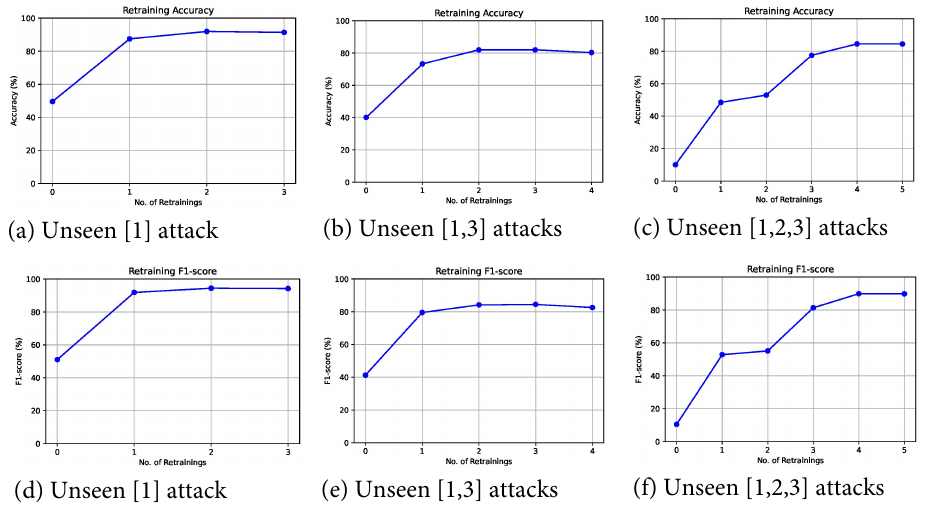}
    \caption{Impact of retraining on accuracy and F1-score on UNSW-NB15 datasets}
    \label{fig:4}
\end{figure}

\begin{figure}[!htbp]
    \centering
    \includegraphics[scale = .92]{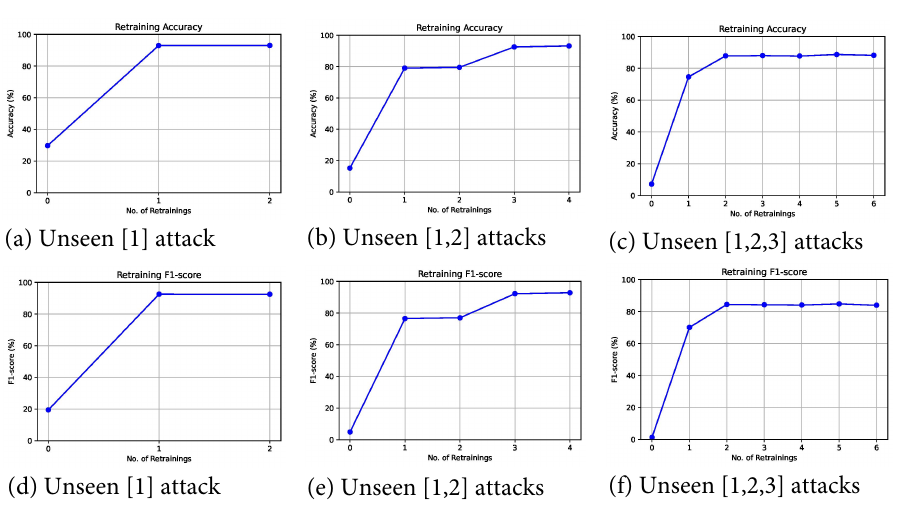}
    \caption{Impact of retraining on accuracy and F1-score on CICDDoS2019 datasets}
    \label{fig:5}
\end{figure}

\begin{figure}[!htbp]
    \centering
    \includegraphics[scale = .90]{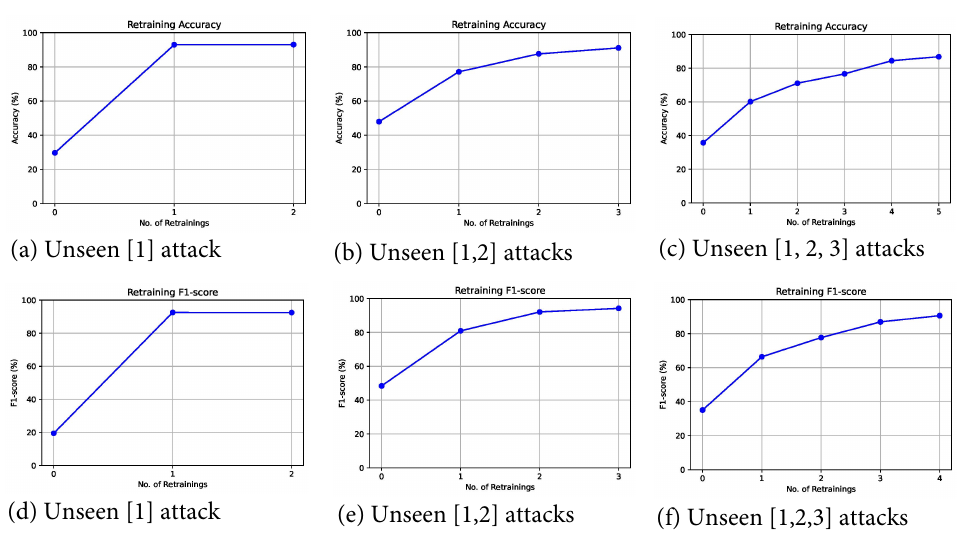}
    \caption{Impact of retraining on accuracy and F1-score on ToN-IoT-Network datasets}
    \label{fig:6}
\end{figure}

We retrain the OCC model at the second level. The model initiates retraining when approximately 1000 unknown points are gathered. These 1000 unknown instances are forwarded to the DBSCAN clustering algorithm, which then forms multiple clusters. We primarily concentrate on the predominant class within the largest cluster. The samples from the dominant class in the largest cluster are added with prior training samples to retrain the second-layer OCC model and the random forest algorithm. On the test dataset, we repeat this process until the clusters incorporate all unknown attack categories for retraining purposes. The necessity for multiple retraining is directly proportional to the quantity of unknown attack samples in the testing set. In this context, we exhibit a retraining progress report that covers scenarios where one, two, and three attack classes are unknown/unseen, utilizing usfAD model (we have chosen usfAD because it shows better results than other methods in detecting known and unknown attack samples across most of the datasets) and four datasets: NSL-KDD, UNSW-NB15, CICDDoS2019, and ToN-IoT-Network, which serve as representatives for other IDS datasets and OCC models. Referencing Figures \ref{fig:3}, \ref{fig:4}, \ref{fig:5}, and \ref{fig:6}, there is a noticeable increasing pattern in both accuracy and F1-score correlating to the varying numbers of unknown attack classes/categories. Herein, we assess the model's performance in learning and detecting various known attacks after retraining.

In this study, we primarily examine how effectively the second-level usfAD and other OCC detection methods learn to identify known attacks after the model retraining is completed. We evaluate the performance of these models in terms of accuracy and average weighted F1-score prior to starting the retraining process in Table \ref{Table6:knownaccclassification} and \ref{Table7:knownf1scoreclassification}. We observe that the models exhibit slightly higher accuracy and F1-score when only one class is unknown. This is because, in such scenarios, the models have been trained with the majority of attack classes, resulting in somewhat better detection performance for known classes even before retraining begins. However, the performance drops when two or three attack classes are unknown, which is understandable as the model gradually learns to recognize these categories through various retraining stages.

\begin{table}[!htbp]
\caption{Accuracy of known attack before retraining second level OCC models}
\label{Table6:knownaccclassification}
\begin{tabular}{|l|l|ll|ll|ll|ll|ll|}
\hline
\multirow{2}{*}{\textbf{Datasets}} & \multicolumn{1}{c|}{\multirow{2}{*}{\textbf{NoUAC}}} & \multicolumn{2}{c|}{usfAD} & \multicolumn{2}{c|}{LOF} & \multicolumn{2}{c|}{IOF} & \multicolumn{2}{c|}{OCSVM} & \multicolumn{2}{c|}{AE} \\ \cline{3-12} 
 & \multicolumn{1}{c|}{} & \multicolumn{1}{l|}{C1} & C2 & \multicolumn{1}{l|}{C1} & C2 & \multicolumn{1}{l|}{C1} & C2 & \multicolumn{1}{l|}{C1} & C2 & \multicolumn{1}{l|}{C1} & C2 \\ \hline
\multirow{3}{*}{NSL-KDD} & UA1 & \multicolumn{1}{l|}{21.88} & \textbf{72.95} & \multicolumn{1}{l|}{25.56} & 59.23 & \multicolumn{1}{l|}{15.44} & 69.87 & \multicolumn{1}{l|}{24.54} & 27.36 & \multicolumn{1}{l|}{\textbf{26.02}} & 67.77 \\ \cline{2-12} 
 & UA2 & \multicolumn{1}{l|}{2.42} & 19.00 & \multicolumn{1}{l|}{\textbf{4.49}} & 22.65 & \multicolumn{1}{l|}{0.70} & 14.33 & \multicolumn{1}{l|}{5.74} & 19.24 & \multicolumn{1}{l|}{3.30} & \textbf{23.07} \\ \cline{2-12} 
 & UA3 & \multicolumn{1}{l|}{0.00} & 2.32 & \multicolumn{1}{l|}{\textbf{19.92}} & 4.53 & \multicolumn{1}{l|}{1.85} & 0.34 & \multicolumn{1}{l|}{3.60} & \textbf{5.60} & \multicolumn{1}{l|}{2.16} & 3.17 \\ \hline
\multirow{3}{*}{ToN-IoT-Network} & UA1 & \multicolumn{1}{l|}{\textbf{60.08}} & \textbf{84.24} & \multicolumn{1}{l|}{52.18} & 73.32 & \multicolumn{1}{l|}{1.08} & 14.13 & \multicolumn{1}{l|}{1.34} & 0.04 & \multicolumn{1}{l|}{0.40} & 0.00 \\ \cline{2-12} 
 & UA2 & \multicolumn{1}{l|}{\textbf{48.24}} & \textbf{48.07} & \multicolumn{1}{l|}{39.67} & 39.46 & \multicolumn{1}{l|}{1.04} & 0.51 & \multicolumn{1}{l|}{3.06} & 0.00 & \multicolumn{1}{l|}{0.42} & 0.40 \\ \cline{2-12} 
 & UA3 & \multicolumn{1}{l|}{\textbf{36.24}} & \textbf{35.75} & \multicolumn{1}{l|}{27.26} & 32.99 & \multicolumn{1}{l|}{0.24} & 0.68 & \multicolumn{1}{l|}{0.04} & 4.57 & \multicolumn{1}{l|}{0.67} & 0.00 \\ \hline
\multirow{3}{*}{UNSW-NB15} & UA1 & \multicolumn{1}{l|}{\textbf{49.93}} & 64.66 & \multicolumn{1}{l|}{37.56} & 59.60 & \multicolumn{1}{l|}{5.12} & \textbf{80.75} & \multicolumn{1}{l|}{25.71} & 31.54 & \multicolumn{1}{l|}{5.64} & 74.09 \\ \cline{2-12} 
 & UA2 & \multicolumn{1}{l|}{\textbf{19.82}} & \textbf{40.32} & \multicolumn{1}{l|}{14.10} & 30.03 & \multicolumn{1}{l|}{2.39} & 4.07 & \multicolumn{1}{l|}{14.68} & 16.06 & \multicolumn{1}{l|}{3.20} & 4.64 \\ \cline{2-12} 
 & UA3 & \multicolumn{1}{l|}{\textbf{10.49}} & \textbf{9.34} & \multicolumn{1}{l|}{7.80} & 6.75 & \multicolumn{1}{l|}{1.37} & 1.30 & \multicolumn{1}{l|}{5.41} & 10.72 & \multicolumn{1}{l|}{1.54} & 2.07 \\ \hline
\multirow{2}{*}{Malmem2022} & UA1 & \multicolumn{1}{l|}{48.93} & 49.69 & \multicolumn{1}{l|}{48.80} & 47.45 & \multicolumn{1}{l|}{50.22} & 50.47 & \multicolumn{1}{l|}{21.94} & 22.31 & \multicolumn{1}{l|}{\textbf{54.21}} & \textbf{54.56} \\ \cline{2-12} 
 & UA2 & \multicolumn{1}{l|}{21.31} & 23.48 & \multicolumn{1}{l|}{23.11} & 25.06 & \multicolumn{1}{l|}{24.66} & 25.58 & \multicolumn{1}{l|}{11.25} & 10.68 & \multicolumn{1}{l|}{\textbf{26.58}} & \textbf{27.76} \\ \hline
\multirow{2}{*}{ISCXURL} & UA1 & \multicolumn{1}{l|}{56.02} & 41.47 & \multicolumn{1}{l|}{56.39} & 42.36 & \multicolumn{1}{l|}{58.21} & 43.84 & \multicolumn{1}{l|}{27.99} & 22.78 & \multicolumn{1}{l|}{\textbf{59.59}} & \textbf{43.14} \\ \cline{2-12} 
 & UA2 & \multicolumn{1}{l|}{18.81} & 35.54 & \multicolumn{1}{l|}{16.20} & 40.94 & \multicolumn{1}{l|}{15.35} & 42.90 & \multicolumn{1}{l|}{9.35} & 21.68 & \multicolumn{1}{l|}{\textbf{16.60}} & \textbf{44.34} \\ \hline
\multirow{3}{*}{CIC-DDoS2019} & UA1 & \multicolumn{1}{l|}{\textbf{28.87}} & 82.79 & \multicolumn{1}{l|}{27.15} & 74.36 & \multicolumn{1}{l|}{14.26} & 81.79 & \multicolumn{1}{l|}{8.70} & 40.99 & \multicolumn{1}{l|}{16.47} & \textbf{86.37} \\ \cline{2-12} 
 & UA2 & \multicolumn{1}{l|}{14.91} & \textbf{21.00} & \multicolumn{1}{l|}{13.92} & 19.88 & \multicolumn{1}{l|}{14.39} & 14.20 & \multicolumn{1}{l|}{7.33} & 5.49 & \multicolumn{1}{l|}{\textbf{17.03}} & 9.62 \\ \cline{2-12} 
 & UA3 & \multicolumn{1}{l|}{7.46} & 10.59 & \multicolumn{1}{l|}{6.87} & 10.17 & \multicolumn{1}{l|}{14.10} & 6.35 & \multicolumn{1}{l|}{3.96} & 5.37 & \multicolumn{1}{l|}{\textbf{8.91}} & \textbf{11.00} \\ \hline
\multirow{3}{*}{CIC-DDoS2017} & UA1 & \multicolumn{1}{l|}{\textbf{36.47}} & \textbf{47.81} & \multicolumn{1}{l|}{27.71} & 22.08 & \multicolumn{1}{l|}{27.59} & 10.78 & \multicolumn{1}{l|}{2.07} & 0.71 & \multicolumn{1}{l|}{34.33} & 15.34 \\ \cline{2-12} 
 & UA2 & \multicolumn{1}{l|}{20.88} & 27.91 & \multicolumn{1}{l|}{9.71} & 26.36 & \multicolumn{1}{l|}{15.38} & 23.85 & \multicolumn{1}{l|}{1.13} & 1.88 & \multicolumn{1}{l|}{\textbf{22.26}} & \textbf{30.51} \\ \cline{2-12} 
 & UA3 & \multicolumn{1}{l|}{11.84} & 14.92 & \multicolumn{1}{l|}{8.26} & 13.94 & \multicolumn{1}{l|}{12.61} & 12.14 & \multicolumn{1}{l|}{2.90} & 4.72 & \multicolumn{1}{l|}{\textbf{20.29}} & \textbf{22.89} \\ \hline
\multirow{3}{*}{ToN-IoT-Linux} & UA1 & \multicolumn{1}{l|}{\textbf{69.04}} & \textbf{77.95} & \multicolumn{1}{l|}{60.90} & 67.30 & \multicolumn{1}{l|}{32.06} & 28.21 & \multicolumn{1}{l|}{2.70} & 5.07 & \multicolumn{1}{l|}{23.86} & 27.43 \\ \cline{2-12} 
 & UA2 & \multicolumn{1}{l|}{\textbf{47.85}} & \textbf{47.60} & \multicolumn{1}{l|}{43.22} & 43.14 & \multicolumn{1}{l|}{19.30} & 15.93 & \multicolumn{1}{l|}{3.12} & 2.68 & \multicolumn{1}{l|}{17.36} & 18.94 \\ \cline{2-12} 
 & UA3 & \multicolumn{1}{l|}{\textbf{25.96}} & \textbf{32.59} & \multicolumn{1}{l|}{25.19} & 26.86 & \multicolumn{1}{l|}{12.58} & 13.16 & \multicolumn{1}{l|}{3.19} & 2.25 & \multicolumn{1}{l|}{11.63} & 12.24 \\ \hline
\multirow{3}{*}{XIIOTID} & UA1 & \multicolumn{1}{l|}{63.03} & 62.25 & \multicolumn{1}{l|}{65.30} & \textbf{85.02} & \multicolumn{1}{l|}{\textbf{69.24}} & 21.03 & \multicolumn{1}{l|}{43.74} & 20.74 & \multicolumn{1}{l|}{49.28} & 44.69 \\ \cline{2-12} 
 & UA2 & \multicolumn{1}{l|}{30.46} & 47.16 & \multicolumn{1}{l|}{\textbf{57.94}} & 39.04 & \multicolumn{1}{l|}{17.24} & \textbf{64.22} & \multicolumn{1}{l|}{8.63} & 41.02 & \multicolumn{1}{l|}{16.79} & 43.44 \\ \cline{2-12} 
 & UA3 & \multicolumn{1}{l|}{14.65} & 23.27 & \multicolumn{1}{l|}{\textbf{31.76}} & \textbf{41.98} & \multicolumn{1}{l|}{11.92} & 17.15 & \multicolumn{1}{l|}{6.99} & 4.95 & \multicolumn{1}{l|}{11.00} & 10.85 \\ \hline
\multirow{3}{*}{CIC-Darknet2020} & UA1 & \multicolumn{1}{l|}{27.18} & 55.73 & \multicolumn{1}{l|}{\textbf{35.68}} & \textbf{64.07} & \multicolumn{1}{l|}{13.50} & 6.44 & \multicolumn{1}{l|}{1.15} & 2.32 & \multicolumn{1}{l|}{29.93} & 23.38 \\ \cline{2-12} 
 & UA2 & \multicolumn{1}{l|}{18.63} & 19.36 & \multicolumn{1}{l|}{22.07} & 29.46 & \multicolumn{1}{l|}{38.73} & 8.71 & \multicolumn{1}{l|}{26.98} & 1.92 & \multicolumn{1}{l|}{\textbf{43.51}} & \textbf{30.68} \\ \cline{2-12} 
 & UA3 & \multicolumn{1}{l|}{10.81} & 14.79 & \multicolumn{1}{l|}{15.61} & 18.48 & \multicolumn{1}{l|}{36.26} & \textbf{46.42} & \multicolumn{1}{l|}{30.22} & 23.42 & \multicolumn{1}{l|}{\textbf{51.34}} & 43.28 \\ \hline
\end{tabular}
\end{table}

\begin{table}[!htbp]
\caption{Average weighted F1-score of known attacks before retraining second level OCC models}
\label{Table7:knownf1scoreclassification}
\begin{tabular}{|l|l|ll|ll|ll|ll|ll|}
\hline
\multirow{2}{*}{\textbf{Datasets}} & \multicolumn{1}{c|}{\multirow{2}{*}{\textbf{NoUAC}}} & \multicolumn{2}{c|}{usfAD} & \multicolumn{2}{c|}{LOF} & \multicolumn{2}{c|}{IOF} & \multicolumn{2}{c|}{OCSVM} & \multicolumn{2}{c|}{AE} \\ \cline{3-12} 
 & \multicolumn{1}{c|}{} & \multicolumn{1}{l|}{C1} & C2 & \multicolumn{1}{l|}{C1} & C2 & \multicolumn{1}{l|}{C1} & C2 & \multicolumn{1}{l|}{C1} & C2 & \multicolumn{1}{l|}{} &  \\ \hline
\multirow{3}{*}{NSL-KDD} & UA1 & \multicolumn{1}{l|}{22.68} & \textbf{73.82} & \multicolumn{1}{l|}{27.77} & 63.38 & \multicolumn{1}{l|}{18.32} & 72.40 & \multicolumn{1}{l|}{\textbf{33.56}} & 37.23 & \multicolumn{1}{l|}{28.18} & 70.72 \\ \cline{2-12} 
 & UA2 & \multicolumn{1}{l|}{2.64} & 19.79 & \multicolumn{1}{l|}{5.48} & 25.16 & \multicolumn{1}{l|}{1.19} & 17.21 & \multicolumn{1}{l|}{\textbf{9.79}} & \textbf{27.60} & \multicolumn{1}{l|}{4.59} & 25.58 \\ \cline{2-12} 
 & UA3 & \multicolumn{1}{l|}{0.00} & 2.57 & \multicolumn{1}{l|}{\textbf{20.40}} & 5.61 & \multicolumn{1}{l|}{3.03} & 0.52 & \multicolumn{1}{l|}{6.49} & \textbf{9.58} & \multicolumn{1}{l|}{3.58} & 4.43 \\ \hline
\multirow{3}{*}{ToN-IoT-Network} & UA1 & \multicolumn{1}{l|}{\textbf{60.60}} & \textbf{85.00} & \multicolumn{1}{l|}{55.87} & 76.47 & \multicolumn{1}{l|}{2.07} & 21.42 & \multicolumn{1}{l|}{2.31} & 0.08 & \multicolumn{1}{l|}{0.79} & 0.00 \\ \cline{2-12} 
 & UA2 & \multicolumn{1}{l|}{\textbf{48.77}} & \textbf{48.54} & \multicolumn{1}{l|}{42.30} & 41.49 & \multicolumn{1}{l|}{2.00} & 1.01 & \multicolumn{1}{l|}{4.89} & 0.00 & \multicolumn{1}{l|}{0.83} & 0.79 \\ \cline{2-12} 
 & UA3 & \multicolumn{1}{l|}{\textbf{36.69}} & \textbf{36.06} & \multicolumn{1}{l|}{28.24} & 35.75 & \multicolumn{1}{l|}{0.47} & 1.31 & \multicolumn{1}{l|}{0.08} & 6.75 & \multicolumn{1}{l|}{1.33} & 0.00 \\ \hline
\multirow{3}{*}{UNSW-NB15} & UA1 & \multicolumn{1}{l|}{\textbf{51.45}} & 65.76 & \multicolumn{1}{l|}{40.37} & 65.24 & \multicolumn{1}{l|}{7.63} & \textbf{83.99} & \multicolumn{1}{l|}{36.36} & 41.71 & \multicolumn{1}{l|}{8.92} & 76.89 \\ \cline{2-12} 
 & UA2 & \multicolumn{1}{l|}{\textbf{20.89}} & \textbf{41.46} & \multicolumn{1}{l|}{15.86} & 32.20 & \multicolumn{1}{l|}{3.85} & 5.67 & \multicolumn{1}{l|}{21.37} & 23.80 & \multicolumn{1}{l|}{5.18} & 7.14 \\ \cline{2-12} 
 & UA3 & \multicolumn{1}{l|}{\textbf{10.90}} & 10.17 & \multicolumn{1}{l|}{8.63} & 7.97 & \multicolumn{1}{l|}{1.99} & 2.26 & \multicolumn{1}{l|}{7.82} & \textbf{16.23} & \multicolumn{1}{l|}{2.41} & 3.47 \\ \hline
\multirow{2}{*}{Malmem2022} & UA1 & \multicolumn{1}{l|}{52.15} & 52.96 & \multicolumn{1}{l|}{52.20} & 51.82 & \multicolumn{1}{l|}{52.64} & 53.12 & \multicolumn{1}{l|}{29.23} & 29.41 & \multicolumn{1}{l|}{\textbf{57.15}} & \textbf{57.45} \\ \cline{2-12} 
 & UA2 & \multicolumn{1}{l|}{23.78} & 25.91 & \multicolumn{1}{l|}{24.89} & 26.83 & \multicolumn{1}{l|}{25.72} & 27.29 & \multicolumn{1}{l|}{14.98} & 14.42 & \multicolumn{1}{l|}{\textbf{28.02}} & \textbf{29.43} \\ \hline
\multirow{2}{*}{ISCXURL} & UA1 & \multicolumn{1}{l|}{60.42} & 44.04 & \multicolumn{1}{l|}{61.83} & 46.72 & \multicolumn{1}{l|}{62.70} & 47.46 & \multicolumn{1}{l|}{40.06} & 32.19 & \multicolumn{1}{l|}{\textbf{64.43}} & \textbf{46.63} \\ \cline{2-12} 
 & UA2 & \multicolumn{1}{l|}{\textbf{19.70}} & 39.20 & \multicolumn{1}{l|}{18.11} & 44.92 & \multicolumn{1}{l|}{17.04} & 46.50 & \multicolumn{1}{l|}{13.96} & 30.65 & \multicolumn{1}{l|}{18.58} & \textbf{47.98} \\ \hline
\multirow{3}{*}{CIC-DDoS2019} & UA1 & \multicolumn{1}{l|}{29.31} & 83.36 & \multicolumn{1}{l|}{\textbf{28.72}} & 78.06 & \multicolumn{1}{l|}{15.38} & 83.22 & \multicolumn{1}{l|}{10.98} & 51.22 & \multicolumn{1}{l|}{17.86} & \textbf{88.87} \\ \cline{2-12} 
 & UA2 & \multicolumn{1}{l|}{\textbf{15.06}} & \textbf{21.39} & \multicolumn{1}{l|}{15.33} & 21.13 & \multicolumn{1}{l|}{14.82} & 15.42 & \multicolumn{1}{l|}{8.84} & 8.01 & \multicolumn{1}{l|}{17.21} & 10.75 \\ \cline{2-12} 
 & UA3 & \multicolumn{1}{l|}{7.55} & 10.77 & \multicolumn{1}{l|}{8.08} & \textbf{11.32} & \multicolumn{1}{l|}{\textbf{14.47}} & 6.67 & \multicolumn{1}{l|}{4.97} & 6.01 & \multicolumn{1}{l|}{9.41} & 11.23 \\ \hline
\multirow{3}{*}{CIC-DDoS2017} & UA1 & \multicolumn{1}{l|}{40.76} & \textbf{51.56} & \multicolumn{1}{l|}{39.30} & 26.50 & \multicolumn{1}{l|}{35.09} & 15.29 & \multicolumn{1}{l|}{2.75} & 0.98 & \multicolumn{1}{l|}{\textbf{47.48}} & 24.37 \\ \cline{2-12} 
 & UA2 & \multicolumn{1}{l|}{24.14} & 31.94 & \multicolumn{1}{l|}{13.64} & 37.90 & \multicolumn{1}{l|}{21.62} & 30.09 & \multicolumn{1}{l|}{1.80} & 2.57 & \multicolumn{1}{l|}{\textbf{33.75}} & \textbf{42.35} \\ \cline{2-12} 
 & UA3 & \multicolumn{1}{l|}{14.77} & 17.53 & \multicolumn{1}{l|}{12.32} & 20.74 & \multicolumn{1}{l|}{18.14} & 17.78 & \multicolumn{1}{l|}{5.39} & 8.54 & \multicolumn{1}{l|}{\textbf{31.02}} & \textbf{34.54} \\ \hline
\multirow{3}{*}{ToN-IoT-Linux} & UA1 & \multicolumn{1}{l|}{\textbf{69.25}} & \textbf{78.14} & \multicolumn{1}{l|}{63.28} & 69.97 & \multicolumn{1}{l|}{43.98} & 40.65 & \multicolumn{1}{l|}{5.22} & 9.51 & \multicolumn{1}{l|}{30.86} & 35.67 \\ \cline{2-12} 
 & UA2 & \multicolumn{1}{l|}{\textbf{47.97}} & \textbf{47.75} & \multicolumn{1}{l|}{44.99} & 44.83 & \multicolumn{1}{l|}{27.93} & 23.89 & \multicolumn{1}{l|}{5.99} & 5.19 & \multicolumn{1}{l|}{23.17} & 24.59 \\ \cline{2-12} 
 & UA3 & \multicolumn{1}{l|}{\textbf{26.08}} & \textbf{32.69} & \multicolumn{1}{l|}{26.18} & 28.02 & \multicolumn{1}{l|}{19.83} & 19.68 & \multicolumn{1}{l|}{6.13} & 4.36 & \multicolumn{1}{l|}{15.75} & 14.50 \\ \hline
\multirow{3}{*}{XIIOTID} & UA1 & \multicolumn{1}{l|}{\textbf{63.85}} & \textbf{63.21} & \multicolumn{1}{l|}{67.27} & 87.60 & \multicolumn{1}{l|}{72.36} & 23.69 & \multicolumn{1}{l|}{54.67} & 26.45 & \multicolumn{1}{l|}{53.94} & 49.10 \\ \cline{2-12} 
 & UA2 & \multicolumn{1}{l|}{\textbf{31.20}} & \textbf{47.94} & \multicolumn{1}{l|}{59.88} & 39.91 & \multicolumn{1}{l|}{18.72} & 66.69 & \multicolumn{1}{l|}{11.53} & 51.42 & \multicolumn{1}{l|}{19.05} & 46.67 \\ \cline{2-12} 
 & UA3 & \multicolumn{1}{l|}{\textbf{15.44}} & \textbf{23.92} & \multicolumn{1}{l|}{32.58} & 43.74 & \multicolumn{1}{l|}{12.66} & 18.38 & \multicolumn{1}{l|}{9.77} & 6.83 & \multicolumn{1}{l|}{12.64} & 12.25 \\ \hline
\multirow{3}{*}{CICDarknet2020} & UA1 & \multicolumn{1}{l|}{31.51} & 61.06 & \multicolumn{1}{l|}{41.20} & \textbf{69.94} & \multicolumn{1}{l|}{21.69} & 10.58 & \multicolumn{1}{l|}{2.23} & 3.84 & \multicolumn{1}{l|}{\textbf{43.72}} & 35.58 \\ \cline{2-12} 
 & UA2 & \multicolumn{1}{l|}{22.44} & 22.29 & \multicolumn{1}{l|}{27.04} & 34.29 & \multicolumn{1}{l|}{52.90} & 14.16 & \multicolumn{1}{l|}{41.38} & 3.74 & \multicolumn{1}{l|}{\textbf{58.41}} & \textbf{44.71} \\ \cline{2-12} 
 & UA3 & \multicolumn{1}{l|}{12.99} & 18.18 & \multicolumn{1}{l|}{19.94} & 23.35 & \multicolumn{1}{l|}{50.23} & 60.52 & \multicolumn{1}{l|}{45.02} & 37.01 & \multicolumn{1}{l|}{\textbf{65.07}} & \textbf{57.95} \\ \hline
\end{tabular}
\end{table}

After collecting a certain number of unknown attacks (1,000 in our simulation), we employ the DBSCAN  \cite{ester1996density} clustering method to create groups of similar kinds of unknown attacks. Here, we focus on the largest cluster (having the largest members) for retraining our second-level OCC and supervised models. In a real-life scenario, one of the unknown attacks among others might form the majority. In our experiment, we notice that in the initial phase of retraining, the most prevalent unknown attack type forms the largest cluster and is utilized for retraining. In subsequent phases, as these attacks become recognized and classified as known, other types of unknown attacks emerge as the largest cluster. Consequently, our attention is centered on the similarity metrics of the largest cluster. 

The effectiveness of this approach depends on the purity of the largest cluster. Therefore, we measure the purity of the largest cluster by dividing the number of members in the dominant class by the total number of members in the cluster. If the number of the dominant class's members is $d$ and the total number of members in the cluster is $n$. The purity score or the largest cluster ($PSL1$) is calculated as follows:

$PSL1 = \frac{d}{n}$

In addition, we evaluate the effectiveness of DBSCAN in clustering members of the same unknown attack category into the largest group. This is measured by calculating the proportion of the dominant class's members in the cluster (denoted as $d$) to the total number of ground truth instances ($T_n$) of that dominant class present in the datasets used for clustering. The formula for this measurement, which we refer to as PSL2, is as follows:

$PSL2 = \frac{d}{T_n}$

The purity is between 0 and 1, where 1 indicates that all members of the cluster belong to a single class (perfect purity), and lower values indicate a more mixed cluster. We explain the $PSL1$ and $PSL2$ as follows:  
\begin{itemize}
    \item PSL1 ($\frac{d}{n}$) measures the concentration of the dominant class within the cluster, indicating how homogeneous the cluster is.
    \item PSL2 ( $\frac{d}{T_n}$) measures the coverage of the dominant class within the cluster compared to its overall presence in the dataset, indicating the clustering effectiveness for that particular class.
\end{itemize}

Using these two metrics, we proposed a new metric for measuring the quality of the largest cluster. We suggest using the harmonic mean of $PSL1$ and $PSL2$ to ensure that both metrics contribute significantly to the result. 
 $PSLC = 2\times \frac{(PSL1\times PSL2)}{(PSL1+PSL2)}$
 
In addition to DBSCAN, we also evaluated the effectiveness of Density Peak Clustering (DPC)  \cite{rodriguez2014clustering} using an implementation from GitHub  \cite{shane2023density}. Table \ref{Table12:puritylargestcluster} and  \ref{Table121:PLSC} show the comparison of the performance of DBSCAN and DPC algorithms based on three metrics: $PSL1$, $PSL2$, and $PSLC$.

Overall, Table \ref{Table121:PLSC} demonstrates that DBSCAN exhibits generally better  $PSLC$ scores than DPC, indicating a higher ability to capture the dominant unknown attack categories within the largest cluster. However, Table \ref{Table12:puritylargestcluster} shows that DPC demonstrates superior performance regarding the purity of the largest cluster, suggesting that its clusters contain a higher proportion of points belonging to the same class. On the other hand, DPC's ability to cover the dominant unknown attack categories within the largest cluster is weaker compared to DBSCAN. 

\begin{table}[!htbp]
\caption{The purity of the largest cluster for some datasets in the retraining process}
\label{Table12:puritylargestcluster}
\resizebox{\columnwidth}{!}{
\begin{tabular}{|l|llllll|llllll|}
\hline
\multicolumn{1}{|c|}{\multirow{3}{*}{Datasets}} & \multicolumn{6}{c|}{PLS1} & \multicolumn{6}{c|}{PLS2} \\ \cline{2-13} 
\multicolumn{1}{|c|}{} & \multicolumn{2}{c|}{UA1} & \multicolumn{2}{c|}{UA2} & \multicolumn{2}{c|}{UA3} & \multicolumn{2}{c|}{UA1} & \multicolumn{2}{c|}{UA2} & \multicolumn{2}{c|}{UA3} \\ \cline{2-13} 
\multicolumn{1}{|c|}{} & \multicolumn{1}{c|}{DBSCAN} & \multicolumn{1}{c|}{DPC} & \multicolumn{1}{c|}{DBSCAN} & \multicolumn{1}{c|}{DPC} & \multicolumn{1}{c|}{DBSCAN} & \multicolumn{1}{c|}{DPC} & \multicolumn{1}{c|}{DBSCAN} & \multicolumn{1}{c|}{DPC} & \multicolumn{1}{c|}{DBSCAN} & \multicolumn{1}{c|}{DPC} & \multicolumn{1}{c|}{DBSCAN} & \multicolumn{1}{c|}{DPC} \\ \hline
\multirow{3}{*}{NSL-KDD} & \multicolumn{1}{l|}{1.00} & \multicolumn{1}{l|}{1.00} & \multicolumn{1}{l|}{1.00} & \multicolumn{1}{l|}{1.00} & \multicolumn{1}{l|}{1.00} & 1.00 & \multicolumn{1}{l|}{\textbf{0.69}} & \multicolumn{1}{l|}{0.31} & \multicolumn{1}{l|}{0.69} & \multicolumn{1}{l|}{0.69} & \multicolumn{1}{l|}{0.69} & 0.69 \\ \cline{2-13} 
 & \multicolumn{1}{l|}{1.00} & \multicolumn{1}{l|}{1.00} & \multicolumn{1}{l|}{\textbf{0.52}} & \multicolumn{1}{l|}{0.46} & \multicolumn{1}{l|}{\textbf{1.00}} & 0.61 & \multicolumn{1}{l|}{\textbf{0.53}} & \multicolumn{1}{l|}{0.44} & \multicolumn{1}{l|}{0.41} & \multicolumn{1}{l|}{\textbf{0.47}} & \multicolumn{1}{l|}{0.60} & \textbf{0.74} \\ \cline{2-13} 
 & \multicolumn{1}{l|}{0.30} & \multicolumn{1}{l|}{\textbf{1.00}} & \multicolumn{1}{l|}{0.51} & \multicolumn{1}{l|}{0.51} & \multicolumn{1}{l|}{0.51} & \textbf{0.99} & \multicolumn{1}{l|}{0.17} & \multicolumn{1}{l|}{\textbf{0.40}} & \multicolumn{1}{l|}{\textbf{0.47}} & \multicolumn{1}{l|}{0.38} & \multicolumn{1}{l|}{\textbf{0.52}} & 0.45 \\ \hline
\multirow{3}{*}{CIC-DDoS2019} & \multicolumn{1}{l|}{\textbf{0.99}} & \multicolumn{1}{l|}{\textbf{0.99}} & \multicolumn{1}{l|}{\textbf{0.99}} & \multicolumn{1}{l|}{\textbf{0.99}} & \multicolumn{1}{l|}{\textbf{0.93}} & 0.78 & \multicolumn{1}{l|}{\textbf{0.97}} & \multicolumn{1}{l|}{0.54} & \multicolumn{1}{l|}{\textbf{0.97}} & \multicolumn{1}{l|}{0.49} & \multicolumn{1}{l|}{\textbf{0.97}} & 0.27 \\ \cline{2-13} 
 & \multicolumn{1}{l|}{0.94} & \multicolumn{1}{l|}{\textbf{1.00}} & \multicolumn{1}{l|}{0.98} & \multicolumn{1}{l|}{\textbf{1.00}} & \multicolumn{1}{l|}{0.87} & \textbf{0.93} & \multicolumn{1}{l|}{\textbf{0.78}} & \multicolumn{1}{l|}{0.53} & \multicolumn{1}{l|}{\textbf{0.90}} & \multicolumn{1}{l|}{0.47} & \multicolumn{1}{l|}{\textbf{0.99}} & 0.42 \\ \cline{2-13} 
 & \multicolumn{1}{l|}{0.94} & \multicolumn{1}{l|}{\textbf{1.00}} & \multicolumn{1}{l|}{0.97} & \multicolumn{1}{l|}{\textbf{0.99}} & \multicolumn{1}{l|}{0.92} & \textbf{1.00} & \multicolumn{1}{l|}{\textbf{0.74}} & \multicolumn{1}{l|}{0.57} & \multicolumn{1}{l|}{\textbf{1.00}} & \multicolumn{1}{l|}{0.88} & \multicolumn{1}{l|}{\textbf{0.82}} & 0.46 \\ \hline
\multirow{3}{*}{UNSW-NB15} & \multicolumn{1}{l|}{\textbf{1.00}} & \multicolumn{1}{l|}{0.94} & \multicolumn{1}{l|}{\textbf{1.00}} & \multicolumn{1}{l|}{\textbf{0.91}} & \multicolumn{1}{l|}{\textbf{0.98}} & 0.91 & \multicolumn{1}{l|}{\textbf{0.96}} & \multicolumn{1}{l|}{0.48} & \multicolumn{1}{l|}{\textbf{0.95}} & \multicolumn{1}{l|}{0.28} & \multicolumn{1}{l|}{\textbf{0.95}} & 0.27 \\ \cline{2-13} 
 & \multicolumn{1}{l|}{\textbf{0.99}} & \multicolumn{1}{l|}{0.90} & \multicolumn{1}{l|}{\textbf{0.89}} & \multicolumn{1}{l|}{0.89} & \multicolumn{1}{l|}{0.85} & \textbf{0.89} & \multicolumn{1}{l|}{\textbf{0.74}} & \multicolumn{1}{l|}{0.30} & \multicolumn{1}{l|}{\textbf{0.80}} & \multicolumn{1}{l|}{0.29} & \multicolumn{1}{l|}{\textbf{0.71}} & 0.29 \\ \cline{2-13} 
 & \multicolumn{1}{l|}{0.49} & \multicolumn{1}{l|}{\textbf{0.93}} & \multicolumn{1}{l|}{0.89} & \multicolumn{1}{l|}{0.91} & \multicolumn{1}{l|}{0.88} & \textbf{0.99} & \multicolumn{1}{l|}{\textbf{0.56}} & \multicolumn{1}{l|}{0.36} & \multicolumn{1}{l|}{\textbf{0.71}} & \multicolumn{1}{l|}{0.32} & \multicolumn{1}{l|}{\textbf{0.62}} & 0.43 \\ \hline
\multirow{3}{*}{ToN-IoT-Linux} & \multicolumn{1}{l|}{\textbf{0.98}} & \multicolumn{1}{l|}{0.95} & \multicolumn{1}{l|}{0.61} & \multicolumn{1}{l|}{\textbf{0.66}} & \multicolumn{1}{l|}{0.38} & \textbf{0.44} & \multicolumn{1}{l|}{\textbf{0.86}} & \multicolumn{1}{l|}{0.48} & \multicolumn{1}{l|}{\textbf{0.86}} & \multicolumn{1}{l|}{0.58} & \multicolumn{1}{l|}{\textbf{0.91}} & 0.54 \\ \cline{2-13} 
 & \multicolumn{1}{l|}{0.80} & \multicolumn{1}{l|}{\textbf{0.88}} & \multicolumn{1}{l|}{\textbf{0.54}} & \multicolumn{1}{l|}{0.48} & \multicolumn{1}{l|}{\textbf{0.52}} & 0.35 & \multicolumn{1}{l|}{\textbf{0.43}} & \multicolumn{1}{l|}{0.42} & \multicolumn{1}{l|}{\textbf{0.82}} & \multicolumn{1}{l|}{0.39} & \multicolumn{1}{l|}{\textbf{0.95}} & 0.42 \\ \cline{2-13} 
 & \multicolumn{1}{l|}{0.46} & \multicolumn{1}{l|}{\textbf{0.81}} & \multicolumn{1}{l|}{\textbf{0.65}} & \multicolumn{1}{l|}{0.58} & \multicolumn{1}{l|}{0.45} & \textbf{0.49} & \multicolumn{1}{l|}{0.24} & \multicolumn{1}{l|}{0.31} & \multicolumn{1}{l|}{\textbf{0.51}} & \multicolumn{1}{l|}{0.28} & \multicolumn{1}{l|}{\textbf{0.66}} & 0.39 \\ \hline
\multirow{3}{*}{ToN-IoT-Network} & \multicolumn{1}{l|}{\textbf{0.97}} & \multicolumn{1}{l|}{0.96} & \multicolumn{1}{l|}{0.74} & \multicolumn{1}{l|}{\textbf{0.81}} & \multicolumn{1}{l|}{0.74} & \textbf{0.83} & \multicolumn{1}{l|}{\textbf{0.96}} & \multicolumn{1}{l|}{0.27} & \multicolumn{1}{l|}{\textbf{0.96}} & \multicolumn{1}{l|}{0.77} & \multicolumn{1}{l|}{\textbf{0.95}} & 0.33 \\ \cline{2-13} 
 & \multicolumn{1}{l|}{0.87} & \multicolumn{1}{l|}{\textbf{0.88}} & \multicolumn{1}{l|}{0.63} & \multicolumn{1}{l|}{\textbf{0.83}} & \multicolumn{1}{l|}{0.70} & \textbf{0.75} & \multicolumn{1}{l|}{\textbf{0.77}} & \multicolumn{1}{l|}{0.31} & \multicolumn{1}{l|}{\textbf{0.97}} & \multicolumn{1}{l|}{0.27} & \multicolumn{1}{l|}{\textbf{0.95}} & 0.93 \\ \cline{2-13} 
 & \multicolumn{1}{l|}{0.68} & \multicolumn{1}{l|}{\textbf{0.93}} & \multicolumn{1}{l|}{\textbf{0.71}} & \multicolumn{1}{l|}{0.63} & \multicolumn{1}{l|}{\textbf{1.00}} & 0.98 & \multicolumn{1}{l|}{\textbf{0.57}} & \multicolumn{1}{l|}{0.39} & \multicolumn{1}{l|}{\textbf{0.80}} & \multicolumn{1}{l|}{0.27} & \multicolumn{1}{l|}{\textbf{0.93}} & 0.90 \\ \hline
\multirow{3}{*}{XIIOTID} & \multicolumn{1}{l|}{0.99} & \multicolumn{1}{l|}{\textbf{1.00}} & \multicolumn{1}{l|}{\textbf{1.00}} & \multicolumn{1}{l|}{\textbf{1.00}} & \multicolumn{1}{l|}{\textbf{0.99}} & 0.43 & \multicolumn{1}{l|}{\textbf{0.65}} & \multicolumn{1}{l|}{0.40} & \multicolumn{1}{l|}{\textbf{0.97}} & \multicolumn{1}{l|}{0.89} & \multicolumn{1}{l|}{\textbf{0.97}} & 0.46 \\ \cline{2-13} 
 & \multicolumn{1}{l|}{\textbf{0.99}} & \multicolumn{1}{l|}{0.85} & \multicolumn{1}{l|}{\textbf{1.00}} & \multicolumn{1}{l|}{0.75} & \multicolumn{1}{l|}{\textbf{0.99}} & 0.61 & \multicolumn{1}{l|}{0.34} & \multicolumn{1}{l|}{\textbf{0.51}} & \multicolumn{1}{l|}{\textbf{0.65}} & \multicolumn{1}{l|}{0.51} & \multicolumn{1}{l|}{\textbf{0.68}} & 0.41 \\ \cline{2-13} 
 & \multicolumn{1}{l|}{0.17} & \multicolumn{1}{l|}{\textbf{0.36}} & \multicolumn{1}{l|}{\textbf{1.00}} & \multicolumn{1}{l|}{0.88} & \multicolumn{1}{l|}{\textbf{1.00}} & 0.54 & \multicolumn{1}{l|}{0.08} & \multicolumn{1}{l|}{\textbf{0.24}} & \multicolumn{1}{l|}{\textbf{0.93}} & \multicolumn{1}{l|}{0.50} & \multicolumn{1}{l|}{\textbf{0.95}} & 0.47 \\ \hline
\end{tabular}
}
\end{table}

\begin{table}[!htbp]
\centering
\caption{The comparison between DBSCAN and DPC in terms of the largest cluster's purity}
\label{Table121:PLSC}
\begin{tabular}{|l|ll|ll|ll|}
\hline
\multicolumn{1}{|c|}{\multirow{2}{*}{Datasets}} & \multicolumn{2}{c|}{UA1} & \multicolumn{2}{c|}{UA2} & \multicolumn{2}{c|}{UA3} \\ \cline{2-7} 
\multicolumn{1}{|c|}{} & \multicolumn{1}{l|}{DBSCAN} & DPC & \multicolumn{1}{l|}{DBSCAN} & DPC & \multicolumn{1}{l|}{DBSCAN} & DPC \\ \hline
\multirow{3}{*}{NSL-KDD} & \multicolumn{1}{l|}{\textbf{0.81}} & 0.48 & \multicolumn{1}{l|}{0.81} & \textbf{0.82} & \multicolumn{1}{l|}{0.81} & \textbf{0.82} \\ \cline{2-7} 
 & \multicolumn{1}{l|}{\textbf{0.70}} & 0.61 & \multicolumn{1}{l|}{0.46} & 0.46 & \multicolumn{1}{l|}{\textbf{0.75}} & 0.67 \\ \cline{2-7} 
 & \multicolumn{1}{l|}{0.22} & \textbf{0.57} & \multicolumn{1}{l|}{\textbf{0.49}} & 0.43 & \multicolumn{1}{l|}{0.51} & \textbf{0.61} \\ \hline
\multirow{3}{*}{CIC-DDoS2019} & \multicolumn{1}{l|}{\textbf{0.98}} & 0.70 & \multicolumn{1}{l|}{\textbf{0.98}} & 0.66 & \multicolumn{1}{l|}{\textbf{0.95}} & 0.40 \\ \cline{2-7} 
 & \multicolumn{1}{l|}{\textbf{0.85}} & 0.70 & \multicolumn{1}{l|}{\textbf{0.94}} & 0.64 & \multicolumn{1}{l|}{\textbf{0.93}} & 0.58 \\ \cline{2-7} 
 & \multicolumn{1}{l|}{\textbf{0.83}} & 0.73 & \multicolumn{1}{l|}{\textbf{0.99}} & 0.94 & \multicolumn{1}{l|}{\textbf{0.87}} & 0.63 \\ \hline
\multirow{3}{*}{UNSW-NB15} & \multicolumn{1}{l|}{\textbf{0.98}} & 0.63 & \multicolumn{1}{l|}{\textbf{0.98}} & 0.43 & \multicolumn{1}{l|}{\textbf{0.97}} & 0.42 \\ \cline{2-7} 
 & \multicolumn{1}{l|}{\textbf{0.84}} & 0.45 & \multicolumn{1}{l|}{\textbf{0.84}} & 0.44 & \multicolumn{1}{l|}{\textbf{0.78}} & 0.44 \\ \cline{2-7} 
 & \multicolumn{1}{l|}{\textbf{0.53}} & 0.52 & \multicolumn{1}{l|}{\textbf{0.79}} & 0.47 & \multicolumn{1}{l|}{\textbf{0.73}} & 0.60 \\ \hline
\multirow{3}{*}{ToN-IoT-Linux} & \multicolumn{1}{l|}{\textbf{0.91}} & 0.64 & \multicolumn{1}{l|}{\textbf{0.71}} & 0.62 & \multicolumn{1}{l|}{\textbf{0.53}} & 0.49 \\ \cline{2-7} 
 & \multicolumn{1}{l|}{0.56} & \textbf{0.57} & \multicolumn{1}{l|}{\textbf{0.65}} & 0.43 & \multicolumn{1}{l|}{\textbf{0.67}} & 0.38 \\ \cline{2-7} 
 & \multicolumn{1}{l|}{0.31} & \textbf{0.44} & \multicolumn{1}{l|}{\textbf{0.57}} & 0.38 & \multicolumn{1}{l|}{\textbf{0.53}} & 0.43 \\ \hline
\multirow{3}{*}{ToN-IoT-Network} & \multicolumn{1}{l|}{\textbf{0.96}} & 0.43 & \multicolumn{1}{l|}{\textbf{0.83}} & 0.79 & \multicolumn{1}{l|}{\textbf{0.83}} & 0.47 \\ \cline{2-7} 
 & \multicolumn{1}{l|}{\textbf{0.82}} & 0.46 & \multicolumn{1}{l|}{\textbf{0.76}} & 0.41 & \multicolumn{1}{l|}{0.81} & \textbf{0.83} \\ \cline{2-7} 
 & \multicolumn{1}{l|}{\textbf{0.62}} & 0.55 & \multicolumn{1}{l|}{\textbf{0.75}} & 0.38 & \multicolumn{1}{l|}{\textbf{0.96}} & 0.94 \\ \hline
\multirow{3}{*}{XIIOTID} & \multicolumn{1}{l|}{\textbf{0.79}} & 0.58 & \multicolumn{1}{l|}{\textbf{0.98}} & 0.94 & \multicolumn{1}{l|}{\textbf{0.98}} & 0.44 \\ \cline{2-7} 
 & \multicolumn{1}{l|}{0.51} & \textbf{0.64} & \multicolumn{1}{l|}{\textbf{0.79}} & 0.61 & \multicolumn{1}{l|}{\textbf{0.81}} & 0.49 \\ \cline{2-7} 
 & \multicolumn{1}{l|}{0.10} & \textbf{0.29} & \multicolumn{1}{l|}{\textbf{0.96}} & 0.64 & \multicolumn{1}{l|}{\textbf{0.97}} & 0.50 \\ \hline
\end{tabular}
\end{table}

Table \ref{Table121:PLSC} provides insights into the strengths and weaknesses of both algorithms for clustering unknown attack categories. The choice between DBSCAN and DPC depends on the relative importance placed on purity and coverage of unknown attack categories within the largest cluster. Considering the overall performance of both algorithms in terms of $PLSC$, $PLS1$ and $PLS2$, we suggested utilizing DBSCAN and analyzed the further performance of the model using the DBSCAN algorithm. Now, we need to analyze the performance of the second-level OCC model and RF after completing retraining. 

Next, we present the accuracy and F1-score of known attacks after retraining the second-level OCC models. With retraining the model, it should be able to identify all previously unseen attacks as known.
From  Table \ref{Table 8: knownattackaccclassificationafter} and \ref{Table 9: knownattackf1scoreclassificationafter}, we find that the average accuracy and weighted F1-score over 5-fold cross-validation, following the completion of retraining the usfAD model and, in some instances, the LOF, IOF and AE models, are notably high for a majority of the datasets. This includes NSL-KDD, ToN-IoT-Network, UNSW-NB15, CIC-DDoS2019, and XIIOTID.

\begin{table}[!htbp]
\caption{Accuracy of known attacks after updating the OCC models with unknown attacks}
\label{Table 8: knownattackaccclassificationafter}
\begin{tabular}{|l|l|ll|ll|ll|ll|ll|}
\hline
\multicolumn{1}{|c|}{\multirow{2}{*}{\textbf{Datasets}}} & \multicolumn{1}{c|}{\multirow{2}{*}{\textbf{NoUAC}}} & \multicolumn{2}{c|}{usfAD} & \multicolumn{2}{c|}{LOF} & \multicolumn{2}{c|}{IOF} & \multicolumn{2}{c|}{OCSVM} & \multicolumn{2}{c|}{AE} \\ \cline{3-12} 
\multicolumn{1}{|c|}{} & \multicolumn{1}{c|}{} & \multicolumn{1}{l|}{C1} & C2 & \multicolumn{1}{l|}{C1} & C2 & \multicolumn{1}{l|}{C1} & C2 & \multicolumn{1}{l|}{C1} & C2 & \multicolumn{1}{l|}{C1} & C2 \\ \hline
\multirow{3}{*}{NSL-KDD} & A1 & \multicolumn{1}{l|}{\textbf{85.87}} & \textbf{92.74} & \multicolumn{1}{l|}{56.20} & 76.07 & \multicolumn{1}{l|}{72.38} & 89.82 & \multicolumn{1}{l|}{49.91} & 61.83 & \multicolumn{1}{l|}{87.94} & 70.23 \\ \cline{2-12} 
 & A2 & \multicolumn{1}{l|}{\textbf{74.41}} & \textbf{89.73} & \multicolumn{1}{l|}{55.91} & 55.42 & \multicolumn{1}{l|}{86.43} & 84.75 & \multicolumn{1}{l|}{49.37} & 48.71 & \multicolumn{1}{l|}{74.26} & 89.32 \\ \cline{2-12} 
 & A3 & \multicolumn{1}{l|}{\textbf{80.26}} & \textbf{85.91} & \multicolumn{1}{l|}{44.09} & 55.73 & \multicolumn{1}{l|}{86.46} & 87.05 & \multicolumn{1}{l|}{47.42} & 48.15 & \multicolumn{1}{l|}{80.55} & 73.33 \\ \hline
\multirow{3}{*}{ToN-IoT-Network} & A1 & \multicolumn{1}{l|}{\textbf{95.97}} & \textbf{97.37} & \multicolumn{1}{l|}{91.67} & 92.95 & \multicolumn{1}{l|}{34.67} & 40.38 & \multicolumn{1}{l|}{20.09} & 22.60 & \multicolumn{1}{l|}{0.67} & 0.02 \\ \cline{2-12} 
 & A2 & \multicolumn{1}{l|}{\textbf{95.64}} & \textbf{95.11} & \multicolumn{1}{l|}{91.44} & 90.57 & \multicolumn{1}{l|}{35.16} & 31.75 & \multicolumn{1}{l|}{19.82} & 11.67 & \multicolumn{1}{l|}{0.91} & 0.81 \\ \cline{2-12} 
 & A3 & \multicolumn{1}{l|}{\textbf{94.77}} & \textbf{91.93} & \multicolumn{1}{l|}{90.18} & 85.56 & \multicolumn{1}{l|}{30.85} & 30.39 & \multicolumn{1}{l|}{11.13} & 16.79 & \multicolumn{1}{l|}{1.20} & 0.02 \\ \hline
\multirow{3}{*}{UNSW-NB15} & A1 & \multicolumn{1}{l|}{76.26} & 61.89 & \multicolumn{1}{l|}{81.26} & 66.43 & \multicolumn{1}{l|}{\textbf{88.14}} & \textbf{84.87} & \multicolumn{1}{l|}{60.58} & 50.56 & \multicolumn{1}{l|}{8.63} & 74.52 \\ \cline{2-12} 
 & A2 & \multicolumn{1}{l|}{64.46} & 72.44 & \multicolumn{1}{l|}{66.38} & 79.85 & \multicolumn{1}{l|}{\textbf{84.28}} & \textbf{84.91} & \multicolumn{1}{l|}{50.55} & 48.21 & \multicolumn{1}{l|}{57.84} & 58.68 \\ \cline{2-12} 
 & A3 & \multicolumn{1}{l|}{59.95} & 65.68 & \multicolumn{1}{l|}{63.43} & 74.25 & \multicolumn{1}{l|}{\textbf{80.46}} & \textbf{83.43} & \multicolumn{1}{l|}{36.56} & 46.92 & \multicolumn{1}{l|}{69.68} & 59.57 \\ \hline
\multirow{2}{*}{Malmem2022} & A1 & \multicolumn{1}{l|}{49.14} & 51.47 & \multicolumn{1}{l|}{46.64} & 51.54 & \multicolumn{1}{l|}{47.00} & 48.27 & \multicolumn{1}{l|}{38.41} & 40.13 & \multicolumn{1}{l|}{\textbf{81.24}} & \textbf{65.86} \\ \cline{2-12} 
 & A2 & \multicolumn{1}{l|}{37.68} & 34.20 & \multicolumn{1}{l|}{35.40} & 31.09 & \multicolumn{1}{l|}{31.03} & 30.43 & \multicolumn{1}{l|}{30.56} & 28.82 & \multicolumn{1}{l|}{\textbf{52.73}} & \textbf{50.70} \\ \hline
\multirow{2}{*}{ISCXURL} & A1 & \multicolumn{1}{l|}{78.35} & \textbf{60.61} & \multicolumn{1}{l|}{\textbf{85.94}} & 59.74 & \multicolumn{1}{l|}{64.40} & 48.87 & \multicolumn{1}{l|}{62.89} & 57.23 & \multicolumn{1}{l|}{59.59} & 43.14 \\ \cline{2-12} 
 & A2 & \multicolumn{1}{l|}{\textbf{58.30}} & 63.36 & \multicolumn{1}{l|}{51.05} & \textbf{67.04} & \multicolumn{1}{l|}{23.23} & 47.53 & \multicolumn{1}{l|}{42.23} & 57.57 & \multicolumn{1}{l|}{49.13} & 44.34 \\ \hline
\multirow{3}{*}{CIC-DDoS2019} & A1 & \multicolumn{1}{l|}{\textbf{94.28}} & \textbf{94.33} & \multicolumn{1}{l|}{87.78} & 89.37 & \multicolumn{1}{l|}{80.29} & 88.56 & \multicolumn{1}{l|}{64.71} & 40.07 & \multicolumn{1}{l|}{78.40} & 86.87 \\ \cline{2-12} 
 & A2 & \multicolumn{1}{l|}{\textbf{94.00}} & \textbf{90.52} & \multicolumn{1}{l|}{87.48} & 82.58 & \multicolumn{1}{l|}{88.58} & 90.06 & \multicolumn{1}{l|}{62.84} & 74.22 & \multicolumn{1}{l|}{92.94} & 80.58 \\ \cline{2-12} 
 & A3 & \multicolumn{1}{l|}{\textbf{88.75}} & \textbf{90.96} & \multicolumn{1}{l|}{81.51} & 83.69 & \multicolumn{1}{l|}{88.40} & 80.68 & \multicolumn{1}{l|}{44.17} & 62.73 & \multicolumn{1}{l|}{79.22} & 87.02 \\ \hline
\multirow{3}{*}{CIC-DDoS2017} & A1 & \multicolumn{1}{l|}{\textbf{70.53}} & \textbf{64.89} & \multicolumn{1}{l|}{23.49} & 20.98 & \multicolumn{1}{l|}{22.91} & 13.17 & \multicolumn{1}{l|}{5.47} & 3.57 & \multicolumn{1}{l|}{35.02} & 22.53 \\ \cline{2-12} 
 & A2 & \multicolumn{1}{l|}{\textbf{56.73}} & \textbf{61.56} & \multicolumn{1}{l|}{21.95} & 22.66 & \multicolumn{1}{l|}{11.23} & 19.87 & \multicolumn{1}{l|}{3.00} & 4.40 & \multicolumn{1}{l|}{31.02} & 31.22 \\ \cline{2-12} 
 & A3 & \multicolumn{1}{l|}{\textbf{48.39}} & \textbf{49.18} & \multicolumn{1}{l|}{19.87} & 18.99 & \multicolumn{1}{l|}{8.74} & 10.13 & \multicolumn{1}{l|}{2.01} & 2.61 & \multicolumn{1}{l|}{27.79} & 30.64 \\ \hline
\multirow{3}{*}{ToN-IoT-Linux} & A1 & \multicolumn{1}{l|}{\textbf{83.13}} & \textbf{82.78} & \multicolumn{1}{l|}{79.69} & 79.63 & \multicolumn{1}{l|}{37.17} & 45.25 & \multicolumn{1}{l|}{18.64} & 19.25 & \multicolumn{1}{l|}{31.55} & 34.43 \\ \cline{2-12} 
 & A2 & \multicolumn{1}{l|}{69.78} & \textbf{75.72} & \multicolumn{1}{l|}{\textbf{70.90}} & 74.26 & \multicolumn{1}{l|}{31.74} & 29.66 & \multicolumn{1}{l|}{14.99} & 15.50 & \multicolumn{1}{l|}{27.91} & 25.52 \\ \cline{2-12} 
 & A3 & \multicolumn{1}{l|}{\textbf{69.15}} & \textbf{69.50} & \multicolumn{1}{l|}{64.54} & 69.21 & \multicolumn{1}{l|}{23.35} & 16.06 & \multicolumn{1}{l|}{12.19} & 11.19 & \multicolumn{1}{l|}{27.64} & 21.58 \\ \hline
\multirow{3}{*}{XIIOTID} & A1 & \multicolumn{1}{l|}{81.61} & 84.72 & \multicolumn{1}{l|}{\textbf{85.80}} & \textbf{90.80} & \multicolumn{1}{l|}{71.03} & 21.51 & \multicolumn{1}{l|}{45.51} & 18.74 & \multicolumn{1}{l|}{62.19} & 45.64 \\ \cline{2-12} 
 & A2 & \multicolumn{1}{l|}{70.07} & 75.37 & \multicolumn{1}{l|}{\textbf{86.57}} & \textbf{78.48} & \multicolumn{1}{l|}{34.45} & 65.26 & \multicolumn{1}{l|}{10.02} & 43.67 & \multicolumn{1}{l|}{20.93} & 58.72 \\ \cline{2-12} 
 & A3 & \multicolumn{1}{l|}{65.86} & 69.28 & \multicolumn{1}{l|}{\textbf{80.79}} & \textbf{73.80} & \multicolumn{1}{l|}{54.83} & 47.10 & \multicolumn{1}{l|}{14.74} & 5.78 & \multicolumn{1}{l|}{27.33} & 19.75 \\ \hline
\multirow{3}{*}{CICDarknet2020} & A1 & \multicolumn{1}{l|}{\textbf{61.86}} & \textbf{70.80} & \multicolumn{1}{l|}{57.99} & 61.55 & \multicolumn{1}{l|}{22.91} & 23.41 & \multicolumn{1}{l|}{9.62} & 13.02 & \multicolumn{1}{l|}{33.12} & 24.28 \\ \cline{2-12} 
 & A2 & \multicolumn{1}{l|}{\textbf{67.65}} & \textbf{49.77} & \multicolumn{1}{l|}{57.83} & 66.02 & \multicolumn{1}{l|}{17.20} & 18.25 & \multicolumn{1}{l|}{13.13} & 11.04 & \multicolumn{1}{l|}{45.65} & 33.10 \\ \cline{2-12} 
 & A3 & \multicolumn{1}{l|}{\textbf{50.29}} & \textbf{63.71} & \multicolumn{1}{l|}{44.85} & 55.55 & \multicolumn{1}{l|}{12.90} & 16.00 & \multicolumn{1}{l|}{11.08} & 11.97 & \multicolumn{1}{l|}{51.34} & 45.99 \\ \hline
\end{tabular}
\end{table}

\begin{table}[!htbp]
\caption{Average weighted F1-score of known attacks after updating the OCC models with unknown attacks}
\label{Table 9: knownattackf1scoreclassificationafter}
\begin{tabular}{|l|l|ll|ll|ll|ll|ll|}
\hline
\multirow{2}{*}{\textbf{Datasets}} & \multicolumn{1}{c|}{\multirow{2}{*}{\textbf{NoUAC}}} & \multicolumn{2}{c|}{usfAD} & \multicolumn{2}{c|}{LOF} & \multicolumn{2}{c|}{IOF} & \multicolumn{2}{c|}{OCSVM} & \multicolumn{2}{c|}{AE} \\ \cline{3-12} 
 & \multicolumn{1}{c|}{} & \multicolumn{1}{l|}{C1} & C2 & \multicolumn{1}{l|}{C1} & C2 & \multicolumn{1}{l|}{C1} & C2 & \multicolumn{1}{l|}{C1} & C2 & \multicolumn{1}{l|}{C1} & C2 \\ \hline
\multirow{3}{*}{NSL-KDD} & A1 & \multicolumn{1}{l|}{84.58} & \textbf{90.15} & \multicolumn{1}{l|}{52.98} & 68.35 & \multicolumn{1}{l|}{73.01} & 86.74 & \multicolumn{1}{l|}{45.62} & 52.16 & \multicolumn{1}{l|}{\textbf{93.11}} & 74.87 \\ \cline{2-12} 
 & A2 & \multicolumn{1}{l|}{74.90} & 86.91 & \multicolumn{1}{l|}{51.06} & 52.62 & \multicolumn{1}{l|}{\textbf{86.01}} & 82.73 & \multicolumn{1}{l|}{46.90} & 45.03 & \multicolumn{1}{l|}{81.31} & \textbf{93.11} \\ \cline{2-12} 
 & A3 & \multicolumn{1}{l|}{79.73} & 84.68 & \multicolumn{1}{l|}{45.47} & 50.95 & \multicolumn{1}{l|}{84.55} & 86.20 & \multicolumn{1}{l|}{44.88} & 45.40 & \multicolumn{1}{l|}{86.72} & 80.82 \\ \hline
\multirow{3}{*}{ToN-IoT-Network} & A1 & \multicolumn{1}{l|}{\textbf{94.86}} & \textbf{96.27} & \multicolumn{1}{l|}{88.75} & 89.85 & \multicolumn{1}{l|}{31.74} & 24.81 & \multicolumn{1}{l|}{14.52} & 8.97 & \multicolumn{1}{l|}{1.32} & 0.04 \\ \cline{2-12} 
 & A2 & \multicolumn{1}{l|}{\textbf{94.54}} & \textbf{94.02} & \multicolumn{1}{l|}{88.59} & 87.84 & \multicolumn{1}{l|}{26.25} & 26.38 & \multicolumn{1}{l|}{9.31} & 5.83 & \multicolumn{1}{l|}{1.78} & 1.59 \\ \cline{2-12} 
 & A3 & \multicolumn{1}{l|}{\textbf{93.76}} & \textbf{90.96} & \multicolumn{1}{l|}{87.60} & 80.76 & \multicolumn{1}{l|}{23.68} & 23.90 & \multicolumn{1}{l|}{5.50} & 7.91 & \multicolumn{1}{l|}{2.35} & 0.04 \\ \hline
\multirow{3}{*}{UNSW-NB15} & A1 & \multicolumn{1}{l|}{72.76} & 55.86 & \multicolumn{1}{l|}{78.58} & 60.71 & \multicolumn{1}{l|}{\textbf{85.54}} & 81.40 & \multicolumn{1}{l|}{52.78} & 40.62 & \multicolumn{1}{l|}{13.95} & \textbf{77.78} \\ \cline{2-12} 
 & A2 & \multicolumn{1}{l|}{62.04} & 66.26 & \multicolumn{1}{l|}{61.38} & 76.04 & \multicolumn{1}{l|}{\textbf{81.95}} & \textbf{81.71} & \multicolumn{1}{l|}{42.13} & 38.56 & \multicolumn{1}{l|}{68.45} & 69.38 \\ \cline{2-12} 
 & A3 & \multicolumn{1}{l|}{57.56} & 63.90 & \multicolumn{1}{l|}{59.25} & 73.44 & \multicolumn{1}{l|}{\textbf{79.20}} & \textbf{80.97} & \multicolumn{1}{l|}{30.91} & 36.69 & \multicolumn{1}{l|}{74.15} & 68.49 \\ \hline
\multirow{2}{*}{Malmem2022} & A1 & \multicolumn{1}{l|}{40.28} & 40.17 & \multicolumn{1}{l|}{37.06} & 39.92 & \multicolumn{1}{l|}{37.63} & 35.04 & \multicolumn{1}{l|}{28.78} & 27.29 & \multicolumn{1}{l|}{\textbf{86.40}} & \textbf{69.47} \\ \cline{2-12} 
 & A2 & \multicolumn{1}{l|}{26.96} & 22.67 & \multicolumn{1}{l|}{24.19} & 17.72 & \multicolumn{1}{l|}{18.38} & 17.21 & \multicolumn{1}{l|}{18.62} & 16.94 & \multicolumn{1}{l|}{\textbf{56.59}} & \textbf{54.09} \\ \hline
\multirow{2}{*}{ISCXURL} & A1 & \multicolumn{1}{l|}{72.89} & \textbf{54.58} & \multicolumn{1}{l|}{\textbf{84.03}} & 52.46 & \multicolumn{1}{l|}{51.03} & 32.55 & \multicolumn{1}{l|}{50.82} & 50.41 & \multicolumn{1}{l|}{64.43} & 46.63 \\ \cline{2-12} 
 & A2 & \multicolumn{1}{l|}{55.40} & 56.79 & \multicolumn{1}{l|}{49.05} & \textbf{58.69} & \multicolumn{1}{l|}{14.34} & 30.63 & \multicolumn{1}{l|}{37.65} & 48.60 & \multicolumn{1}{l|}{\textbf{56.40}} & 47.98 \\ \hline
\multirow{3}{*}{CIC-DDoS2019} & A1 & \multicolumn{1}{l|}{\textbf{93.98}} & \textbf{93.97} & \multicolumn{1}{l|}{86.10} & 87.11 & \multicolumn{1}{l|}{80.25} & 84.89 & \multicolumn{1}{l|}{70.84} & 50.58 & \multicolumn{1}{l|}{86.25} & 89.23 \\ \cline{2-12} 
 & A2 & \multicolumn{1}{l|}{\textbf{93.79}} & \textbf{87.46} & \multicolumn{1}{l|}{86.38} & 76.94 & \multicolumn{1}{l|}{86.56} & 87.62 & \multicolumn{1}{l|}{68.82} & 78.08 & \multicolumn{1}{l|}{93.44} & 83.97 \\ \cline{2-12} 
 & A3 & \multicolumn{1}{l|}{\textbf{84.98}} & \textbf{89.86} & \multicolumn{1}{l|}{77.29} & 81.78 & \multicolumn{1}{l|}{85.66} & 78.28 & \multicolumn{1}{l|}{54.26} & 67.06 & \multicolumn{1}{l|}{82.67} & 87.83 \\ \hline
\multirow{3}{*}{CIC-DDoS2017} & A1 & \multicolumn{1}{l|}{\textbf{59.88}} & \textbf{55.15} & \multicolumn{1}{l|}{14.32} & 13.10 & \multicolumn{1}{l|}{10.16} & 6.96 & \multicolumn{1}{l|}{1.19} & 0.77 & \multicolumn{1}{l|}{48.17} & 32.36 \\ \cline{2-12} 
 & A2 & \multicolumn{1}{l|}{\textbf{45.04}} & \textbf{49.55} & \multicolumn{1}{l|}{16.44} & 13.73 & \multicolumn{1}{l|}{4.99} & 7.93 & \multicolumn{1}{l|}{0.60} & 0.67 & \multicolumn{1}{l|}{43.76} & 43.38 \\ \cline{2-12} 
 & A3 & \multicolumn{1}{l|}{36.90} & 37.53 & \multicolumn{1}{l|}{15.06} & 12.35 & \multicolumn{1}{l|}{4.42} & 4.85 & \multicolumn{1}{l|}{0.17} & 0.41 & \multicolumn{1}{l|}{\textbf{39.66}} & \textbf{43.27} \\ \hline
\multirow{3}{*}{ToN-IoT-Linux} & A1 & \multicolumn{1}{l|}{\textbf{80.98}} & \textbf{78.69} & \multicolumn{1}{l|}{75.86} & 74.77 & \multicolumn{1}{l|}{27.84} & 37.08 & \multicolumn{1}{l|}{10.11} & 8.65 & \multicolumn{1}{l|}{40.99} & 44.44 \\ \cline{2-12} 
 & A2 & \multicolumn{1}{l|}{66.03} & \textbf{74.29} & \multicolumn{1}{l|}{\textbf{66.05}} & 71.22 & \multicolumn{1}{l|}{23.14} & 22.80 & \multicolumn{1}{l|}{5.15} & 5.47 & \multicolumn{1}{l|}{36.45} & 33.80 \\ \cline{2-12} 
 & A3 & \multicolumn{1}{l|}{\textbf{66.44}} & \textbf{66.94} & \multicolumn{1}{l|}{57.15} & 65.23 & \multicolumn{1}{l|}{15.61} & 8.17 & \multicolumn{1}{l|}{3.89} & 3.64 & \multicolumn{1}{l|}{36.33} & 26.77 \\ \hline
\multirow{3}{*}{XIIOTID} & A1 & \multicolumn{1}{l|}{87.30} & 89.70 & \multicolumn{1}{l|}{\textbf{91.14}} & \textbf{93.85} & \multicolumn{1}{l|}{74.88} & 24.21 & \multicolumn{1}{l|}{58.62} & 24.17 & \multicolumn{1}{l|}{72.02} & 49.83 \\ \cline{2-12} 
 & A2 & \multicolumn{1}{l|}{79.67} & 83.30 & \multicolumn{1}{l|}{\textbf{91.23}} & \textbf{86.15} & \multicolumn{1}{l|}{45.11} & 67.58 & \multicolumn{1}{l|}{13.45} & 56.34 & \multicolumn{1}{l|}{23.83} & 67.28 \\ \cline{2-12} 
 & A3 & \multicolumn{1}{l|}{76.32} & 77.09 & \multicolumn{1}{l|}{\textbf{87.40}} & \textbf{77.37} & \multicolumn{1}{l|}{62.62} & 57.84 & \multicolumn{1}{l|}{22.16} & 7.82 & \multicolumn{1}{l|}{35.74} & 26.59 \\ \hline
\multirow{3}{*}{CICDarknet2020} & A1 & \multicolumn{1}{l|}{\textbf{57.32}} & \textbf{60.68} & \multicolumn{1}{l|}{52.89} & 48.46 & \multicolumn{1}{l|}{13.34} & 12.89 & \multicolumn{1}{l|}{3.13} & 4.74 & \multicolumn{1}{l|}{47.73} & 36.70 \\ \cline{2-12} 
 & A2 & \multicolumn{1}{l|}{57.92} & 44.76 & \multicolumn{1}{l|}{50.28} & \textbf{58.50} & \multicolumn{1}{l|}{7.52} & 9.61 & \multicolumn{1}{l|}{4.14} & 3.49 & \multicolumn{1}{l|}{\textbf{60.80}} & 47.60 \\ \cline{2-12} 
 & A3 & \multicolumn{1}{l|}{44.23} & 53.19 & \multicolumn{1}{l|}{39.78} & 47.28 & \multicolumn{1}{l|}{5.09} & 6.88 & \multicolumn{1}{l|}{4.09} & 3.56 & \multicolumn{1}{l|}{\textbf{65.07}} & \textbf{61.00} \\ \hline
\end{tabular}
\end{table}

\subsection{Performance of Supervised Learner at the Second Level in Retraining Process}

\begin{figure}[!htbp]
    \centering
    \includegraphics[scale = .90]{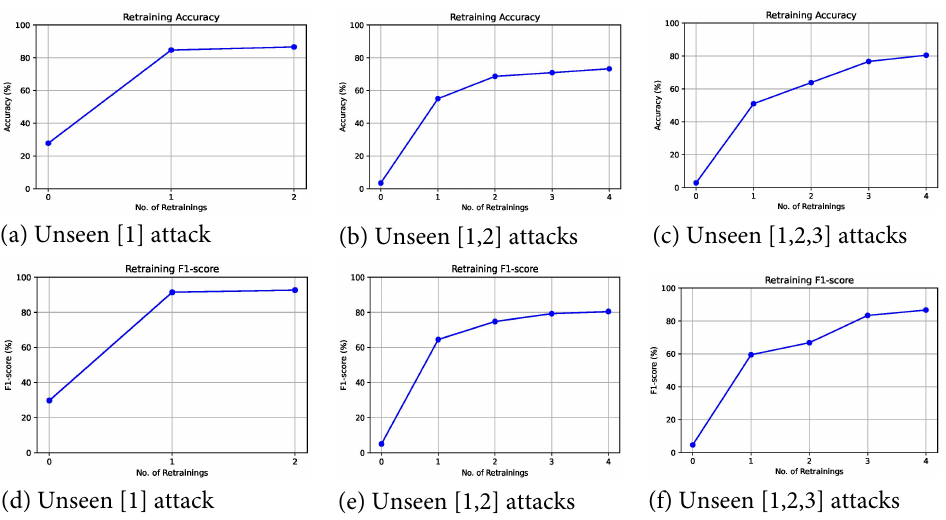}
    \caption{Impact of retraining on accuracy and F1-score on NSL-KDD and UNSW-NB15}
    \label{fig:7}
\end{figure}

\begin{figure}[!htbp]
    \centering
    \includegraphics[scale = .90]{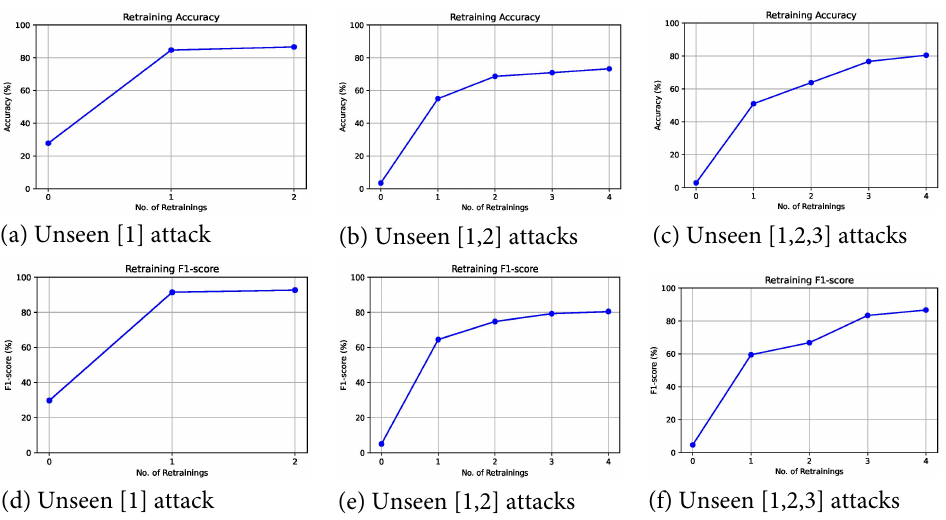}
    \caption{Impact of retraining on accuracy and F1-score on CIC-DDoS2019 and ToN-IoT-Network}
    \label{fig:8}
\end{figure}

The semi-OCC model, mainly the usfAD algorithm is capable of distinguishing network traffic as either known or unknown. However, for the purpose of assisting security experts in setting appropriate security policies, it's crucial to predict the specific family type of network traffic. To do this, we utilize an RF classifier to categorize the known attack types filtered by the second-level OCC model. RF, alongside the semi-OCC model at the second level, is retrained as we accumulate a specific number of unknown attack instances. We need to examine how correctly the RF classifies specific family types of unknown attacks. The F1-scores during the retraining phases of the RF model are depicted in Figures \ref{fig:7} and \ref{fig:8} for NSL-KDD, UNSW-NB15, CIC-DDoS2019, and ToN-IoT-Network datasets as a representative of other datasets. These graphs illustrate scenarios with varying numbers of unseen/unknown attack categories ([1], [1,2], [1,2,3]).  These graphs show that the F1-score of the RF model improves with each retraining cycle. This enhancement indicates the model's growing proficiency in identifying an increased number of unknown attacks after each update with new unknown attack data. 


Tables \ref{Table10:knownatttackaccrfafter} and \ref{Table 11:knownattackf1scorerfafter} present the average accuracy and F1-score of the RF classifier across a 5-fold cross-validation, after its retraining with samples of unknown attacks.

\begin{table}[!htbp]
\caption{Accuracy of known attacks after retraining RF's model}
\label{Table10:knownatttackaccrfafter}
\begin{tabular}{|l|l|ll|ll|ll|ll|ll|}
\hline
\multicolumn{1}{|c|}{\multirow{2}{*}{\textbf{Datasets}}} & \multicolumn{1}{c|}{\multirow{2}{*}{\textbf{NoUAC}}} & \multicolumn{2}{c|}{usfAD} & \multicolumn{2}{c|}{LOF} & \multicolumn{2}{c|}{IOF} & \multicolumn{2}{c|}{OCSVM} & \multicolumn{2}{c|}{AE} \\ \cline{3-12} 
\multicolumn{1}{|c|}{} & \multicolumn{1}{c|}{} & \multicolumn{1}{l|}{C1} & C2 & \multicolumn{1}{l|}{C1} & C2 & \multicolumn{1}{l|}{C1} & C2 & \multicolumn{1}{l|}{C1} & C2 & \multicolumn{1}{l|}{C1} & C2 \\ \hline
\multirow{3}{*}{NSL-KDD} & UA1 & \multicolumn{1}{l|}{\textbf{85.87}} & \textbf{92.74} & \multicolumn{1}{l|}{56.20} & 76.07 & \multicolumn{1}{l|}{72.38} & 89.82 & \multicolumn{1}{l|}{49.91} & 61.83 & \multicolumn{1}{l|}{69.81} & 86.90 \\ \cline{2-12} 
 & UA2 & \multicolumn{1}{l|}{74.41} & \textbf{89.73} & \multicolumn{1}{l|}{55.91} & 55.42 & \multicolumn{1}{l|}{86.43} & 84.75 & \multicolumn{1}{l|}{49.37} & 48.71 & \multicolumn{1}{l|}{\textbf{83.55}} & 67.40 \\ \cline{2-12} 
 & UA3 & \multicolumn{1}{l|}{80.26} & \textbf{85.91} & \multicolumn{1}{l|}{44.09} & 55.73 & \multicolumn{1}{l|}{86.46} & 87.05 & \multicolumn{1}{l|}{47.42} & 48.15 & \multicolumn{1}{l|}{\textbf{81.93}} & 83.37 \\ \hline
\multirow{3}{*}{ToN-IoT-Network} & UA1 & \multicolumn{1}{l|}{\textbf{95.97}} & \textbf{97.37} & \multicolumn{1}{l|}{91.67} & 92.95 & \multicolumn{1}{l|}{34.67} & 40.38 & \multicolumn{1}{l|}{20.09} & 22.60 & \multicolumn{1}{l|}{24.00} & 24.79 \\ \cline{2-12} 
 & UA2 & \multicolumn{1}{l|}{\textbf{95.64}} & \textbf{95.11} & \multicolumn{1}{l|}{91.44} & 90.57 & \multicolumn{1}{l|}{35.16} & 31.75 & \multicolumn{1}{l|}{19.82} & 11.67 & \multicolumn{1}{l|}{23.70} & 21.60 \\ \cline{2-12} 
 & UA3 & \multicolumn{1}{l|}{\textbf{94.77}} & \textbf{91.93} & \multicolumn{1}{l|}{90.18} & 85.56 & \multicolumn{1}{l|}{30.85} & 30.39 & \multicolumn{1}{l|}{11.13} & 16.79 & \multicolumn{1}{l|}{20.43} & 11.30 \\ \hline
\multirow{3}{*}{UNSW-NB15} & UA1 & \multicolumn{1}{l|}{76.26} & 61.89 & \multicolumn{1}{l|}{81.26} & 66.43 & \multicolumn{1}{l|}{\textbf{88.14}} & \textbf{84.87} & \multicolumn{1}{l|}{60.58} & 50.56 & \multicolumn{1}{l|}{83.08} & 76.81 \\ \cline{2-12} 
 & UA2 & \multicolumn{1}{l|}{64.46} & 72.44 & \multicolumn{1}{l|}{66.38} & 79.85 & \multicolumn{1}{l|}{\textbf{84.28}} & \textbf{84.91} & \multicolumn{1}{l|}{50.55} & 48.21 & \multicolumn{1}{l|}{76.09} & 77.92 \\ \cline{2-12} 
 & UA3 & \multicolumn{1}{l|}{59.95} & 65.68 & \multicolumn{1}{l|}{63.43} & 74.25 & \multicolumn{1}{l|}{\textbf{80.46}} & \textbf{83.43} & \multicolumn{1}{l|}{36.56} & 46.92 & \multicolumn{1}{l|}{71.05} & 74.29 \\ \hline
\multirow{2}{*}{Malmem2022} & UA1 & \multicolumn{1}{l|}{49.14} & 51.47 & \multicolumn{1}{l|}{46.64} & 51.54 & \multicolumn{1}{l|}{47.00} & 48.27 & \multicolumn{1}{l|}{38.41} & 40.13 & \multicolumn{1}{l|}{\textbf{50.58}} & \textbf{53.16} \\ \cline{2-12} 
 & UA2 & \multicolumn{1}{l|}{37.68} & 34.20 & \multicolumn{1}{l|}{35.40} & 31.09 & \multicolumn{1}{l|}{31.03} & 30.43 & \multicolumn{1}{l|}{30.56} & 28.82 & \multicolumn{1}{l|}{\textbf{35.57}} & \textbf{32.17} \\ \hline
\multirow{2}{*}{ISCXURL} & UA1 & \multicolumn{1}{l|}{\textbf{78.35}} & \textbf{60.61} & \multicolumn{1}{l|}{85.94} & 59.74 & \multicolumn{1}{l|}{64.40} & 48.87 & \multicolumn{1}{l|}{62.89} & 57.23 & \multicolumn{1}{l|}{65.96} & 47.40 \\ \cline{2-12} 
 & UA2 & \multicolumn{1}{l|}{\textbf{58.30}} & \textbf{63.36} & \multicolumn{1}{l|}{51.05} & 67.04 & \multicolumn{1}{l|}{23.23} & 47.53 & \multicolumn{1}{l|}{42.23} & 57.57 & \multicolumn{1}{l|}{29.37} & 48.77 \\ \hline
\multirow{3}{*}{CIC-DDoS2019} & UA1 & \multicolumn{1}{l|}{\textbf{94.28}} & \textbf{94.33} & \multicolumn{1}{l|}{87.78} & 89.37 & \multicolumn{1}{l|}{80.29} & 88.56 & \multicolumn{1}{l|}{31.56} & 75.37 & \multicolumn{1}{l|}{45.94} & 89.89 \\ \cline{2-12} 
 & UA2 & \multicolumn{1}{l|}{\textbf{94.00}} & \textbf{90.52} & \multicolumn{1}{l|}{87.48} & 82.58 & \multicolumn{1}{l|}{88.58} & 90.06 & \multicolumn{1}{l|}{26.06} & 16.85 & \multicolumn{1}{l|}{43.71} & 65.12 \\ \cline{2-12} 
 & UA3 & \multicolumn{1}{l|}{\textbf{88.75}} & \textbf{90.96} & \multicolumn{1}{l|}{81.51} & 83.69 & \multicolumn{1}{l|}{88.40} & 80.68 & \multicolumn{1}{l|}{71.22} & 24.44 & \multicolumn{1}{l|}{84.15} & 54.82 \\ \hline
\multirow{3}{*}{CIC-DDoS2017} & UA1 & \multicolumn{1}{l|}{\textbf{70.53}} & \textbf{64.89} & \multicolumn{1}{l|}{23.49} & 20.98 & \multicolumn{1}{l|}{22.91} & 13.17 & \multicolumn{1}{l|}{5.47} & 3.57 & \multicolumn{1}{l|}{17.76} & 10.17 \\ \cline{2-12} 
 & UA2 & \multicolumn{1}{l|}{\textbf{56.73}} & \textbf{61.56} & \multicolumn{1}{l|}{21.95} & 22.66 & \multicolumn{1}{l|}{11.23} & 19.87 & \multicolumn{1}{l|}{3.00} & 4.40 & \multicolumn{1}{l|}{10.23} & 16.11 \\ \cline{2-12} 
 & UA3 & \multicolumn{1}{l|}{\textbf{48.39}} & \textbf{49.18} & \multicolumn{1}{l|}{19.87} & 18.99 & \multicolumn{1}{l|}{8.74} & 10.13 & \multicolumn{1}{l|}{2.01} & 2.61 & \multicolumn{1}{l|}{7.46} & 9.26 \\ \hline
\multirow{3}{*}{ToN-IoT-Linux} & UA1 & \multicolumn{1}{l|}{\textbf{83.13}} & \textbf{82.78} & \multicolumn{1}{l|}{79.69} & 79.63 & \multicolumn{1}{l|}{37.17} & 45.25 & \multicolumn{1}{l|}{18.64} & 19.25 & \multicolumn{1}{l|}{43.55} & 47.59 \\ \cline{2-12} 
 & UA2 & \multicolumn{1}{l|}{\textbf{69.78}} & \textbf{75.72} & \multicolumn{1}{l|}{70.90} & 74.26 & \multicolumn{1}{l|}{31.74} & 29.66 & \multicolumn{1}{l|}{14.99} & 15.50 & \multicolumn{1}{l|}{33.91} & 33.86 \\ \cline{2-12} 
 & UA3 & \multicolumn{1}{l|}{\textbf{69.15}} & \textbf{69.50} & \multicolumn{1}{l|}{64.54} & 69.21 & \multicolumn{1}{l|}{23.35} & 16.06 & \multicolumn{1}{l|}{12.19} & 11.19 & \multicolumn{1}{l|}{26.35} & 29.53 \\ \hline
\multirow{3}{*}{XIIOTID} & UA1 & \multicolumn{1}{l|}{\textbf{90.66}} & \textbf{96.80} & \multicolumn{1}{l|}{83.56} & 96.22 & \multicolumn{1}{l|}{80.72} & 82.26 & \multicolumn{1}{l|}{51.00} & 57.76 & \multicolumn{1}{l|}{77.77} & 84.80 \\ \cline{2-12} 
 & UA2 & \multicolumn{1}{l|}{\textbf{88.50}} & \textbf{85.13} & \multicolumn{1}{l|}{88.03} & 78.55 & \multicolumn{1}{l|}{76.13} & 81.36 & \multicolumn{1}{l|}{41.88} & 45.11 & \multicolumn{1}{l|}{52.34} & 75.46 \\ \cline{2-12} 
 & UA3 & \multicolumn{1}{l|}{\textbf{85.22}} & \textbf{88.51} & \multicolumn{1}{l|}{84.05} & 76.44 & \multicolumn{1}{l|}{70.85} & 75.66 & \multicolumn{1}{l|}{30.45} & 40.30 & \multicolumn{1}{l|}{43.49} & 46.28 \\ \hline
\multirow{3}{*}{CIC-Darknet2020} & UA1 & \multicolumn{1}{l|}{\textbf{61.86}} & \textbf{70.80} & \multicolumn{1}{l|}{57.99} & 61.55 & \multicolumn{1}{l|}{57.99} & 61.55 & \multicolumn{1}{l|}{9.62} & 13.02 & \multicolumn{1}{l|}{17.68} & 20.10 \\ \cline{2-12} 
 & UA2 & \multicolumn{1}{l|}{\textbf{67.65}} & \textbf{49.77} & \multicolumn{1}{l|}{57.83} & 66.02 & \multicolumn{1}{l|}{57.83} & 66.02 & \multicolumn{1}{l|}{13.13} & 11.04 & \multicolumn{1}{l|}{17.25} & 14.88 \\ \cline{2-12} 
 & UA3 & \multicolumn{1}{l|}{\textbf{50.29}} & \textbf{63.71} & \multicolumn{1}{l|}{44.85} & 55.55 & \multicolumn{1}{l|}{44.85} & 55.55 & \multicolumn{1}{l|}{11.08} & 11.97 & \multicolumn{1}{l|}{11.46} & 16.36 \\ \hline
\end{tabular}
\end{table}

\begin{table}[!htbp]
\caption{Average weighted F1-score of known attacks after retraining RF's model }
\label{Table 11:knownattackf1scorerfafter}
\begin{tabular}{|l|l|ll|ll|ll|ll|ll|}
\hline
\multicolumn{1}{|c|}{\multirow{2}{*}{\textbf{Datasets}}} & \multicolumn{1}{c|}{\multirow{2}{*}{\textbf{NoUAC}}} & \multicolumn{2}{c|}{usfAD} & \multicolumn{2}{c|}{LOF} & \multicolumn{2}{c|}{IOF} & \multicolumn{2}{c|}{OCSVM} & \multicolumn{2}{c|}{AE} \\ \cline{3-12} 
\multicolumn{1}{|c|}{} & \multicolumn{1}{c|}{} & \multicolumn{1}{l|}{C1} & C2 & \multicolumn{1}{l|}{C1} & C2 & \multicolumn{1}{l|}{C1} & C2 & \multicolumn{1}{l|}{C1} & C2 & \multicolumn{1}{l|}{C1} & C2 \\ \hline
\multirow{3}{*}{NSL-KDD} & UA1 & \multicolumn{1}{l|}{\textbf{84.58}} & \textbf{90.15} & \multicolumn{1}{l|}{52.98} & 68.35 & \multicolumn{1}{l|}{73.01} & 86.74 & \multicolumn{1}{l|}{45.62} & 52.16 & \multicolumn{1}{l|}{69.44} & 83.18 \\ \cline{2-12} 
 & UA2 & \multicolumn{1}{l|}{74.90} & \textbf{86.91} & \multicolumn{1}{l|}{51.06} & 52.62 & \multicolumn{1}{l|}{\textbf{86.01}} & 82.73 & \multicolumn{1}{l|}{46.90} & 45.03 & \multicolumn{1}{l|}{81.43} & 67.41 \\ \cline{2-12} 
 & UA3 & \multicolumn{1}{l|}{79.73} & 84.68 & \multicolumn{1}{l|}{45.47} & 50.95 & \multicolumn{1}{l|}{\textbf{84.55}} & \textbf{86.20} & \multicolumn{1}{l|}{44.88} & 45.40 & \multicolumn{1}{l|}{79.55} & 81.31 \\ \hline
\multirow{3}{*}{ToN-IoT-Network} & UA1 & \multicolumn{1}{l|}{\textbf{94.86}} & \textbf{96.27} & \multicolumn{1}{l|}{88.75} & 89.85 & \multicolumn{1}{l|}{31.74} & 24.81 & \multicolumn{1}{l|}{20.09} & 14.52 & \multicolumn{1}{l|}{16.03} & 12.49 \\ \cline{2-12} 
 & UA2 & \multicolumn{1}{l|}{\textbf{94.54}} & \textbf{94.02} & \multicolumn{1}{l|}{88.59} & 87.84 & \multicolumn{1}{l|}{26.25} & 26.38 & \multicolumn{1}{l|}{19.82} & 9.31 & \multicolumn{1}{l|}{16.26} & 14.97 \\ \cline{2-12} 
 & UA3 & \multicolumn{1}{l|}{\textbf{93.76}} & \textbf{90.96} & \multicolumn{1}{l|}{87.60} & 80.76 & \multicolumn{1}{l|}{23.68} & 23.90 & \multicolumn{1}{l|}{11.13} & 5.50 & \multicolumn{1}{l|}{12.18} & 7.85 \\ \hline
\multirow{3}{*}{UNSW-NB15} & UA1 & \multicolumn{1}{l|}{72.76} & 55.86 & \multicolumn{1}{l|}{78.58} & 60.71 & \multicolumn{1}{l|}{\textbf{85.54}} & \textbf{81.40} & \multicolumn{1}{l|}{52.78} & 40.62 & \multicolumn{1}{l|}{78.98} & 73.07 \\ \cline{2-12} 
 & UA2 & \multicolumn{1}{l|}{62.04} & 66.26 & \multicolumn{1}{l|}{61.38} & 76.04 & \multicolumn{1}{l|}{\textbf{81.95}} & \textbf{81.71} & \multicolumn{1}{l|}{42.13} & 38.56 & \multicolumn{1}{l|}{73.18} & 73.74 \\ \cline{2-12} 
 & UA3 & \multicolumn{1}{l|}{57.56} & 63.90 & \multicolumn{1}{l|}{59.25} & 73.44 & \multicolumn{1}{l|}{\textbf{79.20}} & \textbf{80.97} & \multicolumn{1}{l|}{30.91} & 36.69 & \multicolumn{1}{l|}{69.77} & 70.86 \\ \hline
\multirow{2}{*}{Malmem2022} & UA1 & \multicolumn{1}{l|}{40.28} & 40.17 & \multicolumn{1}{l|}{37.06} & 39.92 & \multicolumn{1}{l|}{37.63} & 35.04 & \multicolumn{1}{l|}{28.78} & 27.29 & \multicolumn{1}{l|}{\textbf{41.40}} & \textbf{40.78} \\ \cline{2-12} 
 & UA2 & \multicolumn{1}{l|}{\textbf{26.96}} & \textbf{22.67} & \multicolumn{1}{l|}{24.19} & 17.72 & \multicolumn{1}{l|}{18.38} & 17.21 & \multicolumn{1}{l|}{18.62} & 16.94 & \multicolumn{1}{l|}{24.00} & 18.29 \\ \hline
\multirow{2}{*}{ISCXURL} & UA1 & \multicolumn{1}{l|}{\textbf{72.89}} & \textbf{54.58} & \multicolumn{1}{l|}{84.03} & 52.46 & \multicolumn{1}{l|}{51.03} & 32.55 & \multicolumn{1}{l|}{50.82} & 50.41 & \multicolumn{1}{l|}{53.09} & 30.68 \\ \cline{2-12} 
 & UA2 & \multicolumn{1}{l|}{\textbf{55.40}} & \textbf{56.79} & \multicolumn{1}{l|}{49.05} & 58.69 & \multicolumn{1}{l|}{14.34} & 30.63 & \multicolumn{1}{l|}{37.65} & 48.60 & \multicolumn{1}{l|}{24.20} & 31.97 \\ \hline
\multirow{3}{*}{CIC-DDoS2019} & UA1 & \multicolumn{1}{l|}{\textbf{93.98}} & \textbf{93.97} & \multicolumn{1}{l|}{86.10} & 87.11 & \multicolumn{1}{l|}{80.25} & 84.89 & \multicolumn{1}{l|}{28.35} & 68.86 & \multicolumn{1}{l|}{49.50} & 87.81 \\ \cline{2-12} 
 & UA2 & \multicolumn{1}{l|}{\textbf{93.79}} & \textbf{87.46} & \multicolumn{1}{l|}{86.38} & 76.94 & \multicolumn{1}{l|}{86.56} & 87.62 & \multicolumn{1}{l|}{26.79} & 10.25 & \multicolumn{1}{l|}{47.28} & 65.82 \\ \cline{2-12} 
 & UA3 & \multicolumn{1}{l|}{\textbf{84.98}} & \textbf{89.86} & \multicolumn{1}{l|}{77.29} & 81.78 & \multicolumn{1}{l|}{85.66} & 78.28 & \multicolumn{1}{l|}{66.36} & 26.01 & \multicolumn{1}{l|}{79.41} & 57.30 \\ \hline
\multirow{3}{*}{CIC-DDoS2017} & UA1 & \multicolumn{1}{l|}{\textbf{59.88}} & \textbf{55.15} & \multicolumn{1}{l|}{14.32} & 13.10 & \multicolumn{1}{l|}{10.16} & 6.96 & \multicolumn{1}{l|}{1.19} & 0.77 & \multicolumn{1}{l|}{6.44} & 4.27 \\ \cline{2-12} 
 & UA2 & \multicolumn{1}{l|}{\textbf{45.04}} & \textbf{49.55} & \multicolumn{1}{l|}{16.44} & 13.73 & \multicolumn{1}{l|}{4.99} & 7.93 & \multicolumn{1}{l|}{0.60} & 0.67 & \multicolumn{1}{l|}{4.44} & 5.49 \\ \cline{2-12} 
 & UA3 & \multicolumn{1}{l|}{\textbf{36.90}} & \textbf{37.53} & \multicolumn{1}{l|}{15.06} & 12.35 & \multicolumn{1}{l|}{4.42} & 4.85 & \multicolumn{1}{l|}{0.17} & 0.41 & \multicolumn{1}{l|}{3.34} & 3.87 \\ \hline
\multirow{3}{*}{ToN-IoT-Linux} & UA1 & \multicolumn{1}{l|}{\textbf{80.98}} & \textbf{78.69} & \multicolumn{1}{l|}{75.86} & 74.77 & \multicolumn{1}{l|}{27.84} & 37.08 & \multicolumn{1}{l|}{10.11} & 8.65 & \multicolumn{1}{l|}{31.48} & 35.48 \\ \cline{2-12} 
 & UA2 & \multicolumn{1}{l|}{\textbf{66.03}} & \textbf{74.29} & \multicolumn{1}{l|}{66.05} & 71.22 & \multicolumn{1}{l|}{23.14} & 22.80 & \multicolumn{1}{l|}{5.15} & 5.47 & \multicolumn{1}{l|}{21.13} & 20.88 \\ \cline{2-12} 
 & UA3 & \multicolumn{1}{l|}{\textbf{66.44}} & \textbf{66.94} & \multicolumn{1}{l|}{57.15} & 65.23 & \multicolumn{1}{l|}{15.61} & 8.17 & \multicolumn{1}{l|}{3.89} & 3.64 & \multicolumn{1}{l|}{17.23} & 16.09 \\ \hline
\multirow{3}{*}{XIIOTID} & UA1 & \multicolumn{1}{l|}{\textbf{89.55}} & \textbf{95.34} & \multicolumn{1}{l|}{82.85} & 94.46 & \multicolumn{1}{l|}{78.13} & 78.14 & \multicolumn{1}{l|}{44.26} & 46.94 & \multicolumn{1}{l|}{74.84} & 78.70 \\ \cline{2-12} 
 & UA2 & \multicolumn{1}{l|}{\textbf{88.16}} & \textbf{84.16} & \multicolumn{1}{l|}{86.44} & 76.27 & \multicolumn{1}{l|}{73.40} & 79.12 & \multicolumn{1}{l|}{36.82} & 38.95 & \multicolumn{1}{l|}{45.77} & 72.54 \\ \cline{2-12} 
 & UA3 & \multicolumn{1}{l|}{\textbf{83.88}} & \textbf{85.98} & \multicolumn{1}{l|}{81.90} & 71.10 & \multicolumn{1}{l|}{68.21} & 72.69 & \multicolumn{1}{l|}{24.61} & 36.81 & \multicolumn{1}{l|}{37.80} & 43.30 \\ \hline
\multirow{3}{*}{CIC-Darknet2020} & UA1 & \multicolumn{1}{l|}{\textbf{57.32}} & \textbf{60.68} & \multicolumn{1}{l|}{52.89} & 48.46 & \multicolumn{1}{l|}{52.89} & 48.46 & \multicolumn{1}{l|}{3.13} & 4.74 & \multicolumn{1}{l|}{8.77} & 10.50 \\ \cline{2-12} 
 & UA2 & \multicolumn{1}{l|}{\textbf{57.92}} & \textbf{44.76} & \multicolumn{1}{l|}{50.28} & 58.50 & \multicolumn{1}{l|}{50.28} & 58.50 & \multicolumn{1}{l|}{4.14} & 3.49 & \multicolumn{1}{l|}{8.45} & 7.24 \\ \cline{2-12} 
 & UA3 & \multicolumn{1}{l|}{\textbf{44.23}} & \textbf{53.19} & \multicolumn{1}{l|}{39.78} & 47.28 & \multicolumn{1}{l|}{39.78} & 47.28 & \multicolumn{1}{l|}{4.09} & 3.56 & \multicolumn{1}{l|}{4.34} & 7.64 \\ \hline
\end{tabular}
\end{table}

\subsection{Comparison of Our Model with a State-of-the-art Work}

Soltani et al.  \cite{soltani2023adaptable} proposed Deep open classification (DOC++) for identifying benign, known, and unknown attack categories. By analyzing performance, we have recommended the usfAD model developed by Sunil et al. \cite{aryal2021usfad} to detect known and unknown attacks in cybersecurity across multiple datasets, against DOC++. In this section, we implemented a two-level hierarchical model by excluding the clustering and retraining process to show the effectiveness of usfAD in detecting known and unknown attacks against the existing DOC++. Here, we exclude the clustering and retraining process because our retraining process differs from the existing works. We design an automatic retraining process focusing on a more practical approach for updating the model while encountering zero-day attacks.

\begin{figure}[!htbp]
    \centering
    \includegraphics[scale = .90]{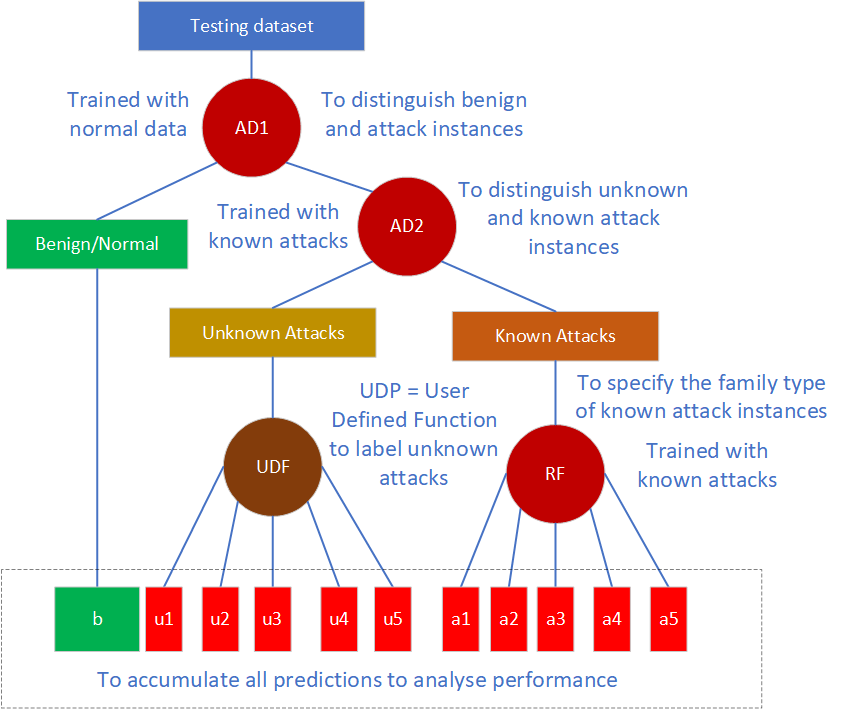}
    \caption{Hierarchical Simulation of the Proposed Adaptive Framework}
    \label{fig:9}
\end{figure}

As illustrated in Figure \ref{fig:9}, at the first level, our model distinguishes between benign and attack samples using an OCC method trained solely on benign instances. For the second-level OCC, we divide the training instances into known and unknown categories, using combination C$\binom{n}{k}$ for CIC-IDS2017 and CIC-IDS2018 datasets (Soltani et al. applied these two datasets), where k represents the number of known attacks and (n-k) is the number of unknown attacks. The second level OCC model is trained on known attacks. The second-level model determines if the instances detected as attacks by the first-level model are Known or unknown. Known instances are then classified using an RF (trained on known attacks during the training phase) to specify the family types of the known attacks. Unknown attacks predicted by the second level OCC, in practice, require labeling by a human expert. In this case, we use the ground truth of the unknown attack groups to label their classes. The accuracy of individual attack categories is measured using the following formula, where x indicates a particular attack type.

$ACC_x = \frac{TP_x}{TP_x+TN_x+FP_x+FN_x}$

Our model's performance was evaluated both with and without including benign predictions. The average 5-fold accuracy of each attack category showed minimal difference between the two scenarios, for both usfAD and LOF. Tables \ref{Table14:comparisoncicids2017} and \ref{Table13: comparisonCICIDS2018} compare the accuracy of our model (including usfAD and LOF) with the existing DOC++ model on the CIC-IDS2017 and CIC-IDS2018 datasets. The results demonstrate that usfAD and LOF outperform DOC++ (a deep learning approach) on both datasets.

\begin{figure}[!htbp]
    \centering
    \includegraphics[scale = .80]{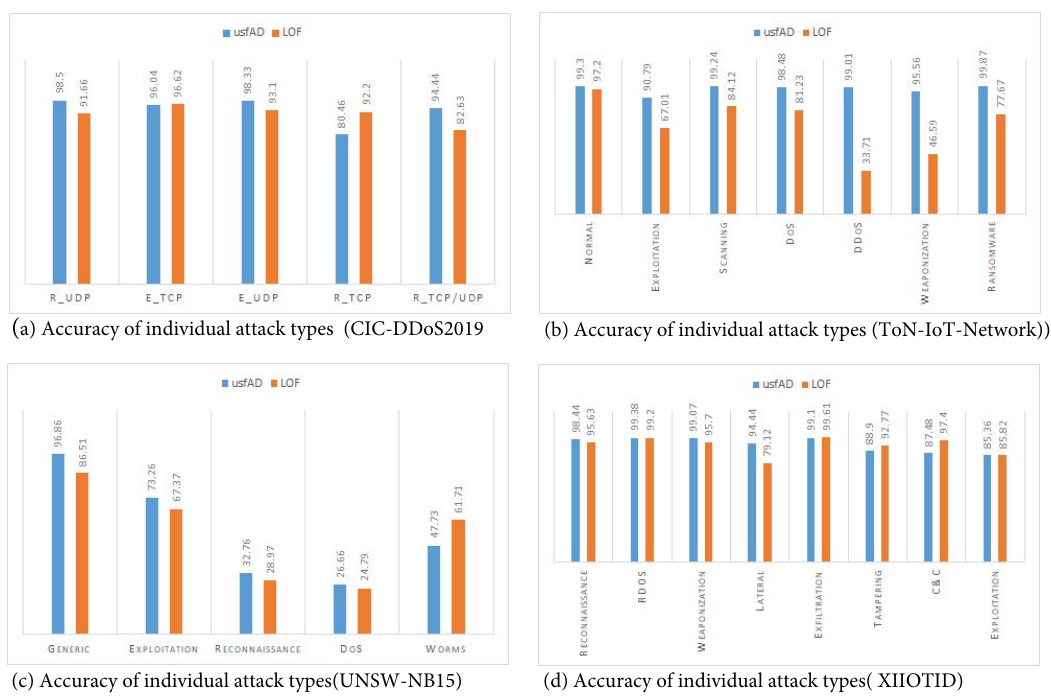}
    \caption{Comparison of accuracy between usfAD and LOF in detecting individual attack families}
    \label{fig:10comp}
\end{figure}

Furthermore, we present the average 5-fold cross-accuracy of our model for other IDS benchmark datasets in Figure  \ref{fig:10comp}. The usfAD exhibits outstanding performance in detecting known and unknown categories across most datasets, highlighting its potential as a viable option for developing practical IDS models.      

\begin{table}[!htbp]
\centering
\caption{Comparison of accuracy between the proposed model and an existing model on CIC-IDS2017}
\label{Table14:comparisoncicids2017}
\begin{tabular}{|l|c|c|c|c|c|c|}
\hline
\multicolumn{1}{|c|}{Attack} & usfAD & \begin{tabular}[c]{@{}c@{}}usfAD\\ (NN)\end{tabular} & DOC++ & \begin{tabular}[c]{@{}c@{}}DOC++\\ (NN)\end{tabular} & LOF & \begin{tabular}[c]{@{}c@{}}LOF\\ (NN)\end{tabular} \\ \hline
Benign & 98.19 & N/A & 50.2 & N/A & 85.05 & N/A \\ \hline
DoS Hulk & 97.74 & 97.72 & 49.01 & 47.08 & 90.32 & 90.35 \\ \hline
DDoS & 98.58 & 98.91 & 47.64 & 56.56 & 96.12 & 96.4 \\ \hline
PortScan & 98.46 & 99.38 & 87.61 & 85.55 & 96.79 & 97.09 \\ \hline
DoS GoldenEye & 97.36 & 97.41 & 54.76 & 50.26 & 98.15 & 98.15 \\ \hline
FTP-Patator & 99.75 & 99.72 & 50.04 & 61.92 & 80.05 & 92.12 \\ \hline
DoS slowloris & 93.06 & 93.49 & 60.16 & 54.46 & 97.31 & 97.97 \\ \hline
DoS   Slowhttptest & 96.28 & 94.57 & 63.09 & 69.13 & 96.51 & 96.45 \\ \hline
SSH-Patator & 95.11 & 98.93 & 51.34 & 52.19 & 69.44 & 91.36 \\ \hline
Botnet & 98.69 & 98.47 & 63.14 & 71.82 & 98.38 & 98.07 \\ \hline
Web Brute   Force & 90.14 & 62.72 & 40.44 & 42.24 & 81.09 & 79.28 \\ \hline
Web XSS & 90 & 50.18 & 37.33 & 38.92 & 86.47 & 78.9 \\ \hline
Infiltration & 90 & 90 & N/A & N/A & 90.91 & 91.94 \\ \hline
Web Sql   Injection & 90 & 79.52 & N/A & N/A & 80 & 83.81 \\ \hline
Heartbleed & 100 & 100 & N/A & N/A & 100 & 100 \\ \hline
\end{tabular}
\end{table}

\begin{table}[!htbp]
\centering
\caption{Comparison of accuracy between the proposed model and an existing model on CIC-IDS2018}
\label{Table13: comparisonCICIDS2018}
\begin{tabular}{|l|c|c|c|c|}
\hline
Benign/Attack Categories & \multicolumn{1}{l|}{usfAD} & \multicolumn{1}{l|}{LOF} & \multicolumn{1}{l|}{\begin{tabular}[c]{@{}l@{}}DOC++\\ (Closed Set)\end{tabular}} & \multicolumn{1}{l|}{\begin{tabular}[c]{@{}l@{}}DOC++ \\ (Open Set)\end{tabular}} \\ \hline
Benign & 97.5 & 93.71 & 88.05 & 48.79 \\ \hline
DDOS   attack-HOIC & 98.8 & 97.91 & N/A & N/A \\ \hline
DDoS   attacks-LOIC-HTTP & 98.95 & 98.89 & 90.04 & 72.16 \\ \hline
DoS   attacks-Hulk & 95.92 & 98.95 & N/A & N/A \\ \hline
Botnet & 94.89 & 99.66 & 92.03 & 53 \\ \hline
Infilteration & 97.45 & 97.17 & 85.89 & 48.43 \\ \hline
SSH-Bruteforce & 99.36 & 99.68 & 92.37 & 42.45 \\ \hline
FTP-BruteForce & 97.99 & 99.89 & 92.41 & 41.13 \\ \hline
DoS   attacks-SlowHTTPTest & 99.86 & 99.9 & N/A & N/A \\ \hline
DoS   attacks-GoldenEye & 94.93 & 98.79 & 93.17 & 54.74 \\ \hline
DoS   attacks-Slowloris & 90.26 & 98.76 & 92.17 & 57.26 \\ \hline
DDOS   attack-LOIC-UDP & 90.06 & 32.66 & N/A & N/A \\ \hline
Brute Force -Web & 89.18 & 91.15 & 88.93 & 40.06 \\ \hline
Brute Force-XSS & 83.91 & 89.13 & N/A & N/A \\ \hline
SQL Injection & 95.56 & 100 & 82.34 & 35.47 \\ \hline
\end{tabular}
\end{table}

\section{Conclusion}
\label{Conclusion}

 In our work, we propose a novel two-stage framework that integrates both binary and multi-classification models. The first stage filters out benign samples using binary classification, reducing the number of samples that proceed to the second stage. The second stage then classifies the remaining malicious samples into unknown and known instances where a supervised learner classifies known attacks into their respective attack categories. At this level, we employ a DBSCAN which clusters together similar types of unknown attacks for retraining the second-level OCC model and supervised learner. Our research reveals that usfAD, a sophisticated OCC detection model developed recently, outperforms other state-of-the-art algorithms in identifying both known and novel attack patterns across most datasets. The efficacy of usfAD is significantly influenced by the accuracy of its initial model, which is designed to distinguish between normal and potential attacks. The retraining phase's effectiveness depends on the precision with which the clustering mechanism can categorize various new attack types. Through our analysis, we have determined that current implementations of the DBSCAN algorithm outshine alternative clustering methods. Despite this, there remains room for enhancement in clustering efficiency to boost group homogeneity. Future developments for usfAD aim to eliminate the need for the transformation of categorical data into numerical formats. This step is crucial, as label encoding may compromise data integrity, and one-hot encoding introduces a computationally demanding increase in data dimensionality.

\section*{Declarations}

\subsection*{Conflict of interest}
The authors have no conflicts of interest to declare that they are relevant to the content of this article.

\section*{Acknowledgments}
This material is based upon work supported by the Air Force Office of Scientific Research under award number FA2386-23-1-4003.

\section*{Author statements}
Md Ashraf Uddin: Conceptualization; Data curation; Implementation, Roles/Writing-original draft; and Writing, Visualization; Formal analysis.
Sunil Aryal: Funding acquisition; Investigation; Methodology; Project administration; Resources; Software; Supervision; Validation;  Roles/Writing-original draft; and Writing - review \& editing.
Mohamed Reda Bouadjenek: Conceptualization; Project administration;
Muna Al-Hawawreh: Review \& editing.
Md. Alamin Talukder: Data curation; Implementation; Visualization.

\bibliographystyle{els-article}
\bibliography{references}

\end{document}